\documentclass{article}
\usepackage[utf8]{inputenc}
\usepackage{amsmath}
\usepackage{amsthm}
\usepackage{graphicx}
\usepackage[unicode=true,
 bookmarks=true,bookmarksnumbered=false,bookmarksopen=false,
 breaklinks=false,pdfborder={0 0 1},backref=false,colorlinks=false]
 {hyperref}
\hypersetup{pdftitle={Time dispersion and quantum mechanics},
 pdfauthor={John Ashmead},
 pdfsubject={Experimental tests of quantum mechanical behavior along the time dimension},
 pdfkeywords={time, quantum electrodynamics, quantum mechanics, relativity, experiments}}

\usepackage{cancel}

\pdfoutput=1

\begin{document}
\title{Time dispersion in quantum electrodynamics}
\author{John Ashmead}
\maketitle
\begin{center}
Visiting Scholar, University of Pennsylvania, USA
\par\end{center}

\begin{center}
\textsl{jashmead@seas.upenn.edu}
\par\end{center}
\begin{abstract}
If we use the path integral approach, we can write quantum electrodynamics
(QED) in a way that is manifestly relativistic. However the path integrals
are confined to paths that are on mass-shell. What happens if we extend
QED by computing the path integrals over all paths in energy momentum
space, not only those on mass-shell? We use the requirement of covariance
to do this in an unambiguous way. This gives a QED where the time/energy
components appear in a way that is manifestly parallel to the space/momentum
components: we have dispersion in time, entanglement in time, full
equivalence of the Heisenberg uncertainty principle (HUP) in time
to the HUP in space, and so on. Entanglement in time has the welcome
side effect of eliminating the ultraviolet divergences. We recover
standard QED in the long time limit. We predict effects at scales
of attoseconds. With recent developments in attosecond physics and
in quantum computing, these effects should be detectable. Since the
predictions are unambiguous and testable the approach is falsifiable.
Falsification would sharpen our understanding of the role of time
in QED. Confirmation would have significant implications for attosecond
physics, quantum computing and communications, and quantum gravity.
\end{abstract}

\section{Introduction}
\begin{quotation}
\textit{\emph{``Look, I don't care what your theory of time is. Just
give me something I can prove wrong.}}'' -- Nathan Gisin at the
2009 Feynman Festival in Olomouc
\end{quotation}
\label{sec:Introduction}

\paragraph{Is quantum electrodynamics fully relativistic?}

\label{par:Is-quantum-electrodynamics}

Quantum electrodynamics (QED) can be developed in a large number of
ways. Perhaps the most common is the Hamiltonian/canonical momentum
approach. Time plays a special role in this formalism (in defining
the canonical momenta) so it is not clear that this approach is completely
relativistic. However the canonical momentum formalism is equivalent
to the Feynman path integral formulation. And in the Feynman path
integral formulation, QED is developed in a relativistic way. 

However perhaps even the Feynman path integral formulation is not
as fully relativistic as it might be. In it the paths are limited
to on mass-shell paths. Consider the simplest possible propagator,
the propagator for a massive spinless field:

\begin{equation}
\Delta\left({p_{\mu}}\right)=\frac{\imath}{{{p^{\mu}}{p_{\mu}}-{m^{2}}+\imath\varepsilon}}
\end{equation}
Consider the infinitesimal $\imath\epsilon$. We construct the Feynman
diagrams by doing integrals in four momentum $\int{{d^{4}}p}$ over
the propagators. The $\imath\epsilon$ identifies one of these four
integrals not as a normal but as a contour integral. Say we make the
contour integral the one over energy. When we look in detail at this
-- done in the text -- we see that for any fixed value of the three
momentum $\vec{p}$ the value of the fourth component $E$ is fixed
by the value of the residues at the poles, typically $E\to E_{\vec{p}}\equiv\sqrt{{m^{2}}+{{\vec{p}}^{2}}}$.
The effect is to fix the paths to only the on-shell paths.

We can generalize the paths to include off-shell paths as well. By
replacing the contour integral with a normal integral we can include
paths that vary in four dimensions. For instance we can write $E$
as $E_{\vec{p}}+\delta E$ and include paths which vary over all values
of $\delta E$. Letting the paths vary in all four dimensions simultaneously
is arguably more in keeping with the ``spirit'' of relativity, more
fully relativistic. 

But of course the question is not whether this is more fully in keeping
with the spirit of relativity but does including off-shell paths in
the path integrals give a more accurate description of nature? 

Our goal here is to put this question in a way that is falsifiable
with current technology. 

\paragraph{If dispersion in time/energy is real, why has it not already been
seen?}

\label{par:If-dispersion-real}

QED has been confirmed to extraordinary precision in a wide variety
of experiments, to the point where there is a wikipedia page on ``Precision
Tests of QED''. If such dispersion in energy (and therefore time)
is present, wouldn't we have already seen indications of this? 

The most obvious estimate of the scale at which such effects should
be seen is the Bohr radius $a_{0}$ divided by $c$; the time it would
take a photon to cross an atom. This is of order attoseconds: $\frac{{a_{0}}}{c}\approx.177a$s.
This is at the edge of current experimental technology so technically
within reach. But it is small enough that associated effects are unlikely
to be seen if not specifically looked for.

Factors that make it less likely that dispersion in time would be
seen by accident include:
\begin{enumerate}
\item Calculations in QED are normally done by taking the limit as time
goes to $\pm\infty$. This will naturally tend to obscure effects
at attosecond scale.
\item Averaging over many interactions -- i.e. shining beams against targets
-- will tend to average out effects in time.
\item It is not something which is expected, so therefore less likely to
be seen. The effects of dispersion in time might be hiding within
the error bars in some existing data sets.
\end{enumerate}

\paragraph{Objective}

\label{par:Objective}

What we are going to do here is to treat include off-shell and on-shell
paths on the same basis when computing the Feynman diagrams and see
what breaks. Do we encounter an unavoidable contradiction on the one
hand? or can we formulate experimental tests of this idea on the other?

Our objective is to force the question; to extend the paths in QED
off-shell in a way that is:
\begin{enumerate}
\item Manifestly covariant,
\item Consistent with observation and experiment,
\item Self-consistent,
\item Has no free parameters,
\item Falsifiable with current technology.
\end{enumerate}

\paragraph{Literature}

\label{par:literature}

The work here has its starting point in the path integral approach
as originated by Stueckelberg and Feynman \cite{Stueckelberg:1941aa,Stueckelberg:1941la,Feynman:1948,Feynman:1949sp,Feynman:1949uy,Feynman:1950rj}
and as further developed in \cite{Feynman:1961kp,Schulman:1981um,Rivers:1987ma,Swanson:1992ju,Kashiwa:1997xt,Huang:1998bd,Zinn-Justin:2005nx,Kleinert:2009hw,Feynman:2010bt}.
This work is specifically part of the Relativistic Dynamics approach
as developed by Horwitz, Fanchi, Piron, Land, Collins, and others
\cite{Horwitz:1973ys,Fanchi:1978aa,Fanchi:1993aa,Fanchi:1993ab,Land:1996aj,Horwitz:2005ix,Fanchi:2011aa,Horwitz:2015jk}.

We are also much indebted to general reviews of the role of time in
quantum mechanics: \cite{Dirac:1958ty,Pauli:1980wd,Schulman:1997fd,Zeh:2001xb,Muga:2002ft,Muga:2008vv,Callender:2017rt}.

And we have taken considerable advantage of the extraordinary literature
for QED. References particularly helpful here include \cite{Feynman:1998wv,Bjorken:1965el,Bjorken:1965mo,Sakurai:1967tl,Ramond:1990ab,Kaku:1993xj,Greiner:1994gf,Peskin:1995rv,Weinberg:1995hu,Weinberg:1995rl,Huang:1998bd,Greiner:2000vl,Itzykson:2005cv,McMahon:2008gt,Zee:2010oy,Klauber:2013tg,Lancaster:qy,Schwartz:2014tx,Schwichtenberg:2020vg}. 

We use the path integral formalism here. Texts on QED typically include
a chapter on path integrals. There is also a considerable literature
on them in their own right, as \cite{Schulman:1981um,Rivers:1987ma,Khandekar:1993ci,Swanson:1992ju,Kashiwa:1997xt,Zinn-Justin:2005nx,Kleinert:2009hw,Feynman:2010bt}.

In previous work we have looked at time dispersion in the single particle
case \cite{Ashmead_2019} (paper A) and at the specific problems created
in doing time-of-arrival measurements \cite{Ashmead:2021aa} (paper
B). The investigation here extends this work to QED. The extension
to QED is necessary to extend the results to high energies/short times,
critical for falsification.

\paragraph{Overview}

\label{par:Overview}
\begin{figure}
\includegraphics[scale=0.5]{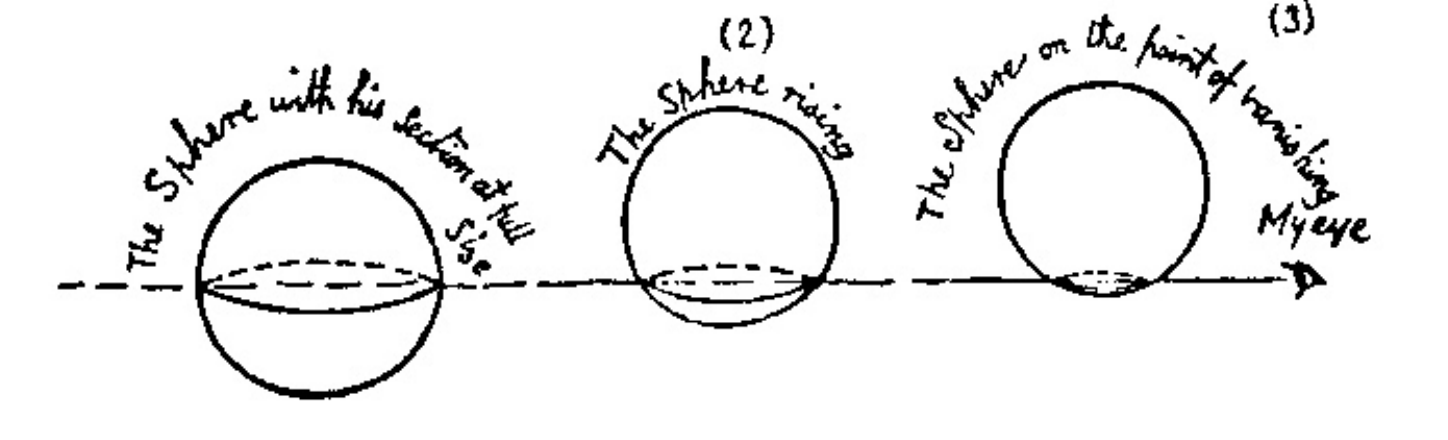}

\caption{Flatland: A Romance of Many Dimensions -- Edwin A. Abbott \cite{Abbott:1884fn}}
\end{figure}

In Edwin Abbott's charming 1884 romance Flatland, ``A Sphere'' --
a visitor to Flatland -- explains how three dimensions work to ``A
Square'', an inhabitant of Flatland. A Sphere uses rotational symmetry
to help take A Square from the idea of a circle or a square to the
idea of a sphere or a cube.

We take a similar approach here. We will start with the established
rules in standard quantum mechanics (SQM): quantum mechanics applied
along the three space dimensions, but time treated classically. We
use covariance to extend  the rules of SQM to include time as an observable
(TQM). 

In the early days of quantum mechanics, classical techniques were
extended to apply to quantum problems. For instance Feynman used the
classical Lagrangian to develop his sum over paths in space. We are
extending his approach to include paths in time as well.

Essentially we are completing the square between special relativity
and quantum mechanics, adding quantum effects to time on the one side,
adding time aspects to quantum mechanics on the other.

\paragraph{Strategy}

\label{par:Strategy}

We use path integrals. These are simple, they require only a few basic
ingredients: paths, a Lagrangian, a procedure for summing over the
paths weighted by the Lagrangian. 

In the single particle case, we promote the paths from three space
dimensions (3D) to time plus the three space dimensions (4D), while
leaving the Lagrangian unchanged. 

In QED we again take the paths -- now seen as successive values of
a field -- and promote them from being fields in three space to being
fields in four. We are again able to keep the Lagrangian and the rest
of the machinery of QED unchanged. In particular, we are able to develop
the Feynman rules in TQM in a way that is clearly parallel to the
rules in SQM. We get manifest covariance by construction.

As a result, conformity to existing results for SQM can be verified
in a straight-forward way. And we can easily pick out experimental
tests to look for the differences. 

The major sections are:
\begin{enumerate}
\item \emph{Time dispersion in QED}. We use the requirement of covariance
to extend the rules of quantum mechanics to include time on the same
basis as space. 
\item \emph{Applications}. We apply these rules to the free case, to the
simplest possible scattering case, to a simple mass correction loop,
and then to the simplest tree diagrams (Møller, Bhabha, and Compton
scattering).
\item \emph{Experimental tests}. We propose a specific experiment to test
the Heisenberg uncertainty principle in time/energy. The experiment
is non-trivial, but appears within reach of current technology. It
is, further, only one of many possible tests. 
\end{enumerate}
One surprising result is that the simple mass correction loop is convergent
without regularization. We might expect that with an additional dimension
to integrate over, the usual loop diagrams would become still more
divergent, perhaps even unrenormalizable. But instead the combination
of dispersion in time and entanglement in time keeps the loop diagrams
finite.

In general, any time dependent system monitored by time sensitive
detectors should show small but definite effects of dispersion in
time. In addition to the Heisenberg uncertainty principle (HUP) in
time/energy we can look at forces of ``anticipation and regret'',
diffraction in time, entanglement in time, corrections to existing
loop and bound state predictions at short times, and so on.

\paragraph{Summary}

\label{par:Summary}

In general, we expect to see the effects of dispersion in time at
scales of attoseconds and less. With recent developments in attosecond
physics and in quantum computing, these effects should now be visible.
The most dramatic are those involving the HUP in time. The hypothesis
is therefore falsifiable in practice.

Since the promotion of time to an operator is done by a straightforward
application of agreed and tested principles of quantum mechanics and
relativity, falsification will have implications for our understanding
of those principles.

Confirmation will have implications for attosecond physics, quantum
computing and communications, and quantum gravity.

\section{Time dispersion in Quantum Electrodynamics}
\begin{quotation}
``The rules of quantum mechanics and special relativity are so strict
and powerful that it's very hard to build theories that obey both.''
-- Frank Wilczek \cite{Wilczek:2008ol}
\end{quotation}
\label{sec:Time-dispersion-in-QED}

In this section we work out the rules for extending quantum electrodynamics
to include time as an operator. We work up the ladder of complexity
till we have all the necessary pieces in place. 
\begin{enumerate}
\item Time dispersion and the single particle. We work out the Schrödinger
equation for a single particle.
\item Spin zero propagator. We work out the Feynman propagator for a spin
zero particle with mass greater than zero.
\item Photon propagator. We work out the Feynman propagator for photons.
\item Dirac propagator. We work out the Feynman propagator for fermions.
\item Interactions. We note that the usual vertex terms are already TQM-ready.
\end{enumerate}
Because of the need to carefully distinguish between SQM and TQM versions
of otherwise familiar objects, some care is required in the notation.
We describe our choices in \ref{sec:Notations}.

\subsection{Time dispersion and the single particle}

\label{subsec:Single-particle} 

We start with the single particle case. The treatment here is largely
based on paper A, but reworked to prepare for QED.

We develop the path integral approach for the single particle: we
use the standard path integral approach, but with the usual paths
generalized from three to four dimensions.

\subsubsection{Clock time}

We start with clock time, defined operationally as what clocks measure:
Alice with a stop watch or perhaps the laboratory clock on the wall
or a carefully tended Cesium clock. This is also referred to as laboratory
time, as in Busch \cite{Busch-2001} and others. We reserve the letter
$\tau$ for this.

Note this is \emph{not} the proper time. For a single particle, the
use of proper time for the particle would give similar results to
those here. However there is no way to extend the proper time approach
to QED: the proper time for each massive particle will be in general
different, while for photons it will be identically zero.

We give an alternate definition of $\tau$ towards the end of this
subsection.

\subsubsection{Paths}

\label{subsec:Paths}

Normally in single particle path integrals the paths vary in space
but not in time. At each clock tick, the path will be assigned a specific
triad of space coordinates. To get the amplitude to go from a starting
point \emph{A} to an endpoint \emph{B} we will consider the set of
all paths from \emph{A} to \emph{B}, weighting each by the action.
We usually do the sum by breaking up the clock time from \emph{A}
to $B$ into \emph{N} time steps with each tick of size:

\begin{equation}
\varepsilon\equiv\frac{T}{N}\label{eq:slice-width-in-clock-time}
\end{equation}
At the end we will take the limit $N\to\infty$. The path is defined
by its space coordinates at each clock tick. To sum over the paths,
we sum over the associated measure:

\begin{equation}
\mathcal{D}\vec{x}\equiv\prod\limits _{n=0}^{N}{d{{\vec{x}}_{n}}}\label{eq:slice-metric}
\end{equation}
Now we extend the paths to include time:

\begin{equation}
\pi\left({{\vec{x}}_{\tau}}\right)\to\pi\left({{t_{\tau}},{{\vec{x}}_{\tau}}}\right)\label{eq:slice-measure-1}
\end{equation}
We refer to the time dimension used in this way as coordinate time
$t$, with its properties defined with respect to space by covariance.

The resulting paths are in four dimensions. They curve around in time,
so can dart into the future or the past. To be sure, the sum over
the paths is in general dominated by the classical paths, whose behavior
is more sedate. 

We extend the measure to include the sum over 4D paths:

\begin{equation}
\mathcal{D}x\equiv\prod\limits _{n=0}^{N}{d{t_{n}}d{{\vec{x}}_{n}}}\label{eq:4D-slice-measure}
\end{equation}

\subsubsection{Kernel}

\label{subsubsec:Kernel}

Our primary object is to compute the kernel to go from \emph{A} to
\emph{B}. This is given by the sum over all paths, weighted by the
action, defined as the integral of the Lagrangian along each path:

\begin{equation}
{K_{\tau}}\left({x'';x'}\right)=\int{\mathcal{D}{x_{\tau}}\exp\left({\imath\int\limits _{0}^{\tau}{d\tau'\mathcal{L}\left[{{x_{\tau}},{{\dot{x}}_{\tau}}}\right]}}\right)}\label{eq:4D-kernel}
\end{equation}

\paragraph{Choice of Lagrangian}

\label{par:Choice-of-Lagrangian}

We need a Lagrangian which is manifestly covariant, which correctly
models the behavior of a particle in an electro-magnetic field, and
which works equally well in 3D and 4D. We will use the following Lagrangian,
which we have from Goldstein \cite{Goldstein:1950dn} and also from
Feynman \cite{Feynman:1998wv}:

\begin{equation}
\mathcal{L}\left[{{x_{\tau}},{{\dot{x}}_{\tau}}}\right]=-\frac{1}{2}m{{\dot{x}}^{\mu}}{{\dot{x}}_{\mu}}-q{{\dot{x}}^{\mu}}{A_{\mu}}\left(x\right)-\frac{m}{2}\label{eq:4D-Lagrangian-1}
\end{equation}

\paragraph{Convergence of the calculation}

\label{par:Convergence-of-slices}

Path integrals are normally computed by starting at a specific time,
then integrating slice-by-slice. For this to make sense, the individual
integrals have to converge. 

Convergence is normally forced by adding small convergence factors,
i.e. rewriting the mass as $m\to m+\imath\epsilon$. Such tricks are
not usable here as they generally break covariance. However if we
are integrating against a Gaussian test function (GTF), these tricks
are not needed in the first place. The GTF itself will keep each step
convergent. GTFs are completely general: by using Morlet wavelet analysis
we can decompose any normalizable wavelet into sums over GTFs (further
discussed in \ref{sec:Gaussian-Test-Functions}). 

This give us convergence.

\paragraph{Result}

\label{par:single-Result}

In paper A \cite{Ashmead_2019} we derive the explicit form of the
free kernel:

\begin{equation}
{K_{\tau}}\left({x;x'}\right)=-\imath\frac{{m^{2}}}{{4{\pi^{2}}{\tau^{2}}}}{e^{-\frac{{\imath m}}{{2\tau}}{{\left({x-x'}\right)}^{2}}-\imath\frac{m}{2}\tau}}\label{eq:4D-kernel-in-coordinate-space}
\end{equation}
or in momentum space:

\begin{equation}
{K_{\tau}}\left({p;p'}\right)=\exp\left({\imath\frac{{{E^{2}}-{{\vec{p}}^{2}}-{m^{2}}}}{{2m}}\tau}\right){\delta^{4}}\left({p-p'}\right)\label{eq:single-4D-kernel}
\end{equation}
This matches the non-relativistic kernel found in introductory quantum
mechanics textbooks except that it now includes paths in time. From
the momentum form, we can see that deviations off-shell will be punished.

\subsubsection{Schrödinger equation}

\label{subsec:Schr=0000F6dinger-equation}

One normally derives the path integral formula from the Schrödinger
equation, see for instance Schulman or Kleinert \cite{Schulman:1981um,Kleinert:2009hw}.
However we can also start with the path integral formula and get the
Schrödinger equation by taking the short time limit of the path integral
expression and running their derivations ``in reverse''. We get:

\begin{equation}
\imath\frac{{\partial{\psi_{\tau}}}}{{\partial\tau}}\left({t,\vec{x}}\right)=-\frac{1}{{2m}}\left({\left({\imath{\partial_{\mu}}-q{A_{\mu}}\left({t,\vec{x}}\right)}\right)\left({\imath{\partial^{\mu}}-q{A^{\mu}}\left({t,\vec{x}}\right)}\right)-{m^{2}}}\right){\psi_{\tau}}\left({t,\vec{x}}\right)\label{eq:single-FS/T-x}
\end{equation}
or in momentum space:

\begin{equation}
-2m\imath\frac{{\partial{\psi_{\tau}}}}{{\partial\tau}}=\left({\left({{p_{\mu}}-q{A_{\mu}}}\right)\left({{p^{\mu}}-q{A^{\mu}}}\right)-{m^{2}}}\right){\psi_{\tau}}\label{eq:single-FS/T-p}
\end{equation}
This is formally identical to the Feynman-Stueckelberg equation in
the Relativistic Dynamics literature \cite{Feynman:1948,Fanchi:1993aa,Fanchi:1993ab,Land:1996aj,Horwitz:2015jk}. 

While the equation \ref{eq:single-FS/T-x} is formally the same as
the Feynman-Stueckelberg equation, the interpretation and use of the
equation here is distinct. We will therefore refer to this as the
FS/T: the Feynman-Stueckelberg equation in the TQM context. 

Note also the resemblance to the non-relativistic Schrödinger equation:

\begin{equation}
\imath\frac{{\partial\psi}}{{\partial\tau}}=\frac{1}{{2m}}{\left({\overrightarrow{p}-q\vec{A}}\right)^{2}}\psi\label{eq:non-rel-schroedinger-equation}
\end{equation}
The only difference is that we have added a term that represents dispersion
in time:

\begin{equation}
-\frac{1}{{2m}}\left({\left({\imath{\partial_{t}}-q\Phi\left({t,\vec{x}}\right)}\right)\left({\imath{\partial^{t}}-q\Phi\left({t,\vec{x}}\right)}\right)}\right){\psi_{\tau}}\left({t,\vec{x}}\right)\label{eq:coordinate-time-term}
\end{equation}
We will modify the FS/T slightly as part of the extension to QED below.

\subsubsection{Long, slow approximation}

\label{subsubsec:Long,-slow-approximation}

If the dependence on clock time is weak, we get the familiar Klein-Gordon
equation with minimal substitution:

\begin{equation}
\left({\imath{\partial_{\mu}}-q{A_{\mu}}}\right)\left({\imath{\partial^{\mu}}-q{A^{\mu}}}\right)\psi-{m^{2}}\psi=0\label{eq:klein-gordon-equivalent}
\end{equation}
Is it reasonable to assume that the dependence on clock time is weak?
That is that:

\begin{equation}
\imath\frac{{\partial\psi}}{{\partial\tau}}\approx0\label{eq:weak-dependence-on-clock-time}
\end{equation}
or more specifically that the expectation value of dependence on clock
time is small:

\begin{equation}
\left\langle \psi\right|\imath\frac{\partial}{{\partial\tau}}\left|\psi\right\rangle \approx0\label{eq:meaning-of-weak-dependence-on-clock-time}
\end{equation}

\paragraph{Effects of dependence on clock time of order picoseconds}

To see the relevant scale, we estimate the clock frequency $\varpi_{p}$:
\begin{equation}
\varpi_{p}\sim-\frac{{{E^{2}}-{{\vec{p}}^{2}}-{m^{2}}}}{{2m}}\label{eq:clock-frequency-nr}
\end{equation}
We are using $\varpi$ rather than $\omega$ for the clock frequency
to distinguish it clearly from the usual frequency $\omega$. We will
modify the definition of $\varpi$ slightly below, again as part of
the extension to QED (equation \ref{eq:clock-frequency}). 

In the non-relativistic case $E$ is of order mass plus kinetic energy:
\begin{equation}
E\sim m+\frac{{{\vec{p}}^{2}}}{{2m}}\label{eq:Non-relativistic-energy}
\end{equation}
so we have: 
\begin{equation}
{E^{2}}-{{\vec{p}}^{2}}-{m^{2}}\sim{\left({m+\frac{{{\vec{p}}^{2}}}{{2m}}}\right)^{2}}-{{\vec{p}}^{2}}-{m^{2}}={\left({\frac{{{\vec{p}}^{2}}}{{2m}}}\right)^{2}}\label{eq:ke-squared}
\end{equation}
This is just the kinetic energy, squared. In an atom the kinetic energy
is of order the binding energy: 
\begin{equation}
\frac{{{\vec{p}}^{2}}}{{2m}}\sim eV\label{eq:ke-in-ev}
\end{equation}
So the numerator is of order $eV$ squared. But the denominator is
of order $MeV$. Therefore we can estimate the clock frequency $\varpi_{p}$
as: 

\begin{equation}
\varpi_{p}\sim\frac{\left(eV\right)^{2}}{{MeV}}\sim{10^{-6}}eV\label{eq:estimate-clock-frequency}
\end{equation}
Energies of millionths of an electron volt ${10^{-6}eV}$ correspond
to times of order millions of attoseconds $10^{6}as$ or picoseconds,
a million times longer than the natural time scale of the effects
we are looking at. Therefore the long, slow approximation (LSA) is
reasonable.

Over long times, the clock frequency term will tend to reinforce on-shell
components of the wave function with respect to the off-shell components.
It is not so much that the off-shell components vanish, it is that
averaged over nanoseconds, as by a slow detector, the off-shell components
will average out to approximately zero. We then get what may look
like a long, slow collapse of the wave function.

Over short times, we will treat the effects of the dependence on clock
time as relatively less significant.

\subsubsection{Meaning of laboratory time}

\label{subsec:Meaning-of-laboratory-time}

We now have two different kinds of time in play: coordinate time and
clock time. We can reduce the ontological overhead of TQM by combining
them. To do this, we take the clock time as the average over coordinate
time over the rest of the universe $\mathcal{U}$:

\begin{equation}
\tau\equiv\langle\mathcal{U}|t\left|\mathcal{U}\right\rangle \label{eq:clock-time-as-expectation-of-coordinate-time}
\end{equation}
We are effectively dividing the wave function of the universe into
two parts, the small part we are focused on and the large part which
is us, the laboratory, and the rest of the universe. We now re-define
the clock time as the expectation of the coordinate time of the large
part.

Therefore the properties of the clock time are those associated with
an expectation value over an Avogadro's number of particles. In particular,
it does not go backwards, as such fluctuations are wildly unlikely
for the usual statistical dynamics reasons. As with a crowd, composed
of individuals, but with the dynamics of the crowd very different
from the dynamics of the individual. So we have:
\begin{enumerate}
\item Defined clock time in terms of laboratory clocks.
\item Defined the extension of paths to coordinate time using clock time
and covariance.
\item Worked out rules for quantum mechanics with coordinate time. Coordinate
time is now an operator in the same way as the three space dimensions
are operators.
\item Then turned around and defined clock time as the coordinate time operator
applied to the laboratory -- and the rest of universe, if it comes
to that.
\end{enumerate}
With this, laboratory time is not only not an operator, it is not
even a parameter, it is merely an expectation value of the fundamental
operator $t$. Therefore Pauli's theorem \cite{Pashby:2014wu,Pauli:1980wd}
does not apply to it. We will continue to use clock time as a short
hand for equation \ref{eq:clock-time-as-expectation-of-coordinate-time}.

This is a significant variation of the work here from the literature
in the Relativistic Dynamics program. In that, the parameter we have
been calling clock time is an additional parameter which is introduced
because various other parts of the problem then become more tractable.
Here it is fixed: defined operationally by clocks and defined theoretically
as the average over the coordinate time. This eliminates $\tau$ as
a degree of freedom. The fewer the degrees of freedom the more falsifiable.

In TQM we have only one time, the coordinate time. The clock time
is derivative, useful as scaffolding to get the analysis started,
but dispensable once the analysis is in place. The clock time applies
in full force only to macroscopic ensembles. The coordinate time represents
the underlying reality.

\subsubsection{Choice of laboratory frame}

\label{subsec:Choice-of-laboratory-frame}

So we understand what is meant by laboratory time in Alice's lab.
But what if Bob is working in his laboratory moving at relativistic
speeds relative to Alice's? Whose time should we use?

If the speeds are not too great, we can argue the effects will be
of second order so may be neglected on a first attack.

However in the interests of achieving a clean and complete treatment
we note we can define an invariant reference frame, to whose judgments
both Alice and Bob must defer. (This is analogous to the way we can
work in the center-of-mass frame, take advantage of the resulting
simplicity, and then transform back to a specific laboratory frame
at the end.)

In \cite{Weinberg:1972un} Weinberg shows we can use Einstein's equations
of general relativity to define an appropriate energy-momentum tensor
of local spacetime. See \ref{sec:rest-frame-vacuum} for specifics.

Since this is an energy-momentum tensor, we can use it to define a
``local rest frame of spacetime'' or $\mathcal{V}$ (for vacuum)
frame. We take $\mathcal{V}$ as the required common frame. The defining
laboratory time is therefore the clock time in this frame. Alice and
Bob can agree on this, then perform the necessary Lorentz transforms
from and to their respective frames confident they will make the same
physical predictions.

\subsection{Spin zero propagator}

\label{subsec:Spin-zero-propagator}

Having established a foundation in the single particle case, we extend
TQM to the case of a massive spin zero particle. This is the core
case for managing the transition from single particle quantum mechanics
to QED. The photon and fermion cases will turn out to be relatively
straightforward extensions of this.

We start with the SQM form. We use as a starting point the careful
and detailed treatment in Klauber's text \cite{Klauber:2013tg}, but
adapt his notation and techniques to the requirements of TQM. We give
only the key ``twists and turns''.

For SQM and then for TQM, we look at:
\begin{enumerate}
\item the free solutions and their associated Fock space, 
\item the field operators constructed as sums over the free solutions, 
\item the propagator constructed as a sum over the field operators.
\end{enumerate}

\subsubsection{Spin zero propagator in SQM}

\label{subsubsec:Neutral-spinless-particles-SQM}

We start with the Lagrangian for SQM:

\begin{equation}
\mathcal{L}^{S}\equiv{{\partial_{\tau}}\phi{\partial^{\tau}}\phi-\nabla\phi\nabla\phi-{m^{2}}{\phi^{\text{2}}}}\label{eq:spin-zero-sqm-lag}
\end{equation}
The corresponding Euler-Lagrange equation is the Klein-Gordon equation:

\begin{equation}
\left({{\partial_{\tau}}{\partial^{\tau}}-{\nabla^{2}}+{m^{2}}}\right)\phi_{\tau}\left({x}\right)=0\label{eq:spinzero-sqm-kg-x}
\end{equation}
The free solutions of this are:

\begin{equation}
\phi_{\tau}^{\left({\vec{k}}\right)}\left({\vec{x}}\right)\sim\exp\left(-\imath{\omega_{\vec{k}}}\tau+{\imath\vec{k}\cdot\vec{x}}\right),{\omega_{\vec{k}}}\equiv\sqrt{{m^{2}}+{{\vec{k}}^{2}}}
\end{equation}

\paragraph{Fock space}

The corresponding spin zero Fock space is built up in the usual way
as appropriately symmetrized combinations of the free single particle
solutions:

\begin{equation}
{\phi_{\vec{k}}}\left({\vec{x}}\right)=\frac{1}{{\sqrt{V}}}\exp\left({\imath\vec{k}\cdot\vec{x}}\right)
\end{equation}
We are using box normalization to a volume $V$ here. This is useful
for dimension checking. We will shift back and forth freely between
box and continuous normalization.

We use the occupation number representation for Fock space:

\begin{equation}
\left|{\left\{ {n_{\vec{k}}}\right\} }\right\rangle \label{eq:spinzero-fock-sqm}
\end{equation}
where $n$ is an integer from zero to infinity and the wave functions
are fully symmetric. The creation and annihilation operators are defined
by their effects on Fock space:

\begin{equation}
{a_{\vec{k}}}\left|{n_{\vec{k}}}\right\rangle =\sqrt{{n_{\vec{k}}}}\left|{{n_{\vec{k}}}-1}\right\rangle ,a_{\vec{k}}^{\dag}\left|{n_{\vec{k}}}\right\rangle =\sqrt{{n_{\vec{k}}}+1}\left|{{n_{\vec{k}}}+1}\right\rangle 
\end{equation}
with the 3D commutators being:

\begin{equation}
\left[{{a_{\vec{k}}},a_{\vec{k}'}^{\dag}}\right]={\delta^{3}}\left({\vec{k}-\vec{k}'}\right)\label{eq:3D commutators}
\end{equation}
All other commutators are zero. We make \emph{no} use of the usual
interpretation of Fock space in terms of harmonic oscillators. The
creation and annihilation operators are defined entirely by their
effects on Fock space.

\paragraph{Field Operators}

\label{par:spinzero-sqm-Field-Operators}

Now we build up the spin zero field operators as sums over the free
single particle solutions in the interaction picture. We have the
sums over the positive frequency components on the left and negative
frequency components on the right:

\begin{equation}
\phi_{\tau}^{S}\left({\vec{x}}\right)=\sum\limits _{\vec{k}}{\frac{1}{{\sqrt{2V{\omega_{\vec{k}}}}}}\left({{a_{\vec{k}}}{\operatorname{e}^{-\imath{\omega_{\vec{k}}}\tau+\imath\vec{k}\cdot\vec{x}}}+a_{\vec{k}}^{{\text{\ensuremath{\dagger}}}}{\operatorname{e}^{\imath{\omega_{\vec{k}}}\tau-\imath\vec{k}\cdot\vec{x}}}}\right)}
\end{equation}
We mark SQM parts with a superscript $S$. The normalization factor
$\frac{1}{{\sqrt{2{\omega_{\vec{k}}}}}}$ corresponds to the convention
of normalizing beams to energy/volume (see for instance Feynman \cite{Feynman:1998wv}).

\paragraph{Feynman propagator}

\label{par:spinzero-sqm-Feynman-propagator}

The SQM Feynman propagator is defined as the time-ordered vacuum expectation
value of two of these field operators. This definition is key to evaluating
the $S$ matrix as a sum over Feynman diagrams: 

\begin{equation}
\imath\Delta_{{\tau_{x}}{\tau_{y}}}^{S}\left({\vec{x}-\vec{y}}\right)\equiv\left\langle {0\left|{T\left\{ {\phi_{{\tau_{x}}}^{S}\left({\vec{x}}\right),\phi_{{\tau_{y}}}^{S}\left({\vec{y}}\right)}\right\} }\right|0}\right\rangle 
\end{equation}
$T$ is the time-ordering operator; if $\tau_{y}<\tau_{x}$, then
the $y$ operator is on the right and vice versa. We break $\phi$
up into its positive and negative frequency parts:

\begin{equation}
\phi_{\tau}^{S}\left(\vec{x}\right)={\phi_{\tau}^{S+}}\left(\vec{x}\right)+{\phi_{\tau}^{S-}}\left(\vec{x}\right)\label{eq:positive and negative halves}
\end{equation}
In a vacuum expectation value, the only non-zero terms are those with
an annihilation operator $a$ on the left and a creation operator
$a^{\dag}$ on the right. As a result most of the terms vanish.

When ${\tau_{x}}>{\tau_{y}}$ the only non-zero term is:

\begin{equation}
\left\langle 0\right|{\phi_{x}^{S+}}\left(\vec{x}\right){\phi_{y}^{S-}}\left(\vec{y}\right)\left|0\right\rangle \label{eq:when tau greater than zero}
\end{equation}
By taking advantage of:

\begin{equation}
0=-\left\langle 0\right|{\phi_{y}^{S-}}\left(\vec{y}\right){\phi_{x}^{S+}}\left(\vec{x}\right)\left|0\right\rangle \label{eq:normal zero}
\end{equation}
we can rewrite this in terms of the commutator:

\begin{equation}
\imath\Delta_{xy}^{S+}\left({\vec{x}-\vec{y}}\right)=\left\langle 0\right|\left[{{\phi_{x}^{S+}}\left(\vec{x}\right),{\phi_{y}^{S-}}\left(\vec{y}\right)}\right]\left|0\right\rangle \label{eq:Feynman prop plus in terms of commutator}
\end{equation}
We expand the operators and use the commutators. To simplify the calculations,
we shift from discrete sums to continuous integrals by replacing $\Sigma\to\int,V\to{\left({2\pi}\right)^{3}}$.
We use the vacuum product $\left\langle {0}\mathrel{\left|{\vphantom{00}}\right.\kern-\nulldelimiterspace}{0}\right\rangle =1$.
We are left with a pure number:

\begin{equation}
\imath\Delta_{xy}^{S+}\left({\vec{x}-\vec{y}}\right)=\frac{1}{{{\left({2\pi}\right)}^{3}}}\int{d\vec{k}\frac{{e^{-\imath{\omega_{\vec{k}}}\tau_{xy}+\imath\vec{k}\cdot\left({\vec{x}-\vec{y}}\right)}}}{{2{\omega_{\vec{k}}}}}}\label{eq:3D Feynman prop as pure number}
\end{equation}
The same development on the negative frequency side gives:

\begin{equation}
\imath\Delta_{xy}^{S-}\left({\vec{x}-\vec{y}}\right)=\frac{1}{{{\left({2\pi}\right)}^{3}}}\int{d\vec{k}\frac{{e^{\imath{\omega_{\vec{k}}}\tau_{xy}-\imath\vec{k}\cdot\left({\vec{x}-\vec{y}}\right)}}}{{2{\omega_{\vec{k}}}}}}\label{eq:3D Feynman negative prop}
\end{equation}
We could also show this by interchanging $x\leftrightarrow y$.

We see clearly here that time and space are being treated differently:
there is no integral over the energy coordinate $\omega$; the value
of $\omega_{\vec{k}}$ is fixed by $\vec{k}$ rather than being allowed
to roam. This implies no dispersion in $\omega$ -- and therefore
none in time.

Combining the positive and negative parts we get the full spin zero
propagator in SQM:

\begin{equation}
\imath{\Delta_{xy}^{S}}\left({\vec{x}-\vec{y}}\right)=\frac{1}{{{\left({2\pi}\right)}^{3}}}\int{{d^{3}}\vec{k}\frac{{\operatorname{e}^{-\imath{\omega_{\vec{k}}}\tau_{xy}-\imath\vec{k}\cdot\left({\vec{x}-\vec{y}}\right)}}}{{2{\omega_{\vec{k}}}}}\theta\left(\tau_{xy}\right)}+\frac{{\operatorname{e}^{\imath{\omega_{\vec{k}}}\tau_{xy}+\imath\vec{k}\cdot\left({\vec{x}-\vec{y}}\right)}}}{{2{\omega_{\vec{k}}}}}\theta\left(-\tau_{xy}\right)\label{eq:spinzero-sqm-propagator}
\end{equation}
We refer to this as the ``unpacked form''. Effectively it carries
positive frequency components into the future; negative into the past. 

\begin{figure}
\includegraphics[scale=0.75]{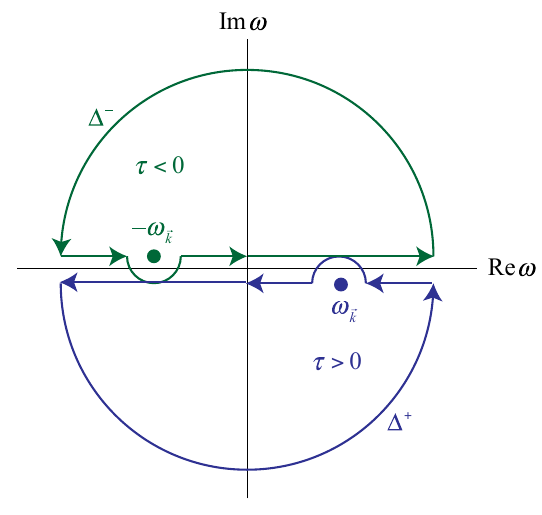}

\caption{Contour integral for Feynman propagators -- after Klauber \cite{Klauber:2013tg} }
\end{figure}

Now we turn this into something that is more covariant in appearance.
We start with the positive frequency side. We replace the integrand
with a contour integral: 

\begin{equation}
\frac{{e^{-\imath{\omega_{\vec{k}}}\tau}}}{{2{\omega_{\vec{k}}}}}\theta\left(\tau\right)=\frac{\imath}{{2\pi}}\int\limits _{-\infty}^{\infty}{d\omega\frac{{e^{-\imath\omega\tau}}}{{\left({\omega-{\omega_{\vec{k}}}+\imath\varepsilon}\right)\left({\omega+{\omega_{\vec{k}}}}\right)}}}
\end{equation}
with a pole at $\omega={\omega_{\vec{k}}-\imath\epsilon}$. This gives
the positive frequency side as:

\begin{equation}
\imath\Delta_{xy}^{S+}\left({\vec{x}-\vec{y}}\right)=\frac{{\imath}}{{{\left({2\pi}\right)}^{4}}}\int\limits _{-\infty}^{\infty}{d{\omega}\int{d\vec{k}}}\frac{{e^{-\imath{\omega}\tau_{xy}+\imath\vec{k}\cdot\left({\vec{x}-\vec{y}}\right)}}}{{\left({{\omega}-{\omega_{\vec{k}}}+\imath\varepsilon}\right)\left({{\omega}+{\omega_{\vec{k}}}}\right)}}\label{eq:positive tau contour integral}
\end{equation}
The negative frequency part can be replaced by a contour integral
in the same way. We take the pole at $\omega=-{\omega_{\vec{k}}}+\imath\varepsilon$:

\begin{equation}
\frac{{e^{\imath{\omega_{\vec{k}}}\tau}}}{{2{\omega_{\vec{k}}}}}\theta\left({-\tau}\right)=\frac{\imath}{{2\pi}}\int\limits _{-\infty}^{\infty}{d\omega\frac{{e^{-\imath\omega\tau}}}{{\left({\omega-{\omega_{\vec{k}}}}\right)\left({\omega+{\omega_{\vec{k}}}-\imath\varepsilon}\right)}}}
\end{equation}
Inserting this back in the previous expression we get:

\begin{equation}
\imath\Delta_{xy}^{S-}\left({\vec{x}-\vec{y}}\right)=\frac{{\imath}}{{{\left({2\pi}\right)}^{4}}}\int\limits _{-\infty}^{\infty}{d{\omega}\int{d\vec{k}}}\frac{{e^{-\imath{\omega}\tau_{xy}+\imath\vec{k}\cdot\left({\vec{x}-\vec{y}}\right)}}}{{\left({{\omega}-{\omega_{\vec{k}}}}\right)\left({{\omega}+{\omega_{\vec{k}}}-\imath\varepsilon}\right)}}\label{eq:negative tau contour integral}
\end{equation}
We could also just flip $x\leftrightarrow y$ again. Combined form
spelled out:

\begin{equation}
\imath\Delta_{xy}^{S}\left({\vec{x}-\vec{y}}\right)=\frac{\imath}{{{\left({2\pi}\right)}^{4}}}\int{{d\omega d}\vec{k}}\frac{{e^{-\imath{\omega}\tau_{xy}+\imath\vec{k}\cdot\left({\vec{x}-\vec{y}}\right)}}}{{{\omega^{2}}-\vec{k}^{2}-{m^{2}}+\imath\varepsilon}}\label{eq:spinzero-prop-3D}
\end{equation}
The momentum space part is:

\begin{equation}
\imath\Delta_{\omega}^{S}\left({\vec{k}}\right)=\frac{\imath}{{{\omega^{2}}-{{\vec{k}}^{2}}-{m^{2}}+\imath\varepsilon}}\label{eq:spinzero-sqm-prop-p}
\end{equation}
This is the propagator for the original Klein-Gordon equation. If
we apply the Klein-Gordon equation (\ref{eq:spinzero-sqm-kg-x}) in
momentum space we have:

\begin{equation}
\left({{\omega^{2}}-{{\vec{k}}^{2}}-{m^{2}}}\right)\Delta_{\omega}^{S}\left({\vec{k}}\right)=1
\end{equation}
The $+\imath\epsilon$ gives the propagator Feynman boundary conditions.

\subparagraph{Energy is only a virtual 4th dimension}

\label{par:energy-only-virtual-fourth-dimension}

So we can see that the propagator includes only on-shell paths. For
instance, fixing the momentum $\vec{k}$ at any point fixes the energy
$\omega$. The integral over $\omega$ sees only the poles in the
contour integral; it does not see the whole of $\omega$ space. Given
this, the usual practice of referring to the particles associated
with these propagators as virtual is correct. Our goal in building
TQM is, in a certain sense, to make them real.

\subsubsection{Spin zero propagator in TQM }

\label{subsubsec:spin-zero-TQM}

We do the same thing for TQM, replacing all 3D functions with 4D functions.
The dependence on clock time and the normalization will require a
bit of thought. We start with the TQM plane waves. They have four
coordinates rather than three:

\begin{equation}
{\phi_{w,\vec{k}}}\left({t,\vec{x}}\right)=\frac{1}{{\sqrt{TV}}}\exp\left({-\imath wt+\imath\vec{k}\cdot\vec{x}}\right)
\end{equation}
The $T$ is the length of a box in time, starting well before anything
interesting happens, and finishing well after everything interesting
is done. It represents box normalization in time. The Fock space is
built up of appropriately symmetrized products of these. It is given
in the occupation representation by:

\begin{equation}
\left|{\left\{ {n_{k}}\right\} }\right\rangle \label{eq:spinzero-fock-tqm}
\end{equation}
where $n$ is an integer from zero to infinity and the wave functions
are fully symmetric. 

The creation and annihilation operators are defined by their effects
in Fock space:

\begin{equation}
{a_{k}}\left|{n_{k}}\right\rangle =\sqrt{{n_{k}}}\left|{{n_{k}}-1}\right\rangle ,a_{k}^{\dag}\left|{n_{\vec{k}}}\right\rangle =\sqrt{{n_{k}}+1}\left|{{n_{k}}+1}\right\rangle 
\end{equation}
with 4D commutators:

\begin{equation}
\left[{{a_{k}},a_{k'}^{\dag}}\right]={\delta^{4}}\left({k-k'}\right)\label{eq:4D commutator}
\end{equation}
All other commutators are zero. Again, per discussion above, we make
\emph{no} use of the usual interpretation in terms of harmonic oscillators.

\paragraph{Field Operators}

\label{par:spinzero-tqm-Field-Operators}

Now we extend the SQM field operator to TQM. For a first cut we take
this as:

\begin{equation}
{\phi_{\tau}}\left({t,\vec{x}}\right)=\sum\limits _{w,\vec{k}}{\frac{1}{{\sqrt{TV}}}\frac{1}{{\sqrt{2{\omega_{\vec{k}}}}}}\left({{a_{w,\vec{k}}}{\operatorname{e}^{-\imath{\varpi_{k}}\tau-\imath wt+\imath\vec{k}\cdot\vec{x}}}+a_{w,\vec{k}}^{\dag}{\operatorname{e}^{\imath{\varpi_{k}}\tau+\imath wt-\imath\vec{k}\cdot\vec{x}}}}\right)}\label{eq:4D single operator}
\end{equation}

The two questions here are:
\begin{enumerate}
\item What should we use for $\varpi_{k}$?
\item And what should we use for $\frac{1}{{\sqrt{2{\omega_{\vec{k}}}}}}$?
This depends to a considerable extent on the answer to the first question,
so we will tackle that first.
\end{enumerate}

\subparagraph{Dependence on clock time}

\label{par:spinzero-dependence-on-clock-time}

We need a way to define the dependence on clock time -- the clock
energy -- in a way that works equally well for massive spin zero
particles, fermions, and photons. This is a non-trivial problem. For
instance, the previous formula for the single particle $\varpi_{\ensuremath{p}}$
has a $1/m$ factor which makes it unsuitable for use with photons.
It is also unclear how best to extend the single particle approach
to the Dirac equation: for instance, should antiparticles use the
same sign for $\varpi_{p}$ as particles do? We require an approach
which lets us treat all kinds of particles uniformly.

To do this, recall we are using the local rest frame of spacetime
as our reference point. What if we argue that the momentum of a particle
should be understood not as an absolute but as relative to the average
four momentum of the vacuum $\mathcal{P}$? 

We do this by replacing the particle's four momentum $k$ with its
four momentum relative to $\mathcal{P}$; $k\to k-\mathcal{P}$. With
this ansatz we rewrite the Klein-Gordon equation as:

\begin{equation}
\left({{{\left({k-\mathcal{P}}\right)}^{2}}-{m^{2}}}\right)\psi=0\label{eq:KG with vacuum-1}
\end{equation}
By working in the rest frame of local spacetime $\mathcal{V}$ (as
above) we reduce spacetime's four momentum to just its energy $\mathcal{P}\to\left({\mathcal{E},\vec{0}}\right)$.
We replace its energy $\mathcal{E}$ with its complementary time operator
$\mathcal{E}\to\imath\frac{\partial}{{\partial{\tau_{V}}}}$. The
laboratory time is defined as $\tau\equiv\left\langle t\right\rangle $
 so is \emph{also} going opposite to the time of the vacuum so $\imath\frac{\partial}{{\partial{\tau_{V}}}}\to-\imath\frac{\partial}{{\partial\tau}}$:

\begin{equation}
\left({{k^{2}}-{m^{2}}}\right)\psi=-2w\imath\frac{\partial}{{\partial\tau}}\psi\label{eq:spinzero-FS/T}
\end{equation}
We refer to this approach as the Machian hypothesis. We give a more
detailed treatment in \ref{subsec:vacuum-4D-Schr=0000F6dinger}. This
is the FS/T for the single particle case (equation \ref{eq:single-FS/T-p})
with the substitution $m\to w$ or in coordinate space $m\to\imath\frac{\partial}{\partial t}$.
Looking forwards, we will see this gives reasonable results for photons
and fermions as well. In energy momentum space we can write the clock
time dependence of the wave function as $\psi\sim\exp\left({-\imath{\varpi_{k}}\tau}\right)$
to get:

\begin{equation}
\imath\frac{\partial}{{\partial\tau}}\psi={\varpi_{k}}\psi,{\varpi_{k}}\equiv-\frac{{{w^{2}}-{{\vec{k}}^{2}}-{m^{2}}}}{{2w}}\label{eq:clock-frequency}
\end{equation}
In general we expect that off-shell components will tend to average
out: $\left\langle w\right\rangle \approx{\omega_{\vec{k}}}$. Therefore
we can expand the clock frequency in terms of $\delta w\equiv w-\omega_{\vec{k}}$:

\begin{equation}
{\varpi_{k}}\approx-\delta w+\frac{{{\left({\delta w}\right)}^{2}}}{{2{\omega_{\vec{k}}}}}\label{eq:spinzero-estimate}
\end{equation}
Using $\omega$ as the complementary variable to $\tau$, we write
in this equation in terms of clock frequency $\omega$ and coordinate
energy $w$:

\begin{equation}
\left({{w^{2}}-{{\vec{k}}^{2}}-{m^{2}}+2\omega w}\right)\psi=0\label{eq:spinzero-tqm-kg-omega}
\end{equation}
From the analysis of the LSA above, we expect that the $2\omega w$
term will have little effect at short times, only coming into its
own at longer times, i.e. on the ``legs'' of the Feynman diagrams. 

For the rest of the text, we will treat the Machian hypothesis as
a formal hypothesis useful for achieving a consistent treatment. In
\ref{subsec:Implications-for-quantum-gravity} we take a quick look
at some of the implications of treating this hypothesis as real.

\subparagraph{Normalization}

\label{par:spinzero-tqm-Normalization}

There are two requirements for the normalization. The first is that
the resulting propagator should be a propagator for (\ref{eq:spinzero-tqm-kg-omega}),
that in momentum space it should look like:

\begin{equation}
\imath{\Delta_{\omega}}\left(k\right)\sim\frac{\imath}{{{w^{2}}-{{\vec{k}}^{2}}-{m^{2}}+2\omega w}}
\end{equation}
The second is that it should obey Feynman boundary conditions, specifically
its dependence on clock time should go as:

\begin{equation}
\exp\left({-\imath{\varpi_{k}}\tau}\right)\theta\left(\tau\right)+\exp\left({\imath{\varpi_{k}}\tau}\right)\theta\left({-\tau}\right)
\end{equation}
This is required so that the construction of the $S$ matrix in TQM
will go in parallel to the construction in SQM. These two requirements
strongly constrain the normalization. We can meet them if we take:

\begin{equation}
\frac{1}{{\sqrt{2{\omega_{\vec{k}}}}}}\to\frac{1}{{\sqrt{2w}}}
\end{equation}
We therefore take as the TQM operator:

\begin{equation}
{\phi_{\tau}}\left({t,\vec{x}}\right)=\sum\limits _{w,\vec{k}}{\frac{1}{{\sqrt{TV}}}\frac{1}{{\sqrt{2w}}}\left({{a_{w,\vec{k}}}{\operatorname{e}^{-\imath{\varpi_{k}}\tau-\imath wt+\imath\vec{k}\cdot\vec{x}}}+a_{w,\vec{k}}^{\dag}{\operatorname{e}^{\imath{\varpi_{k}}\tau+\imath wt-\imath\vec{k}\cdot\vec{x}}}}\right)}
\end{equation}

The differences from the original guess are the normalization $\frac{1}{{\sqrt{2w}}}$
and the precise definition of $\varpi$.

\paragraph{Propagator}

\label{par:spinzero-tqm-Propagator}

We can now derive the unpacked form of the propagator in close parallel
to the derivation for SQM. The propagator is defined by:

\begin{equation}
\imath{\Delta_{xy}}\left({x-y}\right)=\left\langle 0\left|T\left\{ {{\phi_{x}}\left(x\right),{\phi_{y}}\left(y\right)}\right\} \right|0\right\rangle 
\end{equation}
We are using the same conventions and approach as for SQM, but generalizing
$\vec{x}\to x,\vec{y}\to y$.

We again break the wave function into its positive and negative frequency
parts:

\begin{equation}
{\phi_{x}}\left(x\right)=\phi_{x}^{+}\left(x\right)+\phi_{x}^{-}\left(x\right)
\end{equation}
As with SQM, for $\tau_{x}>\tau_{y}$ most terms are zero. We are
left with only:

\begin{equation}
\left\langle 0\left|\phi_{x}^{+}\left(x\right)\phi_{y}^{-}\left(y\right)\right|0\right\rangle 
\end{equation}
We rewrite this in terms of the commutator:

\begin{equation}
\imath\Delta_{xy}^{+}\left({x-y}\right)=\left\langle 0\left|\left[{\phi_{x}^{+}\left(x\right),\phi_{y}^{-}\left(y\right)}\right]\right|0\right\rangle 
\end{equation}
which we write in turn as the integral:

\begin{equation}
\imath\Delta_{xy}^{+}\left({x-y}\right)=\frac{1}{{{\left({2\pi}\right)}^{4}}}\int{{d^{4}}k\frac{{e^{-\imath{\varpi_{k}}\tau_{xy}-\imath k\left({x-y}\right)}}}{2w}}
\end{equation}
As with SQM, for $\tau_{x}<\tau_{y}$ we can get the results for the
propagator by interchanging $x\leftrightarrow y$:

\begin{equation}
\imath\Delta_{xy}^{-}\left({x-y}\right)=\frac{1}{{{\left({2\pi}\right)}^{4}}}\int{{d^{4}}k\frac{{e^{\imath{\varpi_{k}}\tau_{xy}+\imath k\left({x-y}\right)}}}{2w}}
\end{equation}
We flip the sign of $k$ to line this up with the positive frequency
side. Since $\varpi_{k}$ is odd in $w$, it flips sign as well:

\begin{equation}
\imath\Delta_{xy}^{-}\left({x-y}\right)=-\frac{1}{{{\left({2\pi}\right)}^{4}}}\int{{d^{4}}k\frac{{e^{-\imath{\varpi_{k}}\tau_{xy}-\imath k\left({x-y}\right)}}}{2w}}
\end{equation}
The result is the full propagator:

\begin{equation}
\imath{\Delta{}_{xy}}\left({x-y}\right)=\frac{1}{{{\left({2\pi}\right)}^{4}}}\int{{d^{4}}k\frac{{\operatorname{e}^{-\imath{\varpi_{k}}\tau_{xy}-\imath k\left({x-y}\right)}}}{2w}\theta\left(\tau\right)}-\frac{{\operatorname{e}^{-\imath{\varpi_{k}}\tau_{xy}-\imath k\left({x-y}\right)}}}{2w}\theta\left({-\tau}\right)\label{eq:spinzero-tqm-prop-unpacked-x}
\end{equation}
The second term differs in overall sign and in the sign of the clock
frequency from the second term for SQM. 

We now rewrite the integrand in terms of a contour integral over $\omega$.
We do this first to match as closely as possible the development in
SQM and secondly to let us write the $S$ matrix expansion in a way
that gives us conservation not only of coordinate energy but also
of clock energy. We use the representations of the Heaviside unit
step function:

\begin{equation}
\begin{gathered}\theta\left(\tau\right)=\mathop{\lim}\limits _{\varepsilon\to{0^{+}}}-\frac{1}{{2\pi\imath}}\int\limits _{-\infty}^{\infty}{\frac{1}{{\omega+\imath\varepsilon}}}{e^{-\imath\omega\tau}}d\omega\\
\theta\left({-\tau}\right)=\mathop{\lim}\limits _{\varepsilon\to{0^{+}}}\frac{1}{{2\pi\imath}}\int\limits _{-\infty}^{\infty}{\frac{1}{{\omega-\imath\varepsilon}}}{e^{-\imath\omega\tau}}d\omega
\end{gathered}
\end{equation}
We write the propagator in terms of $\omega$:

\begin{equation}
\exp\left({-\imath{\varpi_{k}}\tau}\right)\theta\left(\tau\right)-\exp\left({-\imath{\varpi_{k}}\tau}\right)\theta\left({-\tau}\right)=\frac{\imath}{{2\pi}}\int\limits _{-\infty}^{\infty}{d\omega\exp\left({-\imath\omega\tau}\right)}\left({\frac{1}{{\omega-{\varpi_{k}}+\imath\varepsilon}}+\frac{1}{{\omega-{\varpi_{k}}-\imath\varepsilon}}}\right)\label{eq:key-prop}
\end{equation}
which implies:

\begin{equation}
\imath{\Delta_{\omega}}\left(k\right)=\frac{1}{2w}\left({\frac{\imath}{{\omega-{\varpi_{k}}+\imath\varepsilon}}+\frac{\imath}{{\omega-{\varpi_{k}}-\imath\varepsilon}}}\right)\label{eq:spinzero-tqm-packed}
\end{equation}
We multiply out the $2w$ in the denominator to get:

\begin{equation}
\imath{\Delta_{\omega}}\left(k\right)=\frac{\imath}{{{w^{2}}-{{\vec{k}}^{2}}-{m^{2}}+2w\omega+2w\imath\varepsilon}}+\frac{\imath}{{{w^{2}}-{{\vec{k}}^{2}}-{m^{2}}+2w\omega-2w\imath\varepsilon}}\label{eq:spinzero-tqm-packed-p}
\end{equation}
where both sides are inverses of equation \ref{eq:spinzero-tqm-kg-omega},
with Feynman boundary conditions. 

\subparagraph{Conservation of clock energy in TQM}

\label{par:Conservation-of-clock-energy}

The unpacked form makes the physical meaning more transparent: the
normalization is obvious, the direction in time is obvious, the fact
that the expression is a relativistic invariant is obvious. 

But the packed form does show the propagator as a function of $\omega$,
the Fourier transform of the clock time. The packed form always travels
with an implicit:

\begin{equation}
\exp\left({-\imath\omega\tau}\right)
\end{equation}
which will be used in the inverse Fourier transform back to clock
space. 

This has considerable practical advantages. At a typical vertex, if
we have for example an incoming external line, an outgoing line, and
a exchanged photon the associated packed propagators will give factors
of:

\begin{equation}
\exp\left({\imath{\varpi_{out}}\tau}\right)\exp\left({\mp\imath{\omega_{\gamma}}\tau}\right)\exp\left({-\imath{\varpi_{in}}\tau}\right)
\end{equation}
where the factors of $\imath\epsilon$ make sure the sign of the photon
part is correct. If we have an overall integral over clock time to
$\mp\infty$, then we get integrals of the form:

\begin{equation}
\int\limits _{-\infty}^{\infty}{d\tau}\exp\left({\imath{\varpi_{out}}\tau}\right)\exp\left({\mp\imath{\omega_{\gamma}}\tau}\right)\exp\left({-\imath{\varpi_{in}}\tau}\right)
\end{equation}
associated with each vertex. And after we have done all these integrals
we have conservation of clock energy at each vertex plus an overall
conservation of clock energy for the diagram. In SQM:

\begin{equation}
\delta\left({\sum{\Omega_{out}}-\sum{\Omega_{in}}}\right)
\end{equation}
in TQM:

\begin{equation}
\delta\left({\sum{\varpi_{out}}-\sum{\varpi_{in}}}\right)
\end{equation}
This works for both SQM and TQM; the math is the same. This is helpful
in practical calculations.

\emph{But it is also a crutch}: it depends in a critical way on being
able to take the limits of the integral over clock time to infinity,
and therefore limits the applicability of the $S$ matrix in SQM to
long times. If the limits are for short times, then the conservation
of clock energy will be at best approximate. This is troubling in
SQM. 

But it is not a problem in TQM: in TQM only the \emph{coordinate energy}
is real, the clock energy -- like its companion clock time -- is
ultimately a statistical variable. If it fluctuates a bit here and
there, well that is to be expected when you are dealing with statistical
variables.

We finesse this problem here by looking only at diagrams where the
limits of the clock time integrals may be taken to $\pm\infty$ but
which still allow a direct comparison of TQM to SQM.

\subsubsection{Long and short time scales}

\label{subsec:Long-and-short}The TQM propagator is significantly
more complex than the SQM equivalent. The problem is that we are treating
time at two different levels: the low level quantum realm where coordinate
time is fully symmetric with space and the higher level macroscopic
realm of clock time, laboratories, and observers. The propagator,
like its complementary equation, is a bridge between two disparate
realms of analysis. In practice this can be difficult to work with. 

From the LSA the clock term:
\begin{enumerate}
\item is small: $\left|\varpi_{k}\right|<<\omega_{\vec{k}}$, 
\item averages to zero: $\left\langle \varpi_{k}\right\rangle \approx0$,
and 
\item only takes effect over longer terms i.e. picoseconds. 
\end{enumerate}
At short times we expect it will not play much of a role, Therefore
it is convenient to split the analysis into long (picoseconds) and
short (attosecond) times (with femtosecond times left for negotiation).

\paragraph{Long time scales}

\label{par:Long-time-scales}

It will normally take quite a few picoseconds for a wave packet to
get from the interaction zone to the detector, arguably enough time
for the clock frequency to play a significant role in shaping the
wave packet. For instance, if there is decoherence en route, the on-shell
terms will be preferentially favored over the off-shell. And if the
detector itself is not sensitive to sub-picosecond changes, the detector
will be unlikely to see off-shell components of the wave function. 

We have therefore a natural way to understand how the wave packet
evolves from what is initially a fully four dimensional wave packet
(as it leaves the interaction zone) to what appears to be on-shell
at it registers at a detector.

\paragraph{Short time scales}

\label{par:Short-time-scales}

In SQM the combination of clock frequency and clock time give a clear
direction in time: $\omega>0\Leftrightarrow\theta\left(\tau\right),\omega<0\Leftrightarrow\theta\left({-\tau}\right)$.
But in TQM, at sub-picosecond times, we have $\exp\left({-\imath{\varpi_{k}}\tau}\right)\approx1$
and the clock frequency approximately zero, as likely to be negative
as positive. The clock time/clock frequency pair no longer provides
reliable directionality in time.

\emph{Nor should it.} 

The clock time is defined as the expectation over the coordinate time,
only valid at longer times and for statistical assemblies. Backported
to extremely short times and small numbers of individual particles,
the use of clock time is suspect. Just as it is improper to infer
from the macroscopic behavior of a gas the details of the motion of
a specific molecule within it.

Recall our fundamental hypothesis, that coordinate time is to be completely
defined by covariance and the rules for the three space dimensions.
In SQM the expectation of the three momentum gives the direction in
space: $\left\langle {p_{x}}\right\rangle >0$ implies we are going
in the positive $x$ direction, $\left\langle {p_{x}}\right\rangle <0$
that we are going in the negative $x$ direction, and so on. Therefore
if we have $\left\langle w\right\rangle >0$ we should be going forwards
in time; if $\left\langle w\right\rangle <0$ backwards in time. 

(Admittedly this latter case is perhaps less often seen in the laboratory.
For discussions of what this might look like in practice see Schulman
\cite{Schulman:1999aa} and also Greenberger and Svozil \cite{Greenberger-2005}).

So at short times, the clock time/clock frequency should have little
or nothing to do with the direction in time. That should be defined
by the wave function itself. 

Therefore at short times, we approximate the exponential of the clock
time as one. Since the left side is forwards in clock time and the
right side backwards, we add the two to get the short-time propagator.
(We take $\theta\left(0\right)\equiv1/2$ to have exactly $\theta\left(\tau\right)+\theta\left({-\tau}\right)=1$).

We therefore take as our propagator for attosecond times:

\begin{equation}
\imath\Delta_{\omega}^{{\text{A}}}\left(k\right)\approx\frac{\imath}{{{w^{2}}-{{\vec{k}}^{2}}-{m^{2}}}}\label{eq:spinzero-atto-prop}
\end{equation}
With space time form:

\begin{equation}
\imath\Delta_{\tau}^{{\text{A}}}\left(x\right)=\frac{\imath}{{{\left({2\pi}\right)}^{5}}}\int{{d^{4}}kd\omega\frac{{\operatorname{e}^{-\imath\omega\tau-\imath kx}}}{{{w^{2}}-{{\vec{k}}^{2}}-{m^{2}}}}}
\end{equation}
We will refer to this as the attosecond propagator, tagging it by
a superscript $A$ to make this clear. We expect it will start to
fail at picosecond and greater times.

So at attosecond times, we have \emph{no} imposed direction in time.
We have no dependence in the propagator on clock time, not even via
an $\imath\epsilon$. And we have a natural map to the SQM propagator:
$w\to\omega$.

The attosecond propagator directly addresses the question posed in
the introduction: what do we get if we apply the replacement  ${\omega_{\vec{k}}}\to w$
to the Feynman propagators? And replace the contour integrals with
real ones (by dropping the $\imath\epsilon$'s)? 

\paragraph{Quantum energy and quantum time}

\label{par:Quantum-energy-time}

We define the ``quantum energy'' as the difference between the coordinate
energy and the energy value (the classical energy) associated with
the parallel SQM calculation. For free particles this is $\delta w\equiv w-\omega_{\vec{k}}$.
More generally we can write the quantum energy as the coordinate energy
less the value expected from SQM: $\delta E\equiv E-{E^{S}}$. This
latter definition works within Feynman diagrams as well, where $E^{S}$
is the ``virtual energy'' or the energy associated with a virtual
particle.

We define in parallel the ``quantum time'' as the difference between
the coordinate time and the clock time: $\delta t\equiv t-\tau$. 

We are primarily focused on the quantum energy here, but the quantum
time has its uses as well. Both serve as measures of the difference
between TQM and SQM. And in that sense summarize the effect we are
looking for.

\paragraph{Summary}

\label{par:long-and-short-Summary}

At sub-picosecond times we can use the attosecond time propagator.
On the legs, at longer times, we will use the FS/T equation (\ref{eq:spinzero-FS/T})
and the associated single particle solutions.

This division makes sense when we are looking at high speed scattering
experiments, where the interaction zone is at attosecond scale, but
then the products of the interactions take journeys that can be nanoseconds
or longer. 

In more complex cases we may need to fall back on the full propagator.

Next we develop TQM versions of photon and fermion propagators. The
polarization and spin parts will turn out to be relatively minor complications
from a TQM point-of-view; we have just navigated the trickier parts
of the analysis. Once these are ready, we will turn to applications.

\subsection{Photon propagator}

\label{subsec:photon-propagator} 

We use the same approach here as for spin zero case. The addition
of polarization turns out to be an inessential complication from the
point of view of TQM.

\subsubsection{Photons in SQM}

\label{subsec:Photons-in-SQM}

\paragraph{Fock space}

\label{par:photons-sqm-Fock-space}

We have as the basis functions:

\begin{equation}
\frac{{1}}{{\sqrt{2V{\rm \omega}_{\vec{k}}}}}\varepsilon_{r}^{{\rm \mu}}\left({\vec{k}}\right)\exp\left({-\imath{\rm \omega}_{\vec{k}}{\rm \tau}+\imath\vec{k}\cdot\vec{x}}\right)
\end{equation}

\begin{equation}
{\omega_{\vec{k}}}\equiv\left|{\vec{k}}\right|
\end{equation}
with the polarization vectors:

\begin{equation}
\begin{gathered}{\varepsilon_{1}}={\left({\begin{array}{cccc}
1 & 0 & 0 & 0\end{array}}\right)},{\varepsilon_{2}}={\left({\begin{array}{cccc}
0 & 1 & 0 & 0\end{array}}\right)},\hfill\\
{\varepsilon_{3}}={\left({\begin{array}{cccc}
0 & 0 & 1 & 0\end{array}}\right)},{\varepsilon_{4}}={\left({\begin{array}{cccc}
0 & 0 & 0 & 1\end{array}}\right)}\hfill
\end{gathered}
\label{eq:polarization vectors}
\end{equation}
We build up the associated Fock space from these as above; the only
difference is that the Fock space labels have a polarization index
as well:

\begin{equation}
\left|{\left\{ {m_{r\vec{k}}}\right\} }\right\rangle \label{eq:fock-photon-sqm}
\end{equation}
We have the usual creation and annihilation operators, indexed by
polarization as well as the three space momenta: ${{a}_{r\vec{k}}},a_{r'\vec{k}'}^{\dag}$.
Commutators:

\begin{equation}
\left[{{{a}_{r\vec{k}}},a_{r'\vec{k}'}^{\dag}}\right]={\delta_{rr'}}{\delta^{3}}\left({\vec{k}-\vec{k}'}\right)\label{eq:3D photon operators}
\end{equation}
All other commutators are zero.

\paragraph{Field Operators}

\label{par:photons-sqm-Field-Operators}

The field operators include a sum over the polarization vectors as
well as over the three space momenta:

\begin{equation}
A_{\tau}^{\left(S\right)\mu}\left({\vec{x}}\right)\equiv\sum\limits _{r,\vec{k}}{\frac{1}{{\sqrt{2V{\omega_{\vec{k}}}}}}\varepsilon_{r}^{\mu}\left({\vec{k}}\right)\left(\begin{gathered}{a_{r}}\left({\vec{k}}\right)\exp\left({-\imath{\omega_{\vec{k}}}\tau+\imath\vec{k}\cdot\vec{x}}\right)\hfill\\
+a_{r}^{\dag}\left({\vec{k}}\right)\exp\left({\imath{\omega_{\vec{k}}}\tau-\imath\vec{k}\cdot\vec{x}}\right)\hfill
\end{gathered}
\right)}
\end{equation}

\paragraph{Propagator}

\label{par:photons-sqm-Propagator}

The propagator is again defined as the vacuum expectation value of
the time ordered product of two field operators:

\begin{equation}
\imath D_{21}^{\left(S\right)\mu\nu}\left(\vec{x},\vec{y}\right)\equiv\left\langle 0\right|T\left\{ {{A_{x}^{\mu}}\left(\vec{x}\right),{A_{y}^{\nu}}\left(\vec{y}\right)}\right\} \left|0\right\rangle \label{eq:3D photon propagator}
\end{equation}
And using the same methods as earlier (see also the more detailed
treatment in \cite{Klauber:2013tg}) we get the unpacked SQM photon
propagator:

\begin{equation}
\imath D_{\tau}^{\left(S\right)\mu\nu}\left({\vec{x}}\right)=-\imath{g^{\mu\nu}}\int{\frac{{d\vec{k}}}{{2{\omega_{\vec{k}}}}}\left(\begin{gathered}\exp\left({-\imath{\omega_{\vec{k}}}\tau+\imath\vec{k}\cdot\vec{x}}\right)\theta\left(\tau\right)\hfill\\
+\exp\left({\imath{\omega_{\vec{k}}}\tau-\imath\vec{k}\cdot\vec{x}}\right)\theta\left({-\tau}\right)\hfill
\end{gathered}
\right)}
\end{equation}
We use the same approach as with the SQM spin zero case to rewrite
the propagator in terms of an integral over clock frequency:

\begin{equation}
\imath D_{\tau}^{\left(S\right)\mu\nu}\left({\vec{x}}\right)=\frac{{-\imath{g^{\mu\nu}}}}{{{\left({2\pi}\right)}^{4}}}\smallint d\omega d\vec{k}\frac{{\exp\left({-\imath{\omega}\tau+\imath\vec{k}\cdot\vec{x}}\right)}}{{\omega^{2}-{{\vec{k}}^{2}}+\imath\varepsilon}}\label{eq:prop-photon-sqm}
\end{equation}

\begin{equation}
\imath D_{\omega}^{\left(S\right)\mu\nu}\left({\vec{k}}\right)=\frac{{-\imath{g^{\mu\nu}}}}{{{\omega^{2}}-{{\vec{k}}^{2}}+\imath\varepsilon}}\label{eq:photons-sqm-prop-k}
\end{equation}
 The same comments about the virtual character of the SQM particles
earlier apply here as well.

\subsubsection{Photons in TQM}

\label{subsec:Photons-in-TQM}

From the perspective of TQM, the SQM photon is a hybrid of 3D and
4D approaches: the time coordinate is clock time, but the vector field
is basically a four dimensional object, requiring no adjustment to
promote it to TQM.

In Lorenz gauge the individual components of the vector potential
obey the Klein-Gordon equation. Therefore in TQM the application of
the Machian hypothesis gives:
\begin{equation}
-2w\imath\frac{\partial}{{\partial\tau}}{A^{\nu}}=\left({{w^{2}}-{{\vec{k}}^{2}}}\right){A^{\nu}}\label{eq:photon-FS/T}
\end{equation}
with 4D solutions:

\begin{equation}
A_{\tau}^{\left(k\right)\mu}\left(x\right)=\frac{{\varepsilon_{r}^{\mu}\left(k\right)\exp\left({-\imath{\varpi_{k}}\tau-\imath kx}\right)}}{{\sqrt{2TVw}}}
\end{equation}
and clock frequency as defined above (with the exception of no mass
term):

\begin{equation}
{\varpi_{k}}\equiv-\frac{{{w^{2}}-{{\vec{k}}^{2}}}}{{2w}}\label{eq:4D photon clock frequency}
\end{equation}
Note that the polarization part is the same in the SQM and TQM free
wave functions. 

\paragraph{Fock space}

\label{par:photons-tqm-Fock-space}

In TQM the Fock space is built up from the 4D solutions, appropriately
symmetrized:

\begin{equation}
\left|{\left\{ {m_{rk}}\right\} }\right\rangle \label{eq:fock-photon-tqm}
\end{equation}
The index $r$ to the polarization vectors is unchanged, but the three
vector $\vec{k}$ is promoted to a four vector $w,\vec{k}$. We have
the 4D commutation relations in complete parallel:

\begin{equation}
\left[{{a_{rk}},a_{r'k'}^{\dag}}\right]={\delta_{rr'}}{\delta^{4}}\left({k-k'}\right)
\end{equation}
All other commutators are zero.

\paragraph{Field Operators}

\label{par:photons-tqm-Field-Operators}

The field operator is similarly uncomplicated:

\begin{equation}
A_{\tau}^{\mu}\left({t,\vec{x}}\right)\equiv\sum\limits _{r,w,\vec{k}}{\frac{1}{{\sqrt{2TVw}}}\varepsilon_{r}^{\mu}\left({w,\vec{k}}\right)\left(\begin{gathered}{a_{rk}}\exp\left({-\imath{\varpi_{k}}\tau-\imath wt+\imath\vec{k}\cdot\vec{x}}\right)\hfill\\
+a_{rk}^{\dag}\exp\left({\imath{\varpi_{k}}\tau+\imath wt-\imath\vec{k}\cdot\vec{x}}\right)\hfill
\end{gathered}
\right)}
\end{equation}

\paragraph{Feynman propagator}

\label{par:photons-tqm-Feynman-propagator}

The Feynman propagator is the vacuum expectation value of the time
ordered product of the TQM vector potentials:

\begin{equation}
\imath D_{\tau}^{\mu\nu}\equiv\left\langle 0\left|T\left\{ {A_{\tau}^{\mu}\left(x\right),A_{\tau}^{\nu}\left(y\right)}\right\} \right|0\right\rangle \label{eq:4D photon propagator}
\end{equation}
By the same procedure as for spin zero (equation \ref{eq:spinzero-tqm-prop-unpacked-x}):

\begin{equation}
\imath D_{\tau}^{\mu\nu}\left(x\right)=-\frac{\imath{g^{\mu\nu}}}{{{\left({2\pi}\right)}^{4}}}\int{\frac{{dwd\vec{k}}}{{2w}}\left({\exp\left({-\imath{\varpi_{k}}\tau-\imath kx}\right)\theta\left(\tau\right)-\exp\left(-{\imath{\varpi_{k}}\tau+\imath kx}\right)\theta\left({-\tau}\right)}\right)}\label{eq:photons-prop-tqm-packed-x}
\end{equation}
in momentum space:

\begin{equation}
\imath D_{\tau}^{\mu\nu}\left(k\right)\equiv-\imath\frac{{g^{\mu\nu}}}{{{\left({2\pi}\right)}^{4}}}\left({\frac{{\exp\left({-\imath{\varpi_{k}}\tau}\right)\theta\left(\tau\right)}}{{2w}}-\frac{{\exp\left(-{\imath{\varpi_{k}}\tau}\right)\theta\left({-\tau}\right)}}{{2w}}}\right)
\end{equation}
We use the same approach as with the TQM spin zero case to rewrite
the propagator in terms of an integral over clock frequency:

\begin{equation}
\imath D_{\tau}^{\mu\nu}\left(k\right)\equiv\frac{{-\imath{g^{\mu\nu}}}}{{{\left({2\pi}\right)}^{5}}}\int\limits _{-\infty}^{\infty}{d\omega\frac{{\exp\left({-\imath\omega\tau}\right)}}{{2w}}}\left(\begin{gathered}\frac{\imath}{{{w^{2}}-{{\vec{k}}^{2}}+2w\omega+2w\imath\varepsilon}}\hfill\\
+\frac{\imath}{{{w^{2}}-{{\vec{k}}^{2}}+2w\omega-2w\imath\varepsilon}}\hfill
\end{gathered}
\right)
\end{equation}
In momentum space:

\begin{equation}
\imath D_{\omega}^{\mu\nu}\left(k\right)=\frac{{-\imath{g^{\mu\nu}}}}{{2w}}\left({\frac{\imath}{{{w^{2}}-{{\vec{k}}^{2}}+2w\omega+2w\imath\varepsilon}}+\frac{\imath}{{{w^{2}}-{{\vec{k}}^{2}}+2w\omega-2w\imath\varepsilon}}}\right)\label{eq:photons-tqm-prop-k}
\end{equation}
or: 

\begin{equation}
\imath D_{\omega}^{\mu\nu}\left(k\right)=-\imath{g^{\mu\nu}}{\Delta_{\omega}}\left(k\right)
\end{equation}
So the photon propagator is a polarization wrapper for the spin zero
propagator. This point considerably simplifies the subsequent analysis.
From the arguments above, we have also the attosecond form:

\begin{equation}
\imath D_{\omega}^{\left(A\right)\mu\nu}\left(k\right)=-\imath{g^{\mu\nu}}\frac{\imath}{{{w^{2}}-{{\vec{k}}^{2}}}}
\end{equation}

\subsection{Dirac propagator}

\label{subsec:Dirac-propagator}

We now extend TQM to include the Dirac propagator (Dirac \cite{Dirac:1928}). 

\subsubsection{Dirac propagator in SQM}

\label{subsec:Dirac-propagator-sqm}

\paragraph{Dirac equation}

\label{subsec:Dirac-equation-sqm}

We start by reviewing the Dirac equation in a way that will prepare
for the extension to TQM. Dirac equation in SQM:

\begin{equation}
\left({\imath{\gamma_{o}}{\partial_{\tau}}+\imath\vec{\gamma}\cdot\nabla-m}\right)\psi=0\label{eq:Dirac equation}
\end{equation}
We use standard choices for $\gamma^{\mu}$:

\begin{equation}
{\gamma^{0}}\equiv\left({\begin{array}{cc}
I & 0\\
0 & {-I}
\end{array}}\right),{\gamma^{1,2,3}}\equiv\left({\begin{array}{cc}
0 & {\sigma^{1,2,3}}\\
{-{\sigma^{1,2,3}}} & 0
\end{array}}\right)\label{eq:gamma zero}
\end{equation}
This has positive solutions (particles):

\begin{equation}
\psi_{1,2}^{\left(S\right)\vec{p}}\left({\tau,\vec{x}}\right)={u_{1,2}^{S}}\left({\vec{p}}\right)\exp\left({-\imath{E_{\vec{p}}}\tau+\imath\vec{p}\cdot\vec{x}}\right)
\end{equation}

\begin{equation}
{u_{1}^{S}}=\sqrt{\frac{{{E_{\vec{p}}}+m}}{{2m}}}\left({\begin{array}{c}
1\\
0\\
{\frac{{p^{3}}}{{{E_{\vec{p}}}+m}}}\\
{\frac{{{p^{1}}+\imath{p^{2}}}}{{{E_{\vec{p}}}+m}}}
\end{array}}\right),{u_{2}^{S}}=\sqrt{\frac{{{E_{\vec{p}}}+m}}{{2m}}}\left({\begin{array}{c}
0\\
1\\
{\frac{{{p^{1}}-\imath{p^{2}}}}{{{E_{\vec{p}}}+m}}}\\
{-\frac{{p^{3}}}{{{E_{\vec{p}}}+m}}}
\end{array}}\right)\label{eq:Dirac u}
\end{equation}
and negative solutions (anti-particles):

\begin{equation}
\psi_{3,4}^{\left(S\right)\vec{p}}\left({\tau,\vec{x}}\right)={v_{2,1}^{S}}\exp\left({\imath{E_{\vec{p}}}\tau-\imath\vec{p}\cdot\vec{x}}\right)
\end{equation}

\begin{equation}
{v_{2}^{S}}=\sqrt{\frac{{{E_{\vec{p}}}+m}}{{2m}}}\left({\begin{array}{c}
{\frac{{p^{3}}}{{{E_{\vec{p}}}+m}}}\\
{\frac{{{p^{1}}+\imath{p^{2}}}}{{{E_{\vec{p}}}+m}}}\\
1\\
0
\end{array}}\right),{v_{1}^{S}}=\sqrt{\frac{{{E_{\vec{p}}}+m}}{{2m}}}\left({\begin{array}{c}
{\frac{{{p^{1}}-\imath{p^{2}}}}{{{E_{\vec{p}}}+m}}}\\
{\frac{{-{p^{3}}}}{{{E_{\vec{p}}}+m}}}\\
0\\
1
\end{array}}\right)\label{eq:Dirac negative solutions}
\end{equation}
These solutions define a Fock space:

\begin{equation}
\left|{\left\{ {n_{s\vec{p}}}\right\} }\right\rangle \label{eq:fock-fermion-sqm}
\end{equation}
where ${n_{r\vec{p}}}$ the occupation number is 0 or 1; $s$ indexes
$u_{1},u_{2},v_{2},v_{1}$, and the product wave functions are anti-symmetric
under interchange. As with spin zero, we define creation and annihilation
operators to navigate Fock space. The ${c^{\dag}},c$'s create and
annihilate the particles; the ${d^{\dag}},d$'s the anti-particles.
Their anti-commutation relations are:

\begin{equation}
{\left[{{c_{s\vec{p}}},c_{s'\vec{p}'}^{\dag}}\right]_{+}}={\left[{{d_{s\vec{p}}},d_{s'\vec{p}'}^{\dag}}\right]_{+}}={\delta_{ss'}}{\delta^{3}}\left({\vec{p}-\vec{p}'}\right)\label{eq:anti-commutators}
\end{equation}
with all other anti-commutators zero.

\paragraph{Field Operators}

\label{par:fermions-sqm-Field-Operators}

We define the field operators in parallel with those in the spin zero
case, but with anti-commutators for commutators. The normalization
is given by the $\sqrt{\frac{m}{{V{E_{\vec{p}}}}}}$. $c,c^{\dagger}$
are the annihilation and creation operators for $u_{1,2}^{S}$; $d,d^{\dagger}$
the annihilation and creation operators for $v_{1,2}^{S}$. The resulting
field operator is:

\begin{equation}
{\psi_{\tau}^{S}}\left({\vec{x}}\right)=\sum\limits _{s=1}^{2}{\sum\limits _{\vec{p}}{\sqrt{\frac{m}{{V{E_{\vec{p}}}}}}}\left(\begin{gathered}{c_{s\vec{p}}}{u_{s}^{S}}\left({\vec{p}}\right)\exp\left({-\imath{E_{\vec{p}}}\tau+\imath\vec{p}\cdot\vec{x}}\right)\hfill\\
+d_{s\vec{p}}^{\dag}{v_{s}^{S}}\left({\vec{p}}\right)\exp\left({\imath{E_{\vec{p}}}\tau-\imath\vec{p}\cdot\vec{x}}\right)\hfill
\end{gathered}
\right)}\label{eq:fermion-sqm-spinor-operator}
\end{equation}
We define adjoints $\bar{u}\equiv{u^{\dag}}{\gamma_{0}},\bar{v}\equiv{v^{\dag}}{\gamma_{0}},\bar{\psi}\equiv{\psi^{\dag}}{\gamma_{0}}$:

\begin{equation}
{{\bar{\psi}^{S}}_{\tau}}\left({\vec{x}}\right)=\sum\limits _{s=1}^{2}{\sum\limits _{\vec{p}}{\sqrt{\frac{m}{{V{E_{\vec{p}}}}}}}\left(\begin{gathered}{d_{s\vec{p}}}{{\bar{v}^{S}}_{s}}\left({\vec{p}}\right)\exp\left({-\imath{E_{\vec{p}}}\tau+\imath\vec{p}\cdot\vec{x}}\right)\hfill\\
+c_{s\vec{p}}^{\dag}{{\bar{u}^{S}}_{s}}\left({\vec{p}}\right)\exp\left({\imath{E_{\vec{p}}}\tau-\imath\vec{p}\cdot\vec{x}}\right)\hfill
\end{gathered}
\right)}\label{eq:fermion-sqm-adjoint-operator}
\end{equation}

\paragraph{Dirac propagator}

\label{par:fermions-sqm-Feynman-propagator}

The derivation of the Feynman propagator runs in parallel to the derivation
for spin zero, except for replacing the simple wave functions with
spinors and the commutators with anti-commutators. We define the Feynman
propagator as time-ordered vacuum expectation value:

\begin{equation}
\imath S_{xy}^{S}\left({\vec{x}-\vec{y}}\right)\equiv\left\langle 0\left|T\left\{ {\psi_{{x}}^{S}\left({\vec{x}}\right),{{\bar{\psi}}^{S}}_{{y}}\left({\vec{y}}\right)}\right\} \right|0\right\rangle 
\end{equation}
Inside a vacuum sandwich, only the $c{c^{\dag}},d{d^{\dag}}$ terms
survive. This lets us rewrite the propagator in terms of the anti-commutators
using the same procedure as for the spin zero case. We have the particle
case when $\tau_{x}>\tau_{y}$:

\begin{equation}
\imath S_{{xy}}^{S+}\left({\vec{x}-\vec{y}}\right)=\left\langle 0\left|{\left[{\psi_{x}^{S+}\left({\vec{x}}\right),\bar{\psi}_{y}^{S-}\left({\vec{y}}\right)}\right]_{+}}\right|0\right\rangle 
\end{equation}
and the anti-particle case when $\tau_{y}>\tau_{x}$:

\begin{equation}
\imath S_{xy}^{S-}\left({\vec{x}-\vec{y}}\right)=-\left\langle 0\left|{\left[{\bar{\psi}_{y}^{S+}\left({\vec{y}}\right),\psi_{x}^{S-}\left({\vec{x}}\right)}\right]_{+}}\right|0\right\rangle 
\end{equation}
Then for the first term we have:

\begin{equation}
\imath S_{xy}^{S+}\left({\vec{x}-\vec{y}}\right)=\frac{m}{{{\left({2\pi}\right)}^{3}}}\int{d\vec{p}\sum\limits _{s=1,2}{{u_{s}^{S}}\left({\vec{p}}\right){{\bar{u^{S}}}_{s}}\left({\vec{p}}\right)\frac{{\exp\left({-\imath{E_{\vec{p}}}\tau_{xy}+\imath\vec{p}\cdot\left({\vec{x}-\vec{y}}\right)}\right)}}{{E_{\vec{p}}}}}}
\end{equation}
We use:

\begin{equation}
\frac{{{\cancel{p}}^{S}+m}}{{2m}}=\sum\limits _{s=1,2}{{u_{s}^{S}}\left({\vec{p}}\right){{\bar{u}}_{s}^{S}}\left({\vec{p}}\right)}\label{eq:uu-sum-sqm}
\end{equation}
to get the positive frequency part of the propagator as:

\begin{equation}
\imath S_{xy}^{S+}\left({\vec{x}-\vec{y}}\right)=\frac{1}{{{\left({2\pi}\right)}^{3}}}\int{d\vec{p}\left({{\cancel{p}}^{S}+m}\right){\frac{{\exp\left({-\imath{E_{\vec{p}}}\tau_{xy}+\imath\vec{p}\cdot\left({\vec{x}-\vec{y}}\right)}\right)}}{{2{E_{\vec{p}}}}}}}
\end{equation}
We define ${\cancel{p}^{S}}\equiv{\gamma_{0}}{E_{\vec{p}}}-\vec{\gamma}\cdot\vec{p}$
to distinguish this construction from $\cancel{p}\equiv{\gamma_{0}}E-\vec{\gamma}\cdot\vec{p}$,
needed for TQM.

For the negative frequency part we use:

\begin{equation}
\frac{{{{\cancel{p}}^{S}}-m}}{{2m}}=\sum\limits _{s=1,2}{{v_{s}^{S}}\left({\vec{p}}\right){{\bar{v}^{S}}_{s}}\left({\vec{p}}\right)}
\end{equation}
to get the negative frequency part of the propagator as:

\begin{equation}
\imath S_{xy}^{S-}\left({\vec{x}-\vec{y}}\right)=-\frac{1}{{{\left({2\pi}\right)}^{3}}}\int{d\vec{p}\left({{\cancel{p}}^{S}-m}\right)\frac{{\exp\left({\imath{E_{\vec{p}}}\tau_{xy}-\imath\vec{p}\cdot\left({\vec{x}-\vec{y}}\right)}\right)}}{{2{E_{\vec{p}}}}}}
\end{equation}
Putting the two pieces back together:

\begin{equation}
\imath{S_{xy}^{S}}\left({\vec{x}-\vec{y}}\right)=\frac{1}{{{\left({2\pi}\right)}^{3}}}\int{d\vec{p}\left(\begin{gathered}\left({{\cancel{p}}^{S}+m}\right)\frac{{\exp\left({-\imath{E_{\vec{p}}}{\tau_{xy}}+\imath\vec{p}\cdot\left({\vec{x}-\vec{y}}\right)}\right)}}{{2{E_{\vec{p}}}}}\theta\left(\tau_{xy}\right)\hfill\\
-\left({{\cancel{p}}^{S}-m}\right)\frac{{\exp\left({\imath{E_{\vec{p}}}{\tau_{xy}}-\imath\vec{p}\cdot\left({\vec{x}-\vec{y}}\right)}\right)}}{{2{E_{\vec{p}}}}}\theta\left({-\tau_{xy}}\right)\hfill
\end{gathered}
\right)}
\end{equation}
This is again the ``unpacked form'':

\begin{equation}
\imath{S_{\tau}^{S}}\left({\vec{x}}\right)=\frac{\imath}{{{\left({2\pi}\right)}^{4}}}\int{d\omega d\vec{p}\frac{{{\cancel{p}}^{S}+m}}{{{\omega^{2}}-{{\vec{p}}^{2}}-{m^{2}}+\imath\varepsilon}}\exp\left({-\imath\omega\tau+\imath\vec{p}\cdot\vec{x}}\right)}\label{eq:prop-sqm-contour}
\end{equation}
As with the spin zero case, once the three space momenta have been
picked, the value of the energy component is forced: there is no dispersion
in energy.

The momentum space form is:

\begin{equation}
\imath S_{\tau}^{S}\left({\vec{p}}\right)=\imath\frac{{{\cancel{p}}^{S}+m}}{{{\omega^{2}}-{{\vec{p}}^{2}}-{m^{2}}+\imath\varepsilon}}\label{eq:fermions-sqm-prop-p}
\end{equation}
or:

\begin{equation}
\imath S_{\omega}^{S}\left({\vec{p}}\right)=\imath\left({{{\cancel{p}}^{S}}+m}\right)\Delta_{\omega}^{S}\left({\vec{p}}\right)\label{eq:fermions-sqm-prop-zero}
\end{equation}

\subsubsection{Dirac propagator in TQM}

\label{subsec:Dirac-particles-in-1}

\paragraph{Dirac equation}

\label{par:fermions-sqm-Dirac-equation}

We now add in the time coordinate. We again apply the Machian hypothesis,
this time to get the TQM form of the Dirac equation:

\begin{equation}
{\left({\cancel{p}-\cancel{\mathcal{\mathcal{P}}}-m}\right){\psi_{\tau}}=0}
\end{equation}
As in the spin zero case:

\begin{equation}
\left({\cancel{p}-m}\right){\psi_{\tau}}=\mathcal{P}{\psi_{\tau}},\mathcal{P}{\psi_{\tau}}\to{\mathcal{\mathcal{M}}}{\psi_{\tau}}\to\imath{\gamma_{0}}\frac{\partial}{{\partial{\tau_{V}}}}{\psi_{\tau}}\to-\imath{\gamma_{0}}\frac{\partial}{{\partial{\tau}}}{\psi_{\tau}}
\end{equation}
The result is the TQM form of the Dirac equation:

\begin{equation}
\left({\cancel{p}-m}\right){\psi_{\tau}}=-\imath{\gamma_{0}}\frac{\partial}{{\partial\tau}}{\psi_{\tau}}\label{eq:fermions-FS/T}
\end{equation}

\paragraph{Dirac propagator}

\label{par:fermions-tqm-Dirac-propagator}

To get the corresponding propagator we use the Machian hypothesis
to form an ansatz for the TQM propagator then plug this ansatz in
the equation to verify it is correct.

First we observe that Dirac's approach had its roots in the Klein-Gordon
equation: it was an attempt to get a version of the Klein-Gordon equation
which did not have negative energy solutions. We can see this directly
from the Dirac equation. Apply ${\cancel{p}+m}$ to it using $\cancel{p}\cancel{p}={p^{2}}$:

\begin{equation}
\left({\cancel{p}+m}\right)\left({\cancel{p}-m}\right)\psi=\left({{p^{2}}-{m^{2}}}\right)\psi=0
\end{equation}
This is still true in TQM:

\begin{equation}
\left({\left({\cancel{p}-\cancel{\mathcal{\mathcal{P}}}}\right)+m}\right)\left({\left({\cancel{p}-\cancel{\mathcal{\mathcal{P}}}}\right)-m}\right)\psi=\left({{{\left({p-\mathcal{P}}\right)}^{2}}-{m^{2}}}\right)\psi=0\label{eq:fermions-tqm-kg}
\end{equation}
We extend (\ref{eq:fermions-sqm-prop-p}) in the obvious way: replacing
3D with 4D objects and replacing $p$ with $p-\mathcal{\mathcal{P}}$:

\begin{equation}
\imath{S_{\tau}}\left(p\right)=\int{d\omega\frac{{\left({\cancel{p}-\cancel{\mathcal{\mathcal{P}}}}\right)+m}}{{\left({{{\left({p-\mathcal{\mathcal{P}}}\right)}^{2}}-{m^{2}}}\right)+\imath\varepsilon}}}\exp\left({-\imath\omega\tau}\right)
\end{equation}
We can see that this propagator is the inverse of the TQM Dirac equation
\ref{eq:fermions-tqm-kg}. 

To see more clearly what this means we again apply the Machian hypothesis,
this time to get $\omega$:

\begin{equation}
\mathcal{\mathcal{P}}\to\left({\mathcal{\mathcal{M}},\vec{0}}\right)\to\left({\imath\frac{\partial}{{\partial{\tau_{V}}}},\vec{0}}\right)\to\left({-\imath\frac{\partial}{{\partial\tau}},\vec{0}}\right)\to\left({-\omega,\vec{0}}\right)
\end{equation}
We get:

\begin{equation}
\imath{S_{\tau}}\left(p\right)=\int{d\omega\frac{{\left({\cancel{p}+{\gamma_{0}}\omega}\right)+m}}{{{E^{2}}-{{\vec{p}}^{2}}-{m^{2}}+2E\omega+{\omega^{2}}+\imath\varepsilon}}}\exp\left({-\imath\omega\tau}\right)
\end{equation}
We get a term proportional to $\omega$ in the numerator and terms
proportional to $\omega,\omega^{2}$ in the denominator. We drop the
$\omega^{2}$ term as on the grounds that if the $E\omega$ term is
small, the $\omega^{2}$ term is going to be exceptionally small.
The term linear in $\omega$ in the numerator will be replaced by
$\varpi$ when we do the contour integral. This in turn will average
to zero to lowest order. We will drop this term for now as well.

Therefore we take as the lowest order TQM propagator for fermions:

\begin{equation}
\imath{S_{\omega}}\left(p\right)=\imath\left(\begin{gathered}\frac{{\left({\cancel{p}+m}\right)}}{{{E^{2}}-{{\vec{p}}^{2}}-{m^{2}}+2\omega E+2E\imath\varepsilon}}\hfill\\
+\frac{{\left({\cancel{p}+m}\right)}}{{{E^{2}}-{{\vec{p}}^{2}}-{m^{2}}+2\omega E-2E\imath\varepsilon}}\hfill
\end{gathered}
\right)\label{eq:fermions-tqm-prop-p}
\end{equation}
or in ``unpacked'' form:

\begin{equation}
\imath{S_{\tau}}\left(p\right)=\frac{\imath}{{{\left({2\pi}\right)}^{4}}}\left({\frac{{\operatorname{e}^{-\imath{\varpi_{p}}{\tau}}}}{{2E}}\theta\left(\tau\right)-\frac{{\operatorname{e}^{-\imath{\varpi_{p}}\tau}}}{{2E}}\theta\left({-\tau}\right)}\right)
\end{equation}
From the arguments above, we have the attosecond form:

\begin{equation}
\imath S_{\omega}^{{\text{A}}}\left(p\right)=\imath\frac{{\left({\cancel{p}+m}\right)}}{{{E^{2}}-{{\vec{p}}^{2}}-{m^{2}}}}
\end{equation}
We have implicitly promoted ${E_{\vec{p}}}\to E$ in the various elements
of the spinor formalism:

\begin{equation}
u^{S}\left({\vec{p}}\right)\to u\left(p\right),v^{S}\left({\vec{p}}\right)\to v\left(p\right),\sqrt{\frac{m}{{E_{\vec{p}}}}}\to\sqrt{\frac{m}{E}}
\end{equation}

\subsection{Vertexes}

\label{subsec:Interactions}

\subsubsection{QED interaction in SQM}

\label{subsec:QED-interaction-SQM}

For interactions, we take the Fock space as the product of the photon
and fermion Fock spaces. Here it is enough to consider one species
of fermion, say electrons. For photon and electrons we have:

\begin{equation}
\left|{\left\{ {n_{s\vec{p}}^{e}}\right\} }\right\rangle \left|{\left\{ {m_{r\vec{k}}^{\gamma}}\right\} }\right\rangle 
\end{equation}

The interaction term is:

\begin{equation}
-e{{\bar{\psi}^{S}}_{\tau}}\left({\vec{x}}\right)A_{\tau}^{\left(S\right)\nu}\left({\vec{x}}\right){\gamma_{\nu}}{\psi_{\tau}^{S}}\left({\vec{x}}\right)
\end{equation}
The dependence on the coordinates is already absorbed into the definition
of the Feynman diagram, so this is reduced to sums over terms of the
form:

\begin{equation}
-e\bar{\psi}^{S}\left({\vec{p}'}\right)A_{\tau}^{\left(S\right)\nu}\left({\vec{k}}\right){\gamma_{\nu}}\psi^{S}\left({\vec{p}}\right)\label{eq:sqm-vertex}
\end{equation}
The $\psi^{S}\left({\vec{p}}\right)$ are built up over sums of the
${u_{s}^{S}}\left({\vec{p}}\right),{v_{s}^{S}}\left({\vec{p}}\right)$.
For an electron $e=-\left|e\right|$.

\subsubsection{QED interaction in TQM}

\label{subsec:QED-interaction-tqm}

The full Fock space in TQM is also a product of the fermion and photon
Fock spaces:

\label{subsec:QED-interaction-TQM}

\begin{equation}
\left|{\left\{ {n_{sp}^{e}}\right\} }\right\rangle \left|{\left\{ {m_{sk}^{\gamma}}\right\} }\right\rangle 
\end{equation}
The vertex term is:

\begin{equation}
-e{{\bar{\psi}}_{\tau}}\left({t,\vec{x}}\right)A_{\tau}^{\nu}\left({t,\vec{x}}\right){\gamma_{\nu}}{\psi_{\tau}}\left({t,\vec{x}}\right)\label{eq:4D vertex}
\end{equation}
Again, the dependence on the coordinates is already absorbed into
the definition of the Feynman diagram, so this is reduced to sums
over terms of the form:

\begin{equation}
-e\bar{\psi}\left({p'}\right)A_{\tau}^{\nu}\left(k\right){\gamma_{\nu}}\psi\left({p}\right)\label{eq:tqm-vertex}
\end{equation}
The $\psi\left({p}\right)$ are built up over sums of the ${u_{s}}\left(p\right),{v_{s}}\left(p\right)$.

\subsubsection{Conservation of momentum at a vertex}

\label{subsec:Conservation-of-momentum-at-vertex}

In SQM the integrals over the 3D plane waves give conservation of
three momentum at each vertex:

\begin{equation}
{\delta^{3}}\left({{{\vec{p}}_{out}}-{{\vec{p}}_{in}}}\right)
\end{equation}
As noted earlier, we get conservation of clock energy at a vertex
from the integrals over clock time. For this to work we need to take
the limit as $\tau\to\pm\infty$. If we are looking at short time
$S$ matrix elements -- legitimate if infrequent -- we will not
get exact conservation of clock energy at each vertex, but only an
approximation thereof, corresponding to \emph{not} taking the limit
in:

\begin{equation}
\mathop{\lim}\limits _{T\to\infty}\int\limits _{-T}^{T}{\frac{{d\tau}}{{2\pi}}\exp\left({-\imath\left({{\omega_{out}}-{\omega_{in}}}\right)\tau}\right)=}\mathop{\lim}\limits _{T\to\infty}\frac{{\sin\left({\left({{\omega_{out}}-{\omega_{in}}}\right)T}\right)}}{{\pi\left({{\omega_{out}}-{\omega_{in}}}\right)}}=\delta\left({{\omega_{out}}-{\omega_{in}}}\right)
\end{equation}
If the time range $T$ is sufficiently small the difference between
the $sin$ and the $\delta$ function could become noticeable. So
we see that in SQM the clock energy is not present on exactly the
same basis as the three space momenta.

To a reasonable first approximation, TQM may be thought of as SQM
with a fourth space dimension. So we get conservation of 4D momentum
at each vertex:

\begin{equation}
{\delta^{4}}\left({{p_{out}}-{p_{in}}}\right)
\end{equation}
If we able to take the limit as clock time goes to infinity, we get
a fifth conservation condition at each vertex and over the entire
Feynman diagram:

\begin{equation}
\delta\left({\sum{\varpi_{out}}-\sum{\varpi_{in}}}\right)
\end{equation}
but we can easily do without this at shorter times.

To keep the comparisons between SQM and TQM as straightforward as
possible, we will finesse this problem by looking primarily at the
scattering of GTFs which only intersect for very short times. We can
therefore extend the integrals to infinity at will since, except over
the short interaction zone, the integrand is zero. 

\subsubsection{Normalization factors}

\label{subsec:vertex-:Normalization-factors}

To go from SQM to TQM we changed the normalization on the field operators
from $\frac{1}{{\sqrt{2{\omega_{\vec{k}}}}}}$ to $\frac{1}{{\sqrt{2w}}}$
and from $\sqrt{\frac{m}{{E_{\vec{p}}}}}$ to $\sqrt{\frac{m}{E}}$. 

At any vertex, the field operators either connect to a propagator
or to an outside line. If they connect to a propagator their normalization
factor joins up with the normalization from the other side to give
normalizations which are accounted for in the calculation of the propagators:

\begin{equation}
\frac{1}{{\sqrt{2{\omega_{\vec{k}}}}}}\frac{1}{{\sqrt{2{\omega_{\vec{k}}}}}}\to\frac{1}{{2{\omega_{\vec{k}}}}},\frac{1}{{\sqrt{2w}}}\frac{1}{{\sqrt{2w}}}\to\frac{1}{{2w}},\dots
\end{equation}
If on the other hand they are connected to an external line the normalization
is passed to the overall definition of the S; we get the familiar
exterior factors, as:

\begin{equation}
\prod\limits _{}^{\begin{subarray}{l}
exterior\\
boson
\end{subarray}}{\sqrt{\frac{1}{{2{\omega_{\vec{k}}}}}}},\prod\limits ^{\begin{subarray}{l}
exterior\\
boson
\end{subarray}}{\sqrt{\frac{1}{{2w}}}},\dots
\end{equation}
In SQM there is a conventional understanding that on the legs the
virtual particles will drop off and we will be left with the true
3D on-shell wave function. But in TQM there is no such transition:
the wave packets on the legs are fully four dimensional in character. 

So ultimately, in TQM if we consider the ``Feynman diagram of the
universe'' we can divide it up into $S$ matrices as we please. Where
the legs of two $S$ matrices meet, the two associated exterior factors
will join to help form the propagator for the leg. There is no physical
transition from the system \emph{under} examination to the system
\emph{doing} the examination; it is a matter of convention and names
where and how we choose to make the divisions.

\section{Applications}
\begin{quotation}
``In other words, we are trying to prove ourselves wrong as quickly
as possible, because only in that way can we find progress.'' --
Richard P. Feynman \cite{Feynman:1965jb}
\end{quotation}
\label{sec:Applications}

With the basic tools built, we apply them to a starter set of cases:
\begin{enumerate}
\item Free particles. The creation, propagation, and detection of wave functions
which are dispersed in time presents specific questions.
\item Spin-zero particle exchange. This serves as the first real test of
how TQM plays out in practice.
\item Simple mass correction loop. We show that renormalization is not needed
in TQM; the combination of dispersion in time and entanglement in
time keep the loop integrals finite. 
\item The three basic second order scattering diagrams in QED:
\begin{enumerate}
\item Møller scattering. Electron-electron scattering by photon exchange. 
\item Bhabha scattering. Electron-positron scattering by photon exchange
and by photon creation/annihilation. 
\item Compton scattering. Electron-photon scattering. 
\end{enumerate}
\end{enumerate}
This starter set is sufficient to expose a fair number of experimental
possibilities; we work out the general Feynman rules in \ref{sec:Feynman-rules}.

\subsection{Free particles}
\begin{quotation}
“nothing can be created out of nothing” -- Lucretius \cite{Lucretius:1955ua}
\end{quotation}
\label{subsec:Free-particles}

To see the effects associated with dispersion in time we need to look
at wave functions which have finite dispersion in time/energy. Plane
waves and $\delta$ functions are equally unsuitable. Gaussian test
functions (\ref{sec:Gaussian-Test-Functions}) are well-suited as
models for this: they are easy to work with and they are completely
general -- any wave function may be represented as a sum over GTFs
using Morlet wavelet analysis. 

In SQM the use of GTFs is a convenience, but in TQM the use of GTFs
is mandatory: the convergence of the path integrals in the single
particle case and in loop diagrams depends on the use of GTFs.

To lay a correct foundation for the examination of scattering problems
in TQM we will need to look at the initial values, the propagation,
and the detection -- the birth, life and death -- of GTFs.

\subsubsection{Initial wave function}

\label{subsec:Birth}

\label{subsec:free-Initial-wave-function}

\label{subsubsection:entropic-estimate}

The first problem is how do we create a GTF of known dispersion in
time/energy? This problem is solved in paper A. There we showed that
if we are given the norm, average momentum, and average dispersion
in momentum we can get a maximum entropy estimate of the corresponding
GTF in energy using the method of Lagrange multipliers. 

As maximum entropy estimates tend to be robust, we have what we need
for falsifiability.

We summarize for use here the results of paper A. We assume we have
a description of the wave function in momentum space with the expectation,
average relativistic mass, and dispersion in relativistic mass defined:

\begin{equation}
\begin{array}{l}
\left\langle 1\right\rangle =1\\
\bar{E}\equiv\left\langle E\right\rangle =\sqrt{m^{2}+\left\langle \vec{p}\right\rangle ^{2}}\\
\left\langle E^{2}\right\rangle =\left\langle m^{2}+\vec{p}^{2}\right\rangle =m^{2}+\left\langle \vec{p}^{2}\right\rangle 
\end{array}\label{eq:constraints-1}
\end{equation}
$\bar{E}$ is the average relativistic mass.

By the method of Lagrange multipliers the corresponding GTF in energy
is:

\begin{equation}
\varphi_{0}\left(E\right)=\sqrt[4]{\frac{1}{\pi\sigma_{E}^{2}}}e^{\imath\left(E-\bar{E}\right)\tau_{0}-\frac{\left(E-\bar{E}\right)^{2}}{2\sigma_{E}^{2}}}\label{eq:wave function-in-energy}
\end{equation}

\begin{equation}
\sigma_{E}^{2}=2{\left({\Delta E}\right)^{2}}=2\left(\left\langle {E^{2}}\right\rangle -{\bar{E}^{2}}\right)
\end{equation}
With this estimate of the GTF in energy we can get the corresponding
GTF in time by taking the Fourier transform. In this case we have:

\begin{equation}
{\varphi_{0}}\left(t\right)\equiv\sqrt[4]{{\frac{1}{{\pi\sigma_{t}^{2}}}}}{e^{-\imath{\bar{E}}t-\frac{{{\left({t-{\tau_{0}}}\right)}^{2}}}{{2\sigma_{t}^{2}}}}},{\sigma_{t}}=\frac{1}{{\sigma_{E}}}
\end{equation}
The dispersions in energy are likely to be of order $eV$, if we are
starting with atomic wave functions. Assuming we are starting with
minimum uncertainty packets (expected) the uncertainty in time will
be the inverse of this:

\begin{equation}
\Delta t=\frac{1}{{\Delta E}}\label{eq:initial-estimate}
\end{equation}
and therefore of order attoseconds. 

We will refer to this as the ``entropic estimate''. It will not
pick up any complex structure: to the entropic estimate, everything
looks like an $s$ state. But it should get the order-of-magnitude
of the time part of the wave function right, which is the critical
marker for falsifiability. If no more specific estimate is available,
it should do.

\subsubsection{Propagation of free wave functions in clock time}

\label{subsec:Life}

In general, the three or four dimensional GTFs in momentum space are
solutions of the corresponding free equation with appropriate choices
of the clock frequency. If we have the wave function as time zero
as a function of momentum then we have it at time $\tau$ later. In
SQM:

\begin{equation}
\varphi_{\tau}^{S}\left({\vec{p}}\right)=\varphi_{0}^{S}\left({\vec{p}}\right)\exp\left({-\imath{E_{\vec{p}}}\tau}\right)
\end{equation}
In TQM:

\begin{equation}
{\varphi_{\tau}}\left({E,\vec{p}}\right)={\varphi_{0}}\left({E,\vec{p}}\right)\exp\left({-\imath{\varpi_{p}}\tau}\right)
\end{equation}
These two results are exact.

The square root in the definition of the relativistic mass for SQM
and the $1/2w$ factor in the definition of the clock frequency for
TQM can make these two expressions difficult to work with analytically.
A quadratic approximation offers useful insight. We will use this
approach when analyzing the loop corrections, for instance. Note that
$\varpi_{p}=0$ when we are exactly on-shell, increasing quadratically
as we get far enough off-shell:

\begin{equation}
{\varpi_{p}}=-\frac{{{{\left({\bar{E}+\delta E}\right)}^{2}}-{{\bar{E}}^{2}}}}{{2\left({\bar{E}+\delta E}\right)}}\approx-\delta E+\frac{{{\left({\delta E}\right)}^{2}}}{{2\bar{E}}}
\end{equation}
We expect therefore that if TQM wave packets start on-shell (per the
entropic estimate) they will tend to stay approximately on-shell subsequently.
This means we can take the initial wave packets in a scattering experiment
as on-shell. We will take advantage of this in the next subsection.

\subsubsection{Detection of the wave function}

\label{subsec:Death}

The issues associated with the detection of a wave function in time
were discussed at length in paper B \cite{Ashmead:2021aa}. To keep
this work self-contained we give a summary.

\paragraph{Effect of convoluted paths}

\label{par:Effect-of-convoluted-paths}

The paths in TQM are significantly more complex than those in SQM.
In particular while paths in SQM can go left and right, up and down,
forwards and back, paths in TQM can also go into the future and into
the past. We expect these excursions will be of order attoseconds
and be centered around the classical or SQM values. (This will keep
us from attracting the unwelcome attention of the time police.) 

When we calculate the paths in TQM we typically calculate the wave
function at the detector as a function of clock time:

\begin{equation}
{\varphi_{\tau}}\left({t,x}\right)
\end{equation}
This gives the sum not of all paths to $t,x$ but only of those paths
with length $\tau$. To get the full amplitude at $t,x$ we will need
to sum over paths of all lengths:

\begin{equation}
\varphi\left({t,x}\right)=\int{d\tau}{\varphi_{\tau}}\left({t,x}\right)
\end{equation}
We are, as required, taking coordinate time as fundamental with clock
time playing a secondary role, here letting us categorize the paths
by length.

\paragraph{The problem of measurement}

\label{subsec:Free-Detection}
\begin{quotation}
``To re-emphasize, from a broader perspective, the main argument
that is being articulated in this section is that quantum measurements,
in the interferometric and polarization domains, can be described
without resorting to the concept of the collapse of the wave function
or the collapse of the probability amplitude.'' -- p185 Duarte and
Taylor \cite{Duarte:2021vi}
\end{quotation}
The fundamental problem here is that ultimately we have to treat the
detectors (and even the observers) as also quantum mechanical systems.
They are, after all, made of atoms, and atoms are unavoidably quantum
systems. There is no such thing as a classical atom -- in classical
mechanics electrons spiral into the nucleus in a minute fraction of
a second due to loss of energy from Larmor radiation \cite{Olsen:2017aa}. 

There is of course an extraordinary literature on the ``problem of
measurement''. This problem is not a central focus of this work.
In paper B, where this problem was a central focus, we took as a starting
point an analysis by Marchewka and Schuss \cite{Marchewka:1998aa,Marchewka:1999db,Marchewka:1999nl,Marchewka:2000ys}
who made a strong argument (from probability conservation) that the
probability current correctly gives the detection rate. 

In a certain sense this merely postpones the problem: to compute the
probability current you must first compute the wave function, which
means that you have to first solve the problem of the interaction
of the particle's wave function with that of the detector -- including
absorption, reflection, loss, emission, emission followed by re-absorption,
and so on. Rather a lot to consider.

However if you are prepared to posit an ideal detector, with no loss
or lag, you can use the probability current at the detector as giving
a reasonable estimate of the detection rate as a function of time.
The beauty of this approach, in the current context, is that it:
\begin{enumerate}
\item is easy to use,
\item should give a reasonable first estimate in general,
\item is typical of what workers often do, so clearly not that bad in practice,
\item and most importantly here, works the same way for SQM and TQM, so
creates no bias. 
\end{enumerate}

\subparagraph{In SQM}

We take a detector in the y-z plane placed at $x=L$. For a simple
GTF headed, say, left to right with momentum $p_{0}$, the detection
rate is:

\begin{equation}
{D_{\tau}^{S}}=\frac{{p_{0}}}{m}{\left.{{\rho_{\tau}}\left(x\right)}\right|_{x=L}}
\end{equation}
${\rho_{\tau}}\left(x\right)$ has a dispersion in space given by
$\sigma_{x}$. But it is the resulting dispersion of the detection
rate $D_{\tau}^{S}$ in clock time that we are interested in here.
This is:

\begin{equation}
\sigma_{\tau}^{S}=\frac{1}{{m{v_{0}}{\sigma_{x}}}}\bar{\tau}
\end{equation}
This is proportional to $\bar{\tau}$ the average time of flight:
the greater the average time of flight the greater the uncertainty
in clock time at the detector. It is inversely proportional to $\sigma_{x}$
because of diffraction (see \ref{subsec:gtf-Spacetime}). And it is
inversely proportional to the speed of the packet: the slower the
packet the more spread out it will be by the time it reaches the detector.
The packet functions like a train; the slower the train the longer
the time from when the engine arrives at the station till the time
when the caboose does.

The corresponding uncertainty in clock time, the number we are most
interested in, is:

\begin{equation}
\Delta_{\tau}^{S}=\frac{1}{{\sqrt{2}}}{\sigma_{\tau}^{S}}\label{eq:free-sqm-uncert}
\end{equation}
We are seeing uncertainty in clock time, but it is \emph{entirely}
the product of dispersion in space. As Busch and other workers have
pointed out, in SQM there is a HUP for time/energy, but it is not
on the same basis at all as the HUP for momentum/space.

\subparagraph{In TQM}

In TQM we will have dispersion in time in addition to the dispersion
in space. Suppose the initial wave function is a product of time and
space parts:

\begin{equation}
{\varphi_{\tau}}\left({t,x}\right)=\varphi_{\tau}^{T}\left(t\right)\varphi_{\tau}^{S}\left(x\right)
\end{equation}
We may be getting the wave function in time from the entropic estimate.
And we are ignoring the more general possibility that time and space
are entangled. Using again the probability current, now in TQM, we
showed that the detection rate is now the product of the time and
space parts:

\begin{equation}
{D_{\tau}}\left(t\right)=D_{\tau}^{S}\rho_{\tau}^{T}\left(t\right)
\end{equation}
where the probability distribution in coordinate time as a function
of clock time is:

\begin{equation}
\rho_{\tau}^{T}\left(t\right)\approx\sqrt{\frac{1}{{\pi\sigma_{\tau}^{\left(T\right)2}}}}{e^{-\frac{1}{{\sigma_{\tau}^{\left(T\right)2}}}{{\left({t-\tau}\right)}^{2}}}},\sigma_{\tau}^{T}\equiv\frac{{\bar{\tau}}}{{m{\sigma_{t}}}}
\end{equation}
$\sigma_{\tau}^{T}$ is proportional to the average time to the detector.
It is inversely proportional to the initial dispersion in coordinate
time $\sigma_{t}$. To get the full detection rate at coordinate time
$t$ we have to sum over the clock time:

\begin{equation}
D\left(t\right)=\int{d\tau D_{\tau}^{S}\rho_{\tau}^{T}\left(t\right)}
\end{equation}
This is a convolution of the $S$ and the $T$ distributions. The
total dispersion squared is the sum of the squares of the dispersions:

\begin{equation}
\sigma_{t}^{2}=\sigma_{\tau}^{\left(S\right)2}+\sigma_{\tau}^{\left(T\right)2}\label{eq:uncertainty-in-dispersions}
\end{equation}
This is a ``nice'' result in that if we compute the dispersion in
clock time in SQM and the dispersion from the time part of TQM we
need only to add them to get the total dispersion. If we monitor the
dispersion of arrival times long enough, we should see -- to whatever
level of statistical certainty is desired -- whether the dispersion
is better described by SQM or by SQM + TQM. 

\paragraph{Slow train problem}

\label{par:Slow-train-problem}

The main problem here is the $1/v_{0}$ factor in the dispersion from
space. From the entropic estimate we expect that the dispersions in
space and time will in general be of the same order, $\sigma_{x}\sim\sigma_{t}$.
But if the wave packet is going at non-relativistic speeds, the effects
of the space part will be far greater than those from time, making
it hard to pick out any effects from TQM.

There are (at least) four possible solutions to the slow train problem:
\begin{enumerate}
\item Use a faster train. Get $v_{0}$ closer to one. This in fact is part
of the motivation for this paper, as we need QED to work with high
speeds.
\item Wait longer for the train. Collect enough data points that even a
small difference becomes statically significant to whatever level
is desired.
\item Break the train up into its individual cars. If we do this, we can
see the dispersion in time on a per car basis, which should make it
stand out more. Or in less figurative language, run the wave packet
through a strong magnetic field. Discussed further below in subsubsection
\ref{par:Samurai-versus-pirate}.
\item Diffract the train. (This is a quantum train.) Use the Heisenberg
uncertainty principle in time. We use a second wave packet, chosen
to be narrow in time, as effectively a single slit in time. The single
slit in time acts as a single slit in space would, diffracting the
particle with correspondingly increased uncertainty in time-of-arrival.
\end{enumerate}

\subsubsection{Summary}

At this point we have achieved a measure of falsifiability. 
\begin{enumerate}
\item We write the TQM prediction of the uncertainty in time at the detector
as $\left({\Delta t}\right)_{D}^{2}$,
\item We write the SQM prediction of the uncertainty in (clock time) at
the detector as $\left({\Delta\tau}\right)_{D}^{\left(S\right)2}$.
\item Our signal is the difference, meaning that part of the uncertainty
in time at the detector which is not accounted for by SQM:
\end{enumerate}
\begin{equation}
\left({\Delta t}\right)_{D}^{\left(T\right)2}\equiv\left({\Delta t}\right)_{D}^{2}-\left({\Delta\tau}\right)_{D}^{\left(S\right)2}\label{eq:time-signal}
\end{equation}
For this to be well-defined, we need a good estimate of the initial
wave function in coordinate time. This is why the entropic estimate
is critical: it gives a simple but robust estimate of the initial
uncertainty in coordinate time.

However this does not yet take us to feasibility. We see that for
non-relativistic wave packets, the SQM prediction for uncertainty
in time at the detector may be much larger than the additional contribution
from TQM; therefore the space contribution may drown out the time
contribution.

To get from falsifiability in principle to falsifiability in practice
we need a way to increase the relative size of the time contribution.

To do this we look next at the simplest possible scattering problem.
This continues our development of TQM and it defines the basic effects
we can use to see its effects.

\subsection{Spin zero scattering}

\label{subsec:Spin-zero-scattering}

\begin{figure}
\includegraphics[scale=0.6]{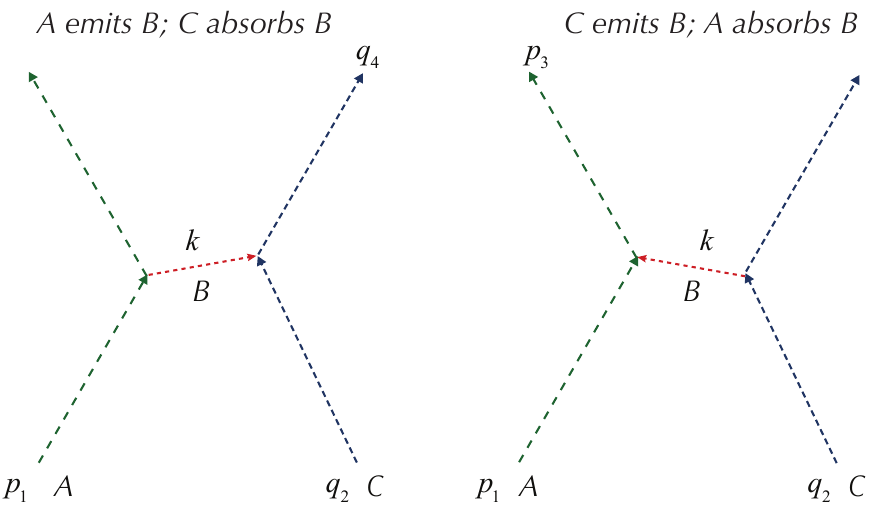}

\caption{An \emph{A} and a \emph{C} particle exchange a \emph{B} particle}
\end{figure}

\subsubsection{Objective}

\label{subsec:-abc-Objective}

We now look at the simplest possible scattering problem. We work with
a toy model consisting of three spin zero particles: $A$, $B$ and
$C$. \emph{$A$} and \emph{$C$} both have mass $m$. They do not
interact directly, but only by exchanging \emph{B}'s. \emph{B}'s have
mass $\mu$. \emph{A}'s and \emph{C}'s are stand-ins for fermions,
\emph{$B$} is a stand-in for photons. We will refer to this as the
ABC model.

The very considerable practical advantage of looking at spinless particles
is that most of the effects that distinguish TQM from SQM have nothing
to do with spin or polarization.

The field operators for A's are\emph{ }for\emph{ SQM:}

\begin{equation}
A_{\tau}^{S}\left({\vec{x}}\right)\equiv\sum\limits _{\vec{p}}{\frac{1}{{\sqrt{2V{E_{\vec{p}}}}}}\left(\begin{gathered}{a_{\vec{p}}}\exp\left({-\imath{E_{\vec{p}}}\tau+\imath\vec{p}\cdot\vec{x}}\right)\hfill\\
+a_{\vec{p}}^{\dag}\exp\left({\imath{E_{\vec{p}}}\tau-\imath\vec{p}\cdot\vec{x}}\right)\hfill
\end{gathered}
\right)}
\end{equation}
and for \emph{TQM: }

\begin{equation}
{A_{\tau}}\left({t,\vec{x}}\right)\equiv\sum\limits _{p}{\frac{1}{{\sqrt{2TVE}}}\left(\begin{gathered}{a_{p}}\exp\left({-\imath{\mathcal{\varpi}_{p}}\tau-\imath px}\right)\hfill\\
+a_{p}^{\dag}\exp\left({+\imath{\mathcal{\varpi}_{p}}\tau+\imath px}\right)\hfill
\end{gathered}
\right)}
\end{equation}
$B$ and $C$'s are the same.

The interaction Lagrangian has the form:

\begin{equation}
V=\lambda\frac{{A^{2}B}}{{2}}+\lambda\frac{{C^{2}B}}{{2}}
\end{equation}
with specific terms of the form:

\begin{equation}
\frac{{\lambda\left({a+{a^{\dag}}}\right)\left({b+{b^{\dag}}}\right)\left({a+{a^{\dag}}}\right)}}{2}+\frac{{\lambda\left({c+{c^{\dag}}}\right)\left({b+{b^{\dag}}}\right)\left({c+{c^{\dag}}}\right)}}{2}
\end{equation}
We will assume $\lambda$ is small, so that perturbation theory makes
sense. We will look at the case where an $A$ and $C$ exchange a
$B$. 

To help make the diagrams more readable, we identify the $A$'s by
$p$'s, $B$'s by $k$'s, and $C$'s by $q$'s. We use ${{\bar{E}}_{1,2,3,4}},\omega$
for the clock energies for SQM. We use ${E_{1,2,3,4}},w$ for the
coordinate energies and $\varpi_{1,2,3,4},\varpi_{k}$ for the clock
energies in TQM. $A$ will start with momentum $p_{1}$, finish with
$p_{3}$; \emph{$C$} will start with $q_{2}$, finished with $q_{4}$;
and the exchanged \emph{$B$} will have momentum $k$.

We will work in the center-of-mass frame. For definiteness we will
assume the particles are approaching each other along the $x$-axis
and are scattered along the $y$-axis. Results for the $z$-axis follow
from symmetry. The coordinate system is given by $\left(x,y,z\right)=r\left({\sin\theta\cos\varsigma,\sin\theta\sin\varsigma,\cos\theta}\right)$.
In the case of GTFs, we will work in the center-of-time frame as well:
arrange for $\tau=0$ to correspond to the time of closest approach
of the opposing GTFs.

We assume we have detectors well off to the left and right, placed
at $x=\pm L$, running in the $y,z$ plane, and time-sensitive. And
further we will assume that the detectors are far enough away that
the position of the detector divided by time-of-flight from the interaction
zone is a good proxy for the velocity and therefore for the momentum.
With this arrangement it is easy to map detector time and position
to momentum. To first order: 

\begin{equation}
{p_{x}}=\gamma m{v_{x}},{p_{y}}=\gamma m{v_{y}},E=\gamma m
\end{equation}
For SQM ${v_{x}}=\frac{L}{\tau},{v_{y}}=\frac{y}{\tau}$; for TQM
${v_{x}}=\frac{L}{t},{v_{y}}=\frac{y}{t}$. We will compute the energy
component indirectly, by using the three velocity to compute $\gamma=\frac{1}{{\sqrt{1-{{\vec{v}}^{2}}}}}.$
We will look first at the case where the initial wave functions are
defined as plane waves; then as GTFs.

We have conservation of clock energy as well, for both SQM and TQM.
For plane waves it is customary to take the limit as the integral
over clock time goes to $\pm\infty$, which in turn gives conservation
of clock energy. As noted earlier, for GTFs we argue that if the interaction
time is small an integration over the interaction will be zero outside
of that time, so that the $\pm$ limits may be extended to $\pm\infty$
without affecting the value of the integral. So we can again take
the limit as clock time goes to $\pm\infty$ and therefore have conservation
of clock energy.

\subsubsection{Plane waves}

\label{subsec:abc-Plane-waves}

We can see the plane wave case as Møller scattering without the spin
or polarization. In terms of the Mandelstam variables \cite{Mandelstam:1958aa}
we are working in $t$-channel, with the energy-momentum of the exchanged
particle being given by $p_{3}-p_{1}$ or $q_{4}-q_{2}$:

\begin{equation}
t={\left({{p_{1}}-{p_{3}}}\right)^{2}}={\left({{q_{2}}-{q_{4}}}\right)^{2}}
\end{equation}
The $S$ matrix is given in each case as a product of the appropriate
$\delta$ functions, normalization factors, and the matrix element.
We start with the matrix element as the interesting bit. In this simple
case, this is $\lambda^{2}$ times the propagator for the exchanged
particle. 

\paragraph{SQM}

\label{par:abc-plane-waves-QM}

For SQM the propagator is:

\begin{equation}
\Delta_{\omega}^{S}\left({\vec{k}}\right)=\frac{1}{{{\omega^{2}}-{{\vec{k}}^{2}}-{\mu^{2}}+\imath\varepsilon}}
\end{equation}
We can compute this from the values of the external momenta and the
$\delta$ functions at each vertex. If the $A$ particle emits the
$B$, then we have:

\begin{equation}
\left({\omega,\vec{k}}\right)=\left({{{\bar{E}}_{3}}-{{\bar{E}}_{1}},{{\vec{p}}_{3}}-{{\vec{p}}_{1}}}\right),\left({{{\bar{E}}_{4}},{{\vec{q}}_{4}}}\right)=\left({{{\bar{E}}_{2}}+\omega,{{\vec{q}}_{2}}+\vec{k}}\right)
\end{equation}
while if the $C$ particle emits the $B$ we have:

\begin{equation}
\left({\omega,\vec{k}}\right)=\left({{{\bar{E}}_{4}}-{{\bar{E}}_{2}},{{\vec{q}}_{4}}-{{\vec{q}}_{2}}}\right),\left({{{\bar{E}}_{3}},{{\vec{p}}_{3}}}\right)=\left({{{\bar{E}}_{1}}+\omega,{{\vec{p}}_{1}}+\vec{k}}\right)
\end{equation}
In both cases we have overall conservation of momentum:

\begin{equation}
\left({{{\bar{E}}_{3}}+{{\bar{E}}_{4}},{{\vec{p}}_{3}}+{{\vec{q}}_{4}}}\right)=\left({{{\bar{E}}_{1}}+{{\bar{E}}_{2}},{{\vec{p}}_{1}}+{{\vec{q}}_{2}}}\right)
\end{equation}
This forces the magnitude of the three vectors for both $A$ and $C$
to be the same before and after the interaction. From conservation
of clock energy:

\begin{equation}
2{m^{2}}+\vec{p}_{1}^{2}+\vec{q}_{2}^{2}=2{m^{2}}+\vec{p}_{3}^{2}+\vec{q}_{4}^{2}
\end{equation}
In the center-of-mass frame:

\begin{equation}
\vec{p}_{1}^{2}=\vec{q}_{2}^{2},\vec{p}_{3}^{2}=\vec{q}_{4}^{2}
\end{equation}
Combining these two results we get:

\begin{equation}
2{m^{2}}+2\vec{p}_{1}^{2}=2{m^{2}}+2\vec{p}_{3}^{2}\Rightarrow\vec{p}_{1}^{2}=\vec{p}_{3}^{2}
\end{equation}
and from this we see there is no clock energy available for the unfortunate
$B$. As the propagator is not being integrated over, we may drop
the $\imath\epsilon$. We have:

\begin{equation}
\Delta_{\omega}^{S}\left({\vec{k}}\right)=-\frac{1}{{{{\vec{k}}^{2}}+{\mu^{2}}}}
\end{equation}
We set $\vec{p}\equiv{{\vec{p}}_{1}}=-{{\vec{q}}_{2}}$ and write: 

\begin{equation}
\vec{k}=\left|{\vec{p}}\right|\left({\cos\left(\theta\right),\sin\left(\theta\right),0}\right)
\end{equation}
so the propagator is:

\begin{equation}
\Delta_{\omega}^{S}\left({\vec{k}}\right)=-\frac{1}{{2{{\vec{p}}^{2}}\left({1-\cos\left(\theta\right)}\right)+{\mu^{2}}}}
\end{equation}
and the matrix element ${\left(-\imath\lambda\right)}^{2}$ times
this:

\begin{equation}
\mathcal{M^{S}}=\frac{{\lambda^{2}}}{{2{{\vec{p}}^{2}}\left({1-\cos\left(\theta\right)}\right)+{\mu^{2}}}}
\end{equation}
 The external factors are:

\begin{equation}
{N^{S}}=\sqrt{\frac{1}{{2{{\bar{E}}_{1}}}}}\sqrt{\frac{1}{{2{{\bar{E}}_{2}}}}}\sqrt{\frac{1}{{2{{\bar{E}}_{3}}}}}\sqrt{\frac{1}{{2{{\bar{E}}_{4}}}}}
\end{equation}
In the center-of-mass frame they are all equal, so there is just one
$\bar{E}$. The $\delta$ functions are:

\begin{equation}
{D^{S}}={\left({2\pi}\right)^{4}}\delta\left({\left({{{\bar{E}}_{3}}+{{\bar{E}}_{4}}}\right)-\left({{{\bar{E}}_{1}}+{{\bar{E}}_{2}}}\right)}\right){\delta^{3}}\left({\left({{{\vec{p}}_{3}}+{{\vec{q}}_{4}}}\right)-\left({{{\vec{p}}_{1}}+{{\vec{q}}_{2}}}\right)}\right)
\end{equation}
The full $S$ matrix is:

\begin{equation}
{S^{S}}={D^{S}}{N^{S}}{\mathcal{M}^{S}}
\end{equation}
The probability is given by the square of this. The four $\delta$
functions make clear that this is a formal object; it will not acquire
meaning until we integrate over the GTFs.

\paragraph{TQM}

\label{par:abc-plane-waves-TQM}

The calculation for the plane wave for TQM is the same. We will take
the SQM values as the starting point. Recall the quantum energy $\delta w$
is defined as the difference between the coordinate energy and the
SQM energy. Since the SQM energy is zero, the quantum energy is just
the coordinate energy $w$ itself.

We have from the entropic estimate that the average energy in the
initial value of a TQM wave function will match that in SQM, so we
have: 

\begin{equation}
{E_{1}}={{\bar{E}}_{1}},{E_{2}}={{\bar{E}}_{2}}\Rightarrow\delta{E_{1}}=0,\delta{E_{2}}=0
\end{equation}
From this we also get $\varpi_{1}=\varpi_{2}=0$. Or to put it another
way, wave functions in TQM start on-shell. We have four $\delta$
functions in momentum plus one in clock energy at each vertex, forcing
the four momenta for the exchanged $B$ in the same way as with SQM.

If the $A$ particle emits the $B$, then we have:

\begin{equation}
{E_{3}}=\bar{E}-w,{{\vec{p}}_{3}}={{\vec{p}}_{1}}-\vec{k};{E_{4}}=\bar{E}+w,{{\vec{q}}_{4}}=-{{\vec{p}}_{1}}+\vec{k}
\end{equation}
while if the $C$ particle emits the $B$ we have:

\begin{equation}
{E_{3}}=\bar{E}+w,{{\vec{p}}_{3}}={{\vec{p}}_{1}}+\vec{k};{E_{4}}=\bar{E}-w,{{\vec{q}}_{4}}=-{{\vec{p}}_{1}}-\vec{k}
\end{equation}
In both cases we have overall conservation of momentum:

\begin{equation}
{E_{3}}+{E_{4}}={E_{1}}+{E_{2}},{{\vec{p}}_{3}}=-{{\vec{q}}_{4}}
\end{equation}
The only remaining ambiguity is whether we have $\varpi_{3}=\varpi_{4}=$
0 or not, whether the two outgoing particles are forced to be on-shell
or not. To check this we expand in powers of $w$.

We have for the two external clock energies if $A$ emits the $B$:

\begin{equation}
{\varpi_{3}}=-\frac{{{{\left({\bar{E}-w}\right)}^{2}}-{{\left({{{\vec{p}}_{1}}-\vec{k}}\right)}^{2}}-{\mu^{2}}}}{{\bar{E}-w}},{\varpi_{4}}=-\frac{{{{\left({\bar{E}+w}\right)}^{2}}-{{\left({{{\vec{p}}_{1}}-\vec{k}}\right)}^{2}}-{\mu^{2}}}}{{\bar{E}+w}}
\end{equation}
and the same thing with $w\to-w$ if $C$ emits the $B$. Therefore
the sum of the two, in either case, is:

\begin{equation}
-\frac{{{{\left({\bar{E}-w}\right)}^{2}}-{{\left({{{\vec{p}}_{1}}-\vec{k}}\right)}^{2}}-{\mu^{2}}}}{2\left({\bar{E}-w}\right)}-\frac{{{{\left({\bar{E}+w}\right)}^{2}}-{{\left({{{\vec{p}}_{1}}-\vec{k}}\right)}^{2}}-{\mu^{2}}}}{2\left({\bar{E}+w}\right)}=0
\end{equation}
We already know, since the incoming lines are on-shell, that:

\begin{equation}
{{\left({\bar{E}}\right)}^{2}}-{{\left({{\vec{p}}_{1}}\right)}^{2}}-{\mu^{2}}=0
\end{equation}
Using this and clearing the fractions we get:

\begin{equation}
2\bar{E}w^{2}=0\Rightarrow w=0
\end{equation}
So here the quantum energy $w$ of the exchanged particle is zero
just as the classical energy is. And therefore the clock energies
$\varpi_{3},\varpi_{4}$ are also zero. And therefore both outgoing
lines are on-shell. To put it another way, an analysis using on-shell
plane waves shows no effects of TQM. We will need to work with GTFs
to see the effects of dispersion in time/energy. 

\subsubsection{Gaussian test functions}

\paragraph{Calculation of the propagation of the GTFs through the diagram}

\label{subsec:abc-Gaussian-test-functions}

\begin{figure}
\includegraphics[scale=0.6]{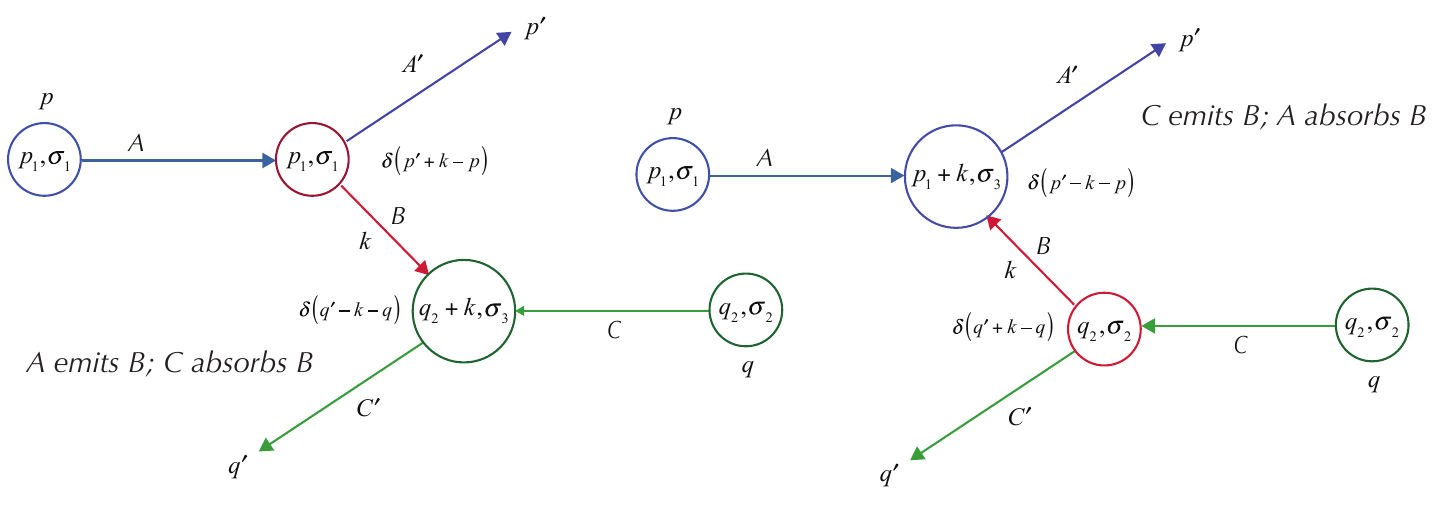}

\caption{Calculation of dispersion in exchange of \emph{B} particle}
\end{figure}

Our initial wave functions will be direct products of GTFs in $x,y,z$
and (in TQM) $t$. The value of the $t$ wave function will be inferred
from the other three parts using the entropic estimate.

For GTFs the calculation is essentially the same for all three or
four components of the momentum, which we will represent as $p,q,p',q'$. 

We will start by assuming we are dealing with fixed input GTFs, functions
of $p,q$ centered on $p_{1},q_{2}$ with dispersions $\sigma_{1},\sigma_{2}$.
We wish to compute the amplitude to measure the outgoing momenta with
values $p',q'$.

We can compute this by integrating over the $S$ matrix from the plane
wave case. We express this in a generic notation:

\begin{equation}
\psi\left({p',q'}\right)=\int{dpdqdk}S\left({p',q';p,q}\right){\varphi_{1}}\left(p\right){\varphi_{2}}\left(q\right)
\end{equation}
First we assume the incoming GTFs are narrow in energy/momentum, that
the dispersions is a small fraction of the corresponding momenta: 

\begin{equation}
\frac{{\sigma_{\left|{\vec{p}}\right|}}}{{\vec{p}}}\ll1
\end{equation}
We refer to this as the ``narrow beam'' approximation. Our GTFs
are more rounded than simple plane waves but each is still relatively
focused on a specific value of the momentum. They inhabit a kind of
halfway house between plane wave and GTF.

This lets us make some useful approximations for the normalizations
and for the matrix elements. In particular we can approximate each
external factor by its average. For SQM we can ignore the variation
caused by the GTF:

\begin{equation}
\sqrt{\frac{m}{{E_{{{\vec{p}}_{1,2}}}}}}\approx\sqrt{\frac{m}{{{\bar{E}}_{1,2}}}}
\end{equation}
for TQM we can ignore both this and variation caused by the quantum
energy ($\delta E\equiv E-E_{\vec{p}}$):

\begin{equation}
\sqrt{\frac{m}{{E_{1,2}}}}\approx\sqrt{\frac{m}{{E_{{{\vec{p}}_{1,2}}}}}}\approx\sqrt{\frac{m}{{{\bar{E}}_{1,2}}}}
\end{equation}
By the same argument, for SQM we replace the technically correct $k_{i}+\delta k_{i}$
in the denominator of the matrix element by $k_{i}$:

\begin{equation}
\frac{\imath}{{{{\left({\vec{k}+\delta\vec{k}}\right)}^{2}}+{\mu^{2}}}}\to\frac{\imath}{{\left\langle {{\vec{k}}^{2}}\right\rangle +{\mu^{2}}}}
\end{equation}
For TQM we do the same:

\begin{equation}
\frac{\imath}{{{{\left({\delta w}\right)}^{2}}-{{\left({\vec{k}+\delta\vec{k}}\right)}^{2}}+{\mu^{2}}}}\to\frac{\imath}{{\left\langle {{\vec{k}}^{2}}\right\rangle +{\mu^{2}}}}
\end{equation}
This means we can pull out from inside the integral a common factor
(the same for SQM and TQM) of:

\begin{equation}
\bar{S}\equiv\prod\limits _{i=1}^{4}{\sqrt{\frac{m}{{{\bar{E}}_{i}}}}}\frac{\imath}{{{{\vec{k}}^{2}}+{\mu^{2}}}}
\end{equation}
 leaving the integral as purely an integral over the dispersions in
the GTFs and the $\delta$ functions at the vertexes.

There are three of these integrals for SQM plus one more -- over
energy -- for TQM:

\begin{equation}
I\left({p'q'}\right)=\int{dpdq}dk\left(\begin{gathered}\delta\left({p'+k-p}\right)\delta\left({q'-k-q}\right)\theta\left(\tau\right)\hfill\\
+\delta\left({p'-k-p}\right)\delta\left({q'+k-q}\right)\theta\left(-\tau\right)\hfill
\end{gathered}
\right)\varphi\left(p\right)\varphi\left(q\right)
\end{equation}
where the sign of $\tau$ depends on whether we are looking at the
case where $A$ emits $B$ and $C$ absorbs it or where $C$ emits
a $B$ and $A$ absorbs it. 

What we have here is a ``stick-and-cloud'' approximation. The plane
wave analysis gives the average result -- the stick -- then the
integrals over the dispersions pull in the quantum effects we are
interested in here, the clouds. With the $\delta$ functions we have
only one real integral to do.

For the first pair, we do the integral over $p$ first. This turns
${\varphi_{1}}\left(p\right)\to{\varphi_{1}}\left({p'+k}\right)$.
Next we do the integral over $k$ and the second $\delta$ function.
This leaves:

\begin{equation}
{\psi^{\left(1\right)}}\left({p',q'}\right)=\int{dq}{\varphi_{1}}\left({p'+k}\right){\varphi_{2}}\left({q'-k}\right)
\end{equation}
This is a convolution of two Gaussians. This results in a single Gaussian
of the form:

\begin{equation}
\exp\left({-\frac{{{\left({\left({p'+q'}\right)-\left({{p_{1}}+{q_{2}}}\right)}\right)}^{2}}}{{2\sigma_{3}^{2}}}}\right)\label{eq:final-distribution}
\end{equation}
with dispersion:

\begin{equation}
\sigma_{3}^{2}=\sigma_{1}^{2}+\sigma_{2}^{3}
\end{equation}
What this is saying is that the resulting amplitude is a GTF centered
on the average value of the total initial momentum, with a total dispersion
which is a Pythagorean sum over the two initial dispersions. The second
pair of $\delta$ functions goes exactly the same way, but in reverse.

At this point we need to check the counting. Each vertex has a factor
of 1/2, for a total of 1/4. We can have the $B$ emit from the first
vertex or the second and from $A$ or from $C$. This is a counter-balancing
factor of four, for a total value of one. This is as expected. Therefore
the final distribution is still given by equation \ref{eq:final-distribution}.

\paragraph{Compare to conservation of energy with plane waves}

We define:
\begin{equation}
\delta p\equiv p'-\bar{p},\delta q\equiv q'-\bar{p}
\end{equation}
giving:

\begin{equation}
\delta p+\delta q=p'+q'-\left({p+q}\right)
\end{equation}
and:

\begin{equation}
\psi\left({\delta p,\delta q}\right)=\exp\left({-\frac{{{\left({\delta p+\delta q}\right)}^{2}}}{{2\sigma_{3}^{2}}}}\right)
\end{equation}
So the incoming total value of the momentum in question has an average
of $\bar{p}$, but is dispersed around that average. The outgoing
value is also dispersed around the average, with the dispersion given
by the Pythagorean sum.

The full outgoing wave function is the product of three of these functions
for SQM; the product of four of these functions -- the three from
SQM and a fourth on the energy axis -- for TQM. Defining:

\begin{equation}
\varphi_{3,4}^{\mu}\left({\delta{p_{\mu}},\delta{q_{\mu}}}\right)\equiv\exp\left({-\frac{{{\left({\delta{p_{\mu}}+\delta{q_{\mu}}}\right)}^{2}}}{{2\left({\sigma_{\mu}^{\left(1\right)2}+\sigma_{\mu}^{\left(2\right)2}}\right)}}}\right)
\end{equation}
we get for SQM:

\begin{equation}
\varphi_{3,4}^{S}\left({\vec{p}',\vec{q}'}\right)=\bar{S}\left({{{\vec{p}}_{3}},{{\vec{q}}_{4}};{{\vec{p}}_{1}},{{\vec{q}}_{2}}}\right)\prod\limits _{i=1}^{3}{\varphi_{3,4}^{i}\left({\delta{p_{i}},\delta{q_{i}}}\right)}
\end{equation}
and for TQM we have the product of the energy part with this:

\begin{equation}
\varphi_{3,4}\left({p',q'}\right)=\varphi_{3,4}^{\mu=0}\left({\delta{p_{0}},\delta{q_{0}}}\right)\varphi_{3,4}^{S}\left({\vec{p}',\vec{q}'}\right)
\end{equation}
For our purposes the SQM part is the carrier; the energy GTF is the
signal.

\paragraph{Distribution in three space for both SQM and TQM}

The distribution in the three space dimensions is the same for both
SQM and TQM. We pick a specific scattering angle $\theta$.

Suppose we have in the $p_{\theta}$ direction the final dispersion
$\hat{\sigma}_{\theta}$:

\begin{equation}
\hat{\sigma}_{\theta}^{2}=\hat{\sigma}_{\theta}^{\left(1\right)2}+\hat{\sigma}_{\theta}^{\left(2\right)2}
\end{equation}
then we have $\sigma_{\theta}$, the space dispersion in the $\theta$
direction, as:

\begin{equation}
\sigma_{\theta}^{2}=\frac{1}{{\hat{\sigma}_{\theta}^{2}}}
\end{equation}
Then from the previous section the effective dispersion in clock time
is:

\begin{equation}
\sigma_{\tau}^{\left(S\right)2}=\frac{1}{{{{\vec{p}}^{2}}\sigma_{\theta}^{2}}}{{\bar{\tau}}^{2}}=\frac{{\hat{\sigma}_{\theta}^{\left(1\right)2}+\hat{\sigma}_{\theta}^{\left(2\right)2}}}{{{\vec{p}}^{2}}}{{\bar{\tau}}^{2}}
\end{equation}

\paragraph{Distribution in coordinate time/energy for TQM}

TQM we have a distribution in energy as well:

\begin{equation}
\rho\left(E\right)\sim\exp\left({-\frac{{{\left({\delta{E_{3}}+\delta{E_{4}}}\right)}^{2}}}{{\left({\sigma_{E}^{\left(1\right)2}+\sigma_{E}^{\left(2\right)2}}\right)}}}\right)
\end{equation}
This is centered on the average in energy. Therefore if $\delta E$
for one of the outgoing legs is positive, the mostly likely value
of $\delta E$ for the other is its negative. But what does $\delta E$
mean in operational terms? The most direct way to determine this is
to run the outgoing particles through a detector that responds to
energy. But to be consistent with the treatment so far, we will convert
this into time-of-arrival measurements.

If $\delta E>0$ then the particle is traveling ``hot'' and will
arrive a bit early at the detector. For a fixed angle $\theta$, if
we have the expected time-of-arrival as $\bar{\tau}$, then we will
measure a $\delta t$ which is ``early'', so negative. The other
particle is most likely to be traveling exactly that amount ``cold''
and therefore to arrive (to first order) exactly $-\delta t$ ``late'',
so positive.

We don't have an absolute distribution for either particle, what we
have is a correlated distribution: if, say, the $C$ particle exactly
on time, then the $A$ particle will have a distribution in time centered
on $\bar{\tau}$, with dispersion:

\begin{equation}
{\rho}\left(t\right)=\exp\left({-\frac{{{\left({\delta t}\right)}^{2}}}{{\sigma_{t}^{2}}}}\right),\sigma_{t}^{2}=\frac{1}{{\sigma_{E}^{2}}},\sigma_{E}^{2}=\sigma_{E}^{\left(1\right)2}+\sigma_{E}^{\left(2\right)2}\label{eq:dispersion-in-time}
\end{equation}
If the $C$ particle is a bit off-center, the $A$ particle's own
distribution will shift in the opposite direction. So we have the
distribution in coordinate time as we leave the interaction zone. 

\paragraph{Diffraction with respect to coordinate time in TQM}

There is one more piece to the puzzle: we have the dispersion in coordinate
time as we leave the interaction zone. But this is not the dispersion
in coordinate time at the detector, some temporal distance in the
particle's future. Referring to the appendix on GTFs , we have the
probability density at the detector as:

\begin{equation}
\rho_{\tau}^{T}\left(t\right)=\sqrt{\frac{1}{{\pi\sigma_{t}^{2}\left({1+\frac{{{\bar{\tau}}^{2}}}{{{{\bar{E}}^{2}}\sigma_{t}^{4}}}}\right)}}}\exp\left({-\frac{{{\left({t-\bar{\tau}}\right)}^{2}}}{{\sigma_{t}^{2}\left({1+\frac{{{\bar{\tau}}^{2}}}{{{{\bar{E}}^{2}}\sigma_{t}^{4}}}}\right)}}}\right)
\end{equation}
with uncertainty in time as:

\begin{equation}
{\left({\Delta t}\right)^{2}}\equiv\left\langle {t^{2}}\right\rangle -{{\bar{\tau}}^{2}}=\frac{{\sigma_{t}^{2}}}{2}\left|{1+\frac{{{\bar{\tau}}^{2}}}{{{{\bar{E}}^{2}}\sigma_{t}^{4}}}}\right|\to\frac{{{\bar{\tau}}^{2}}}{{2{{\bar{E}}^{2}}\sigma_{t}^{2}}}
\end{equation}
so both the TQM part and the SQM part scale as $\bar{\tau}$, as one
would expect intuitively. But the TQM contribution to the uncertainty
in time at the detector is inversely proportional to the initial dispersion
in coordinate time. The smaller we can make the post-interaction $\sigma_{t}$,
the greater the dispersion in time at the detector. How to make $\sigma_{t}$
small?

\paragraph{Single slit in time}

\label{par:Single-slit-in-time}

Suppose we think of the particle $C$ as a gate in time, a single
slit in time, with $A$ being the particle passing through the slit.
We need $C$ narrow in time, so we make its dispersion in energy large,
enough to dominate the sum in (\ref{eq:dispersion-in-time}). We can
do this by setting its original dispersion in three momentum large
and relying on the entropic estimate to set the dispersion in energy
large as well. Now the dispersion in energy in $A$ (post-interaction)
is large and its dispersion in coordinate time is therefore small:
$A$ will be strongly localized in time. We will have the desired
small $\sigma_{t}$.

By judicious manipulation of the uncertainty from SQM -- reducing
it by increasing the velocity in space and by reducing the dispersion
in momentum -- while at the same time using the shortest possible
wave function for $C$, we should be able to use $C$ as a time gate,
and get a workable ratio of:

\begin{equation}
\frac{{\Delta_{t}^{\left(T\right)2}}}{{\Delta_{t}^{\left(T\right)2}+\Delta_{\tau}^{\left(S\right)2}}}
\end{equation}
enough to make the effects of TQM visible if it is correct. Or \emph{prove}
it is false, if it is not correct.

\subsubsection{Scattering of indistinguishable particles}

\label{subsec:Scattering-of-indistinguishable}

\begin{figure}
\includegraphics[scale=0.6]{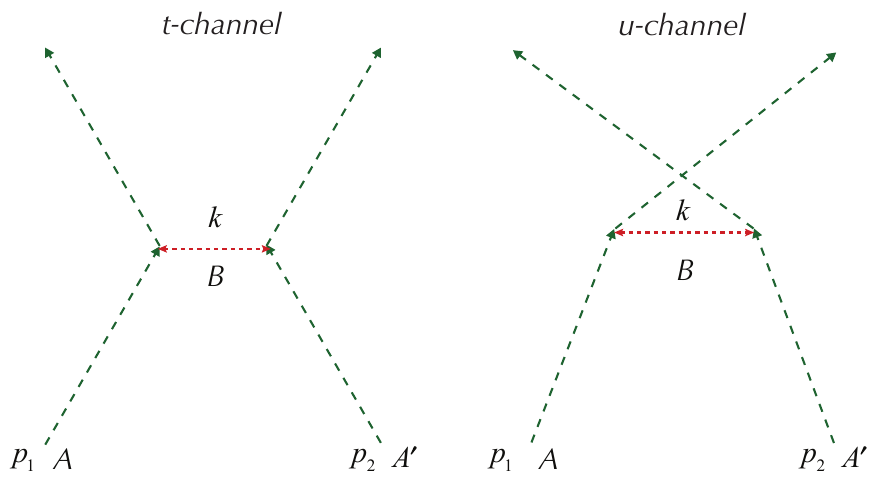}

\caption{Two \emph{A} particles exchange a \emph{B} particle}
\end{figure}

If we replace the $C$ particle with a second $A$ particle we have
two indistinguishable particles to deal with. We therefore cannot
tell whether the particle we detect at location 3 is the one that
started at location 1 or the one that started at location 2. The diagram
on the left is again the $t$-channel as above. The crossed diagram
on the right is associated with the $u$-channel:

\begin{equation}
u={\left({{p_{1}}-{q_{4}}}\right)^{2}}={\left({{p_{2}}-{q_{3}}}\right)^{2}}
\end{equation}

The sum of the $t$ and $u$ channel diagrams gives the total amplitude.
The core matrix element will go as:

\begin{equation}
{\lambda^{2}}\left({\frac{1}{{{{\left({{{\vec{p}}_{3}}-{{\vec{p}}_{1}}}\right)}^{2}}+{\mu^{2}}+\imath\varepsilon}}+\frac{1}{{{{\left({{{\vec{p}}_{3}}-{{\vec{p}}_{2}}}\right)}^{2}}+{\mu^{2}}+\imath\varepsilon}}}\right)
\end{equation}
Since we are dealing with bosons, the relative sign is positive and
the two amplitudes add. We see we have symmetry under the interchange
of $1\leftrightarrow2$ as expected.

By our fundamental hypothesis this rule must apply in TQM for time
and the three space coordinates as a whole. To take advantage of this,
consider a starting wave function which is symmetric under the interchange
of both time and space, but anti-symmetric under each separately.
This is not that hard to arrange: we showed in paper A that if the
left side has, say, large dispersions in time and space, the right
side small -- like one of those comedy duos with a tall wide partner
and a short thin one -- then the total wave function will be the
sum of a part which is symmetric in both time and space and a part
which is anti-symmetric in time and space separately but symmetric
in the combination.

The anti-symmetric part will then give rise to a probability distribution
which has a part symmetric in space and time separately and a part
anti-symmetric in space and in time, but still symmetric in the combination.
For instance it might be symmetric going from $+y,+t\to-y,-t$, but
anti-symmetric going $+y\to-y$. Identical yet opposite considerations
apply to fermions. 

The result is we will see small apparent violations of space (anti-)symmetry
in TQM where it is mandated in SQM. As we suspect that this effect
is of the second or third order of practicality, we do not pursue
it further here. But it is suggestive of the large number of experimental
possibilities that TQM generates.

\subsubsection{Primary effects}

\label{subsec:Primary-effects}

Judging by this simple example, the first order of corrections from
TQM comes less from the propagators and normalization factors; more
from the simple presence of time dispersion in the initial wave functions.
There are three kinds of effects we might look for: 

\paragraph{Increased dispersion in time/energy}

These are expected typically small, of scale attoseconds. However
they are expected present in all cases.

\paragraph{Diffraction in time}

In a scattering experiment, we can use one particle, narrowly focused
in time, to act as a kind of ``single slit in time'' experiment
with respect to another. This provides a direct test of the HUP in
time/energy, and appears to be the most promising of the candidate
experiments.

\paragraph{Entanglement in time/energy}

Anti-symmetry in time is an example of this. These in some ways the
most interesting, but may be a bit subtle for use in a first attack
on the problem.

\subsection{Simple loop diagram}

\label{sec:massloop}

\subsubsection{Ultraviolet divergences}

\label{par:Ultraviolet-divergences}

\begin{figure}
\includegraphics[scale=0.75]{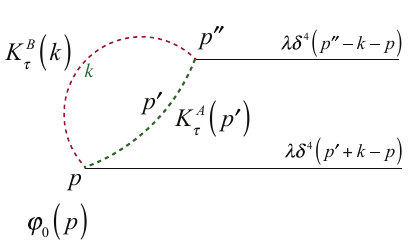}

\caption{Simple mass correction loop}
\end{figure}

In the founding days of QED the problem of the loop divergences was
one of the most troubling. If you look at essentially any loop diagram
in QED, the associated integral diverges. For instance, the loop diagram
for an electron to emit and then absorb a photon is divergent.

We will work, as above, with our simple ABC model which has the same
problem and for the same reason. Consider the amplitude for an $A$
particle to emit a $B$ particle and then absorb it. We assume a starting
momentum for the \emph{$A$} particle of $p$ and integrate over all
possible values of the $B$ particle's four momentum $k$:

\begin{equation}
L=\int{d\omega d\vec{k}\frac{\imath}{{{\omega^{2}}-{{\vec{k}}^{2}}-{\mu^{2}}+\imath\varepsilon}}\frac{\imath}{{{{\left({E-\omega}\right)}^{2}}-{{\left({\vec{p}-\vec{k}}\right)}^{2}}-{m^{2}}+\imath\varepsilon}}}
\end{equation}
This integral does not converge. If we focus on the high momentum
part of the integral, the integrand goes as $1/k$. If we impose a
high momentum cutoff $\Lambda$, the integral diverges logarithmically:

\begin{equation}
\int\limits _{}^{\Lambda}{\frac{{dk}}{k}\sim\ln\left(\Lambda\right)}
\end{equation}
This was very troubling to the founders of QED. This particular form
of the loop shows up as simple corrections to the mass of a particle,
so would be expected small. And these infinities show up in all the
basic loop diagrams.

\subparagraph{Renormalization}

Now suppose we were to compare the value of this loop diagram taken
at a specific value of $p$ to the value of the same loop diagram
taken at a slightly different starting value $p'=p+\delta p$:

\begin{equation}
L'=\int{d\omega d\vec{k}\frac{\imath}{{{k^{2}}-{\mu^{2}}+\imath\varepsilon}}\frac{\imath}{{{{\left({p+\delta p-k}\right)}^{2}}-{m^{2}}+\imath\varepsilon}}}
\end{equation}
The difference between the two loop diagrams goes as:

\begin{equation}
L'-L=\frac{{\partial L}}{{\partial p}}\delta p
\end{equation}
The derivative picks up an extra factor of the \emph{A} propagator
in the denominator:

\begin{equation}
\frac{{\partial L}}{{\partial p}}\to-\frac{{2\imath\left({p-k}\right)}}{{{\left({{{\left({p-k}\right)}^{2}}-{m^{2}}+\imath\varepsilon}\right)}^{2}}}
\end{equation}
causing the integrand to acquire an extra factor of $1/k$:

\begin{equation}
\frac{1}{{k^{4}}}\to\frac{1}{{k^{5}}}
\end{equation}
and thereby making the integral converge. So that while the original
loop is infinite, the differences between nearby loops are finite.
Since we do not actually measure any mass directly, but only by comparison
-- every scale needs a weight on the left side and on the right --
we are not strictly required to compute the absolute value of any
loop but only the difference between it and another. The calculations
are tricky (see for example \cite{Bjorken:1965mo,Peskin:1995rv,Klauber:2013tg,Schwartz:2014tx})
but give not only finite results but extraordinarily good agreement
with experiment. 

The comparison program -- referred to as renormalization -- has
been worked out to a high degree of sophistication. It can be done
to all orders, for not only QED but for the entire Standard Model,
and in a way that is covariant. So we have a procedure that makes
no sense (why are these basic loop diagrams infinite?) yet agrees
perfectly and to many places with experiment. 

This problem would appear to be even more severe in the case of TQM.
TQM has one more dimension to integrate over, so we would expect its
divergences to be linear, perhaps even beyond the reach of a renormalization
program. This would not keep us from using TQM as a way to develop
a program to develop interesting experiments involving time and quantum
mechanics, but it would make clear that TQM could not be an entirely
satisfactory theory in its own right.

\subsubsection{Simple loop calculation}

\label{par:Simple-loop-calculation}

We are not going to show that TQM is fully renormalizable. But we
will show that simple loop diagrams in TQM -- we will focus on the
$A,B$ mass correction diagram -- are in fact finite and do not require
regularization in the first place, provided they are done in a way
that is consistent with the spirit of TQM. It is still true that we
never measure a mass or any other quantity in isolation but only in
comparison with other measurements -- but the associated integrals
are not only finite but small.

The key is to take advantage of two points:
\begin{enumerate}
\item We have to start with a physically realistic wave function, one which
is normalizable itself. Our GTFs will work well for this.
\item In computing each step in a loop diagram we have to take advantage
of the entanglement in time of that step with the previous. This lets
each step take advantage of the finite character of the previous steps
to make sure that it stays finite as well. This entanglement in time
is not available in SQM, since it is a quantum feature not available
to any theory that treats time classically.
\end{enumerate}
We will do the calculation in two stages:
\begin{enumerate}
\item Calculate the value of the loop for a fixed clock time.
\item Take the Fourier transform of the fixed clock time result, to get
the actual mass correction.
\end{enumerate}
The attosecond form of the propagator is not suitable for this. This
is one of the places where that approximation breaks down. But the
full, ``unpacked'' propagator works. We compute the loop from past
to future, so need only the positive time branches of the propagators
for the $A$ and $B$ particles. Since we are only interested in the
questions ``is the loop integral finite?'' and ``does the loop
integral make sense?'' we ignore the squared coupling constant and
an overall factor of $\imath$. We are after the loop, the whole loop,
and nothing but the loop.

We start as usual with a GTF. We again use the narrow beam approximation,
assuming that for all four momentum components $\frac{\sigma}{{\left|p\right|}}\ll1$.The
propagator for the $A$ particle is:

\begin{equation}
K_{\tau}^{A}\left(p\right)=\frac{1}{{2E}}\exp\left({-\imath{\mathcal{\varpi}_{p}}\tau}\right),{\mathcal{\varpi}_{p}}=-\frac{{{E^{2}}-{{\vec{p}}^{2}}-{m^{2}}}}{{2E}}
\end{equation}
The propagator for the \emph{$B$} particle is:

\begin{equation}
K_{\tau}^{B}\left(k\right)=\frac{1}{{2w}}\exp\left({-\imath{\omega_{k}}\tau}\right),{\varpi_{k}}=-\frac{{{w^{2}}-{{\vec{k}}^{2}}-{\mu^{2}}}}{{2w}}
\end{equation}
For fixed clock time the loop integral is:

\begin{equation}
{L_{\tau}}\left(p\right)=\int{{d^{4}}k\hat{K}_{\tau}^{A}\left({p-k}\right)\hat{K}_{\tau}^{B}\left(k\right){{\hat{\varphi}}_{0}}\left(p-p_{0}\right)}
\end{equation}
We are using the sample GTF (\ref{eq:sample-tqm-gtf}) from the GTF
appendix.

In the narrow beam approximation we may estimate the normalization
factor using:

\begin{equation}
E\to{E_{0}\equiv\sqrt{{m^{2}}+\vec{p}_{0}^{2}}},w\to\mu
\end{equation}

Since these factors are in the denominator, this approximation will
make the integral more divergent rather than less. Since they are
now constant, we can pull them out of the integral. For the rest of
the analysis we will work with a re-scaled loop integral $L\to4{E_{0}}\mu L$.

We will take the same step with respect to the factors of clock time
in the exponentials:

\begin{equation}
{\varpi_{k}^{B}}\approx-\frac{{{w^{2}}-{{\vec{k}}^{2}}-{\mu^{2}}}}{{2\mu}},{\varpi_{p}^{A}}\approx-\frac{{{E^{2}}-{{\vec{p}}^{2}}-{m^{2}}}}{{2{E_{0}}}},
\end{equation}
This will tend to make the arguments of the exponentials oscillate
less, again making the integrals more divergent rather than less.

The loop integral is a pair of convolutions in momentum space. If
we shift to coordinate space, these convolutions will turn into the
coordinate space equivalents. The GTF is:

\begin{equation}
{\varphi_{0}}\left(x\right)=\sqrt[4]{{\frac{1}{{{\pi^{4}}\det\left(\Sigma\right)}}}}\exp\left({-\imath{p_{0}}x-\frac{{{\left({x-{x_{0}}}\right)}^{2}}}{{2\Sigma}}}\right),\Sigma={{\hat{\Sigma}}^{-1}}
\end{equation}
The coordinate space forms of the kernels are (compare to \ref{par:TQM-Kernels}):

\begin{equation}
\begin{array}{c}
K_{\tau}^{A}\left({{x_{1}};{x}}\right)=-\imath\frac{{E_{0}^{2}}}{{4{\pi^{2}}{\tau^{2}}}}\exp\left({-\frac{{\imath E_{0}}}{{2\tau}}{{\left({{x_{1}}-{x}}\right)}^{2}}}{+\imath\frac{m^{2}}{2E_{0}}\tau}\right)\hfill\\
K_{\tau}^{B}\left({{x_{1}};{x}}\right)=-\imath\frac{{\mu^{2}}}{{4{\pi^{2}}{\tau^{2}}}}\exp\left({-\frac{{\imath\mu}}{{2\tau}}{{\left({{x_{1}}-{x}}\right)}^{2}}}{+\imath\frac{\mu}{2}\tau}\right)\hfill
\end{array}
\end{equation}
The product equals a single coordinate space kernel: 

\begin{equation}
K_{\tau}^{\left(M\right)}\left({{x_{1}};{x}}\right)=-\imath\frac{{M^{2}}}{{4{\pi^{2}}{\tau^{2}}}}\exp\left({-\imath\frac{{M}}{{2\tau}}{{\left({{x_{1}}-{x}}\right)}^{2}}}{+\imath\frac{M}{2}\tau}\right)
\end{equation}
with a modified mass $M\equiv E_{0}+\mu$, a prefactor $-\imath\frac{1}{{4{\pi^{2}}}}\frac{{E_{0}^{2}{\mu^{2}}}}{{{\tau^{2}}{M^{2}}}}$,
and a post-factor $\exp\left({-\imath\frac{M}{2}\tau+\imath\frac{{m^{2}}}{{2{E_{0}}}}\tau+\imath\frac{\mu}{2}\tau}\right)$.
In the limit as $\mu\to0$ we have $M\to{E_{0}}$.

So the loop term in coordinate space is (without prefactor and post-factor):

\begin{equation}
{L'_{\tau}}\left({x_{1}}\right)=\int{{d^{4}}{x}K_{\tau}^{\left(M\right)}\left({{x_{1}};{x}}\right){\varphi_{0}}\left({x}\right)}
\end{equation}
This is just the integral to advance a wave function of mass $M$
a distance $\tau$ in time, so we have:

\begin{equation}
{L'_{\tau}}\left({x_{1}}\right)=\varphi_{\tau}^{\left(M\right)}\left({x_{1}}\right)
\end{equation}
We have a correction that shows a spread in time, but at the slightly
slower rate associated with the slightly larger mass $M$. The full
correction is greater at shorter clock times, as one would expect.
Now we transform back to momentum space:

\begin{equation}
\begin{array}{c}
{{L'}_{\tau}}\left({p_{1}}\right)=\int{{d^{4}}{p}\hat{K}_{\tau}^{\left(M\right)}\left({{p_{1}};{p}}\right){{\hat{\varphi}}_{0}}\left({p}\right)}\hfill\\
\hat{K}_{\tau}^{\left(M\right)}\left({{p_{1}};{p}}\right)=\exp\left({\imath\frac{{p^{2}-{M^{2}}}}{{2M}}\tau}\right)\delta^{4}\left({{p_{1}}-{p}}\right)\hfill
\end{array}
\end{equation}
So the loop correction for fixed clock time (folding back in the prefactor
and post-factor):

\begin{equation}
{L_{\tau}}\left({p_{1}}\right)=-\imath\frac{1}{{4{\pi^{2}}}}\frac{{{m^{2}}{\mu^{2}}}}{{{M^{2}}{\tau^{2}}}}\exp\left({\imath\frac{{p_{1}^{2}}}{{2M}}{\tau}}+\imath\frac{{m^{2}}}{{2{E_{0}}}}\tau+\imath\frac{\mu}{2}\tau\right){{\hat{\varphi}}_{0}}\left({p_{1}}\right)
\end{equation}
At this point the value of the loop correction at a particular value
of $p$ is independent of the specific shape of the incoming wave
function. We are therefore free to drop the initial wave function
from the analysis. We rewrite the loop without the initial wave function,
but fold back in the trailing factor:

\begin{equation}
{L_{\tau}}\left(p\right)=-\imath\frac{1}{{4{\pi^{2}}}}\frac{{{m^{2}}{\mu^{2}}}}{{{M^{2}}{\tau^{2}}}}\exp\left({-\imath\varpi_{p}^{M}\tau}\right),\varpi_{p}^{M}\equiv-\frac{{p^{2}}}{{2\left({{E_{0}}+\mu}\right)}}-\frac{{m^{2}}}{{2{E_{0}}}}-\frac{\mu}{2}
\end{equation}
The $\varpi_{p}^{M}\to\varpi_{p}$ as $\mu\to0$. The initial GTF
acted in a way parallel to the regularization factors often employed
in SQM, but it comes organically out of the calculation, it is not
supplied by hand.

\subsubsection{Fourier transform}

\label{par:mass-correction}

The actual mass correction is given by the Fourier transform of the
loop integral. Now that we have the loop integral for a specific value
of the clock time we can take the Fourier transform with respect to
$\tau$. We have the value of the Fourier transform of the core element:

\begin{equation}
\mathcal{F}\mathcal{T}\left[{\frac{{\exp\left({-\imath\varpi_{p}^{M}\tau}\right)}}{{\tau^{2}}}}\right]=-\sqrt{\frac{\pi}{2}}\left|{\omega-\varpi_{p}^{M}}\right|
\end{equation}
and therefore of the loop:

\begin{equation}
{{\hat{L}}_{\omega}}\left(p\right)=\imath\frac{1}{{4{\pi^{2}}}}\frac{{{m^{2}}{\mu^{2}}}}{{M^{2}}}\sqrt{\frac{\pi}{2}}\left|{\omega-\varpi_{p}^{M}}\right|
\end{equation}
The loop correction is zero if $\omega=\varpi$. However in TQM this
is a set of measure zero: the 4D wave functions will form a cloud
around the average, so the correction will be proportional to the
uncertainty in the coordinate energy:

\begin{equation}
\left\langle {\left|{\omega-\varpi_{p}^{M}}\right|}\right\rangle \sim\left\langle {\frac{{{\left({\delta E}\right)}^{2}}}{{2\left({{E_{0}}+\mu}\right)}}}\right\rangle \approx\frac{{\sigma_{E}^{2}}}{{4\left({{E_{0}}+\mu}\right)}}
\end{equation}
So although the initial wave function dropped out of the loop calculation,
its influence, like the smile of the Cheshire cat, is still felt. 

\subsubsection{Implications }

\label{par:Implications}

This is a simple result for a toy model. But it was achieved without
artifice. We did not have to add regularizing or convergence factors
by hand to the integrals; the initial wave function acted as the convergence
factor. And the results make sense: the correction is small, and still
smaller when the mass of the $B$ is smaller.

This is perhaps not that surprising. In classical electrodynamics
comparable calculations, as of the self-energy of the electron, suffered
from linear divergences. The transition from classical electrodynamics
to QED softened these divergences from linear to logarithmic. 

TQM represents a further step from the classical to the quantum mechanical
approaches. We are applying quantization procedures used in space
and applying them in time as well. We see a further ``softening''
of the integrals, to the point where they are finite.

Therefore we have addressed our motivating question: is TQM ruled
out as a theory in its own right by not being renormalizable?

And we have provided additional evidence to suggest that TQM is worth
exploring experimentally.

\subsection{Second order-of-magnitude tree diagrams}

\label{subsec:Simple-Feynman-diagrams}

We discuss here the second order tree diagrams. There are three:
\begin{enumerate}
\item Møller scattering: two electrons exchange a photon. 
\item Bhabha scattering: an electron and a positron either exchange a photon
or else annihilate and then recreate themselves. 
\item Compton scattering: a photon and an electron scatter. 
\end{enumerate}
We see photons and fermions as composed of a spin zero component times
a polarization vector or times a spinor. To compute the associated
$S$ matrices, we have to look at the input wave functions, the external
normalization factors, the polarization vectors, and the spinors.
We are primarily looking for additional dispersion in time-of-flight
measurements.

If we use the narrow beam approximation, we will be able to ignore
the effects of the changes to the external normalization factors,
spinors, and polarization vector to lowest order. The SQM part of
the diagram will function, as above, largely as a carrier. The principle
effects of TQM will be:
\begin{enumerate}
\item The general presence of dispersion in time, leading to additional
uncertainty at a detector.
\item Diffraction in time, as in the HUP in time/energy.
\item Entanglement in time, as in the effects of anti-symmetry in the time
part of the wave function.
\end{enumerate}

\subsubsection{Møller scattering}

\label{subsec:M=0000F8ller-scattering}

\begin{figure}
\includegraphics[scale=0.6]{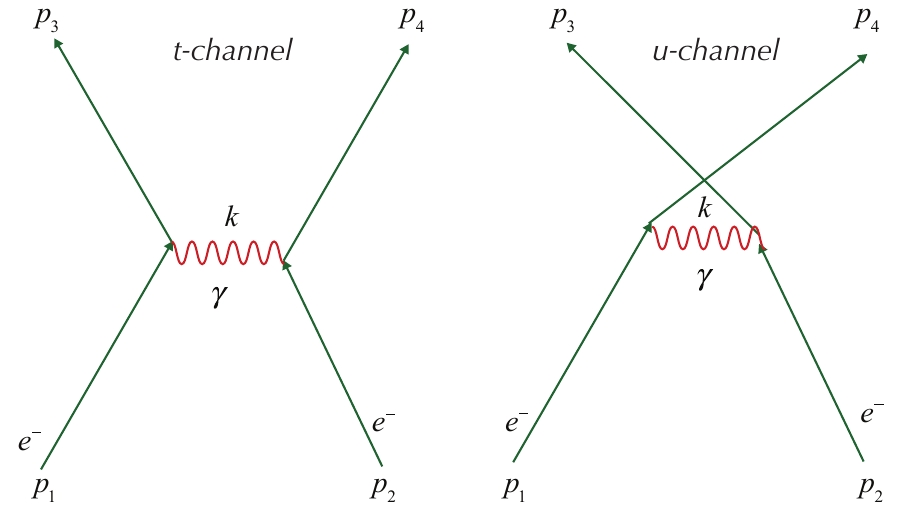}

\caption{Møller scattering}
\end{figure}

Møller scattering is the term for electron-electron scattering via
the exchange of an electron. From our point of view Møller scattering
is a relatively minor generalization of the spin zero scattering problem
in subsection \ref{subsec:Scattering-of-indistinguishable}. The differences
between the SQM and TQM versions of this are in the external factors,
the photon propagator, and the spinors at the vertex. 

We use Mandelstam variables $t,u$ to characterize the interaction:

\begin{equation}
\begin{gathered}t={\left({{p_{1}}-{p_{3}}}\right)^{2}}={\left({{p_{2}}-{p_{4}}}\right)^{2}}\hfill\\
u={\left({{p_{1}}-{p_{4}}}\right)^{2}}={\left({{p_{2}}-{p_{3}}}\right)^{2}}\hfill
\end{gathered}
\end{equation}
The heart of the $S$ matrix is the photon propagator. Here the mapping
from SQM to SQM is simple, if we use the attosecond propagator:

\begin{equation}
\imath D_{\omega}^{\left(S\right)\mu\nu}\left({\vec{k}}\right)=\frac{{-\imath{g^{\mu\nu}}}}{{{\omega^{2}}-{{\vec{k}}^{2}}+\imath\varepsilon}}\to\imath D_{\omega}^{\left(A\right)\mu\nu}\left(k\right)=\frac{{-\imath{g^{\mu\nu}}}}{{{w^{2}}-{{\vec{k}}^{2}}}}\label{eq:moller-photon-propagator}
\end{equation}
We are mapping $\omega\to w$ (coordinate frequency to clock frequency).
In SQM, $\omega\to0$, so the propagator is just:

\begin{equation}
\imath D_{\omega}^{\left(S\right)\mu\nu}\left({\vec{k}}\right)=\frac{{\imath{g^{\mu\nu}}}}{{{\vec{k}}^{2}}}
\end{equation}
The same analysis as in the spin zero case gives same result for the
TQM propagator, so it too is:

\begin{equation}
\imath D_{\omega}^{\left(A\right)\mu\nu}\left(k\right)=\frac{{\imath{g^{\mu\nu}}}}{{{\vec{k}}^{2}}}
\end{equation}
And therefore the values of the $t$ and $u$ variables are the same
for both SQM and TQM:

\begin{equation}
\frac{1}{t}\approx\frac{1}{{t^{S}}},\frac{1}{u}\approx\frac{1}{{u^{S}}}
\end{equation}
where the SQM context merely means we replace coordinate energy $w$
by its average $\omega$, which is zero.

In the external factors we have to map $E_{\vec{p}}\to E$:

\begin{equation}
\sqrt{\frac{m}{{E_{\vec{p}}}}}\to\sqrt{\frac{m}{E}}
\end{equation}
This is where the narrow beam approximation is useful. We are assuming
that the variation in energy and the other components of the four
momentum are no more than, say, 1\% of the average for each component.
Therefore if we are looking for order-of-magnitude changes only, we
can use the SQM external factor for TQM.

The vertex contribution is more complex. To go from the ABC model
to QED in SQM we have:

\begin{equation}
\imath\lambda ABA,\imath\lambda CBC\to-\imath e\bar{\psi}A^{\nu}{\gamma_{\nu}}\psi
\end{equation}
To then go from SQM to TQM we replace $E_{\vec{p}}\to E$ in the spinors:

\begin{equation}
-\imath e\bar{\psi}^{S}\left({\vec{p}'}\right)A^{\nu}\left(\vec{k}\right){\gamma_{\nu}}\psi^{S}\left({\vec{p}}\right)\to-\imath e\bar{\psi}\left({p'}\right)A^{\nu}\left(k\right){\gamma_{\nu}}\psi\left({p}\right)
\end{equation}
The $\psi$'s are sums over the $u,v$'s. The $u,v$'s depend on $E$
via factors of $\sqrt{\frac{{E+m}}{{2m}}},\frac{1}{{E+m}}$. We can
expand these in power series of $\delta E$:

\begin{equation}
\begin{gathered}\sqrt{\frac{{E+m}}{{2m}}}=\sqrt{\frac{{{E_{\vec{p}}}+m}}{{2m}}}+\frac{{\delta E}}{{2m\sqrt{{E_{\vec{p}}}+m}}}-\frac{{{\left({\delta E}\right)}^{2}}}{{16m{{\sqrt{{E_{\vec{p}}}+m}}^{3}}}}+\ldots\hfill\\
\frac{1}{{E+m}}\approx\frac{1}{{{E_{\vec{p}}}+m}}-\delta E\frac{1}{{{\left({{E_{\vec{p}}}+m}\right)}^{2}}}+{\left({\delta E}\right)^{2}}\frac{1}{{{\left({{E_{\vec{p}}}+m}\right)}^{3}}}-\ldots\hfill
\end{gathered}
\end{equation}
In the narrow beam approximation we drop the first order and higher
terms in $\delta E$ so that $u,v\approx{u^{S}},{v^{S}}$. The implication
is that the matrix elements in TQM are the same as in SQM:

\begin{equation}
S\approx S^{S}
\end{equation}
All the usual trace-tricks will work the same way: use the SQM procedures
replacing $E_{\vec{p}}\to E$ throughout, then drop the quantum energy
component $\delta E\equiv E-E_{\vec{p}}$ leaving us right back where
we started with the SQM case.

This implies that the first order effects are, as with the spin zero
case, a function of the GTFs, rather than the propagator, spinors,
or the normalizations.

If we may write the incoming wave functions as the direct product
of a spinor part and a GTF, then we have for any SQM calculation the
same formula for the TQM signal as with the spin zero case. We have
to lowest order reduced the Møller calculation to the ABC one.

\paragraph{Samurai versus pirate}

\label{par:Samurai-versus-pirate}

Now that we are dealing with charged particles we can employ some
additional techniques to help prove TQM effects do not exist. Consider
a charged particle going through a magnetic field. It will have a
radius of curvature given by:

\begin{equation}
r=\frac{{mv}}{{qB}}=\frac{{p_{x}}}{{eB}}
\end{equation}
so the faster it is going, the greater the radius of curvature. So
the larger $p_{x}$, the less the path of the particle is bent. This
provides us a way to address the ``slow train'' problem. We can
start with the FS/T equation for a single particle, using the minimal
substitution to include the magnetic and electric fields. We ignore
spin:

\begin{equation}
\imath E\frac{{\partial{\psi_{\tau}}}}{{\partial\tau}}=-\frac{{\left({\left({{p_{\mu}}-q{A_{\mu}}}\right)\left({{p^{\mu}}-q{A^{\mu}}}\right)-{m^{2}}}\right)}}{2}{\psi_{\tau}}\label{eq:moller-FS/T}
\end{equation}
We can use this equation to calculate the motion of a TQM wave packet
as function of $x,y,t$. We can use the Klein-Gordon equation with
the minimal substitution to calculate the motion of the particle in
the SQM case.

Suppose the particle is going in the $x$ direction and the magnetic
field is in the $z$ direction. The force from the magnetic field
will be in the $y$ direction. The slower particles will have a smaller
radius of curvature, so will be pushed further in the $y$ direction.
They will also arrive later, so if the particles are being bent to
the right and time is being tracked going up, the main trace will
go to the right and up. It will look a bit like a sword trace.

Since the magnetic field has no effect on the time part, the dependence
on $p_{x}$ and on $E$ are separate. If SQM is true, then the dispersion
in time at each $y$ position will be small. If TQM is true, then
the dispersion in time at each $y$ position will be greater, depending
on the specifics significantly greater. For SQM the sword trace will
be narrow, looking like the thin precise scar left by a skilled samurai's
katana. But if TQM is true, then the sword trace will be broader,
looking more like the undisciplined scar left by a pirate's cutlass.

\subsubsection{Bhahba scattering}

\label{subsec:Bhahba-scattering}

In Bhabha scattering we look at an electron scattering from a positron.
They can either do this by exchanging a photon, with results very
like those for the Møller case, or they can interact by annihilating
with emission of a photon, which then decays into an electron-positron
pair.

\begin{figure}
\includegraphics[scale=0.6]{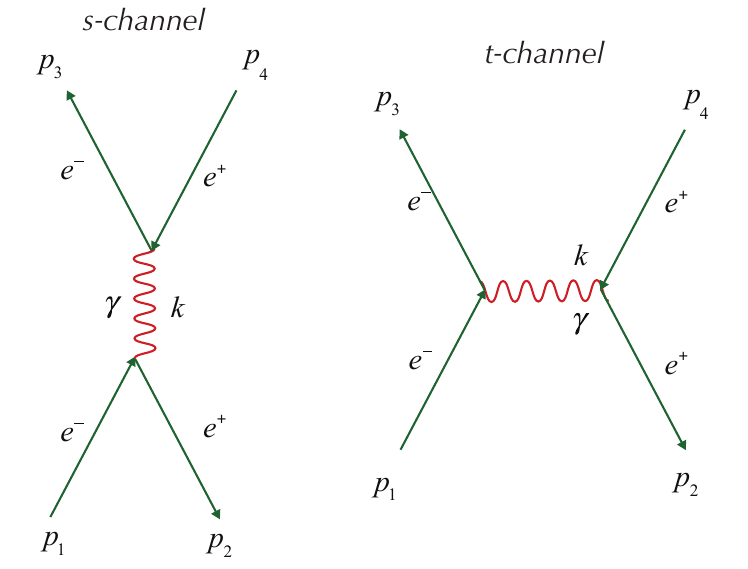}

\caption{Bhabha scattering}
\end{figure}

There are two slight differences from the Møller case for our purposes.
The exchanges are described by the $s$ and $t$ channels:

\begin{equation}
\begin{gathered}s={\left({{p_{1}}+{p_{2}}}\right)^{2}}={\left({{p_{3}}+{p_{4}}}\right)^{2}}\hfill\\
t={\left({{p_{1}}-{p_{3}}}\right)^{2}}={\left({{p_{2}}-{p_{4}}}\right)^{2}}\hfill
\end{gathered}
\end{equation}
This is the first point at which the Machian hypothesis has come up
as a possible issue. Recall this gave us a well-defined value for
the magnitude and sign of the clock frequency. The symmetry between
particle and anti-particle means that the magnitude of the clock frequency
can't plausibly change. The only reasonable possibility is that the
sign changes for anti-matter. However for a normal wave packet the
clock frequency will be averaged over values of the coordinate energy
centered on the on-shell case. Therefore the sign of the clock frequency
will be averaged away.

The other point is that for the first time, the photon propagator
will be carrying a non-zero energy component, in the pair annihilation
/ pair creation branch ($s$-channel):

\begin{equation}
\imath D_{\omega}^{\left(S\right)\mu\nu}\left({\vec{k}}\right)=\frac{{-\imath{g^{\mu\nu}}}}{{{\omega^{2}}-{{\vec{k}}^{2}}+\imath\varepsilon}}\to\imath D_{\omega}^{\left(A\right)\mu\nu}\left(k\right)=\frac{{-\imath{g^{\mu\nu}}}}{{{{\left({\omega+\delta w}\right)}^{2}}-{{\vec{k}}^{2}}}}\approx\frac{{-\imath{g^{\mu\nu}}}}{{{\omega^{2}}-{{\vec{k}}^{2}}}}
\end{equation}
Here again we invoke the narrow beam approximation to argue that the
dependence on $\delta w$ in the denominator of the propagator will
be averaged out in first order and small in second.

And we are left with again the position that all significant dependence
on TQM will be in the GTF part of the wave function, there will be
no first order dependence on the spinor parts.

The results for the GTFs are essentially the same as with the ABC
case: the two final dispersions in energy will be the averages of
the initial dispersions in energy. If one of the two incoming particles
has a much greater dispersion in energy, it will dominate the result.
This particle will then act as a narrow gate in time with respect
to the other, with results already discussed.

\subsubsection{Compton scattering}

\label{subsec:Compton-scattering}

\begin{figure}
\includegraphics[scale=0.6]{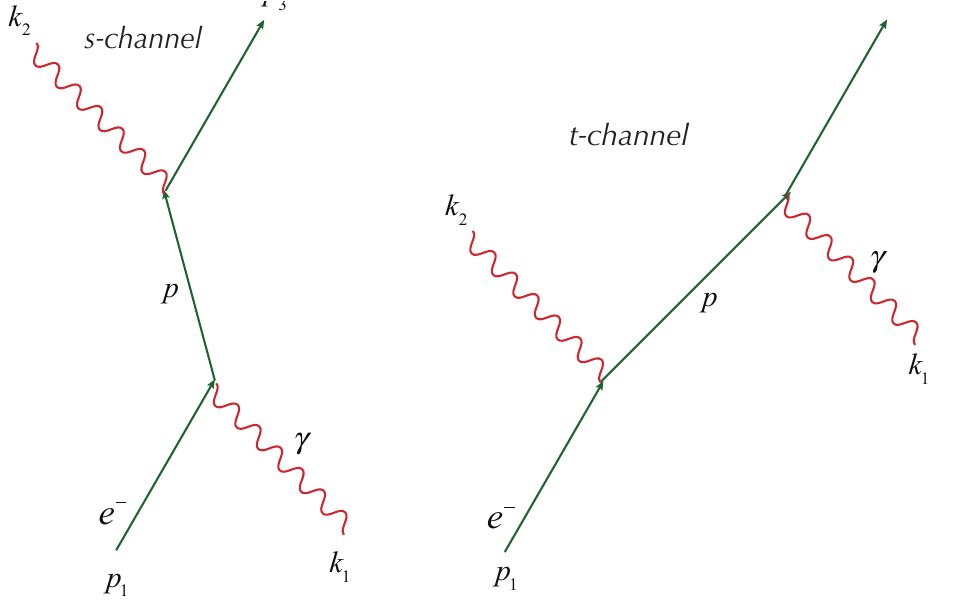}

\caption{Compton scattering}
\end{figure}

The exchanges are described by the $s$ and $t$ channels:

\begin{equation}
\begin{gathered}s={\left({{p_{1}}+{k_{1}}}\right)^{2}}={\left({{p_{3}}+{k_{2}}}\right)^{2}}\hfill\\
t={\left({{p_{1}}-{k_{2}}}\right)^{2}}={\left({{k_{1}}-{p_{3}}}\right)^{2}}\hfill\\
\hfill
\end{gathered}
\end{equation}
This is our first chance to see the fermion propagator in action:

\begin{equation}
\imath S_{\omega}^{S}\left({\vec{p}}\right)=\imath\frac{{\left({\cancel{{p^{S}}}+m}\right)}}{{{E^{\left(S\right)2}}-{{\vec{p}}^{2}}-{m^{2}}}}\to\imath S_{\omega}^{\text{A}}\left(p\right)=\imath\frac{{\left({\cancel{p}+m}\right)}}{{{E^{2}}-{{\vec{p}}^{2}}-{m^{2}}}}
\end{equation}
We see that the situation is not much different than with the two
previous cases. The difference between ${{\not p}^{S}}$ and $\cancel{p}$
drops out in the narrow beam approximation, as does the difference
between $E^{S}$ and $E$ in the denominator. We are left with SQM
as carrier and the TQM GTF in time/energy the signal.

The main point of interest here is that as our technology for creating
short photon pulses is now extremely sophisticated \cite{Plaja:2013,Lin:2018uh},
the changes of using a short pulse of light as a ``narrow gate in
time'' should be good. We discuss this next.

\section{Experimental Tests}

\label{sec:Experiments}
\begin{quotation}
“In so far as a scientific statement speaks about reality, it must
be falsifiable: and in so far as it is not falsifiable, it does not
speak about reality.” -- Karl Popper \cite{Popper:1968ul}
\end{quotation}
Our goal in this investigation has not been to argue that TQM is a
correct extension of QED but rather that it is falsifiable, to give
Gisin and his peers ``something they can prove wrong''.

There is no question that due to the small size of expected effects
the associated experiments will be difficult. But at the same time
there is a compensating variety of experiments: any time dependent
system monitored by time sensitive detectors should show small but
definite effects of dispersion in time.

Here we look at one possible experiment, define a figure of merit
for such experiments, and then review experimental possibilities in
general.

\subsection{Heisenberg uncertainty principle in time/energy}

\label{sec:Ultra-fast-gate}
\begin{verse}
``Ah, but a man's reach should exceed his grasp, Or what's a heaven
for?'' -- Robert Browning
\end{verse}
\begin{figure}
\includegraphics[scale=0.75]{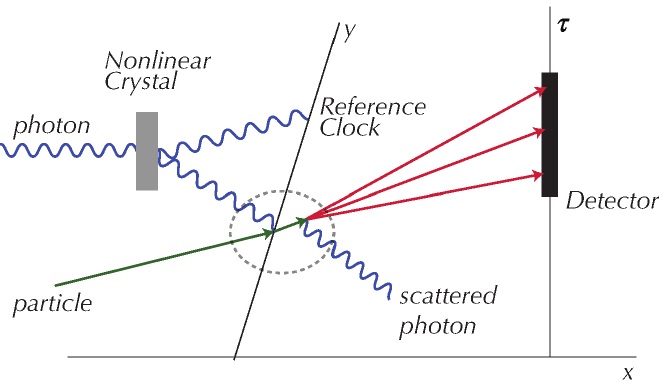}

\caption{Test of Heisenberg uncertainty principle in time}
\end{figure}

We sketch out a simple candidate solution for using the HUP in time/energy
to falsify TQM. 
\begin{enumerate}
\item Suppose we start with a photon with a narrow width in time (as in
Lindner's classic experiment \cite{Lindner:2005vv}). 
\item We send this through a non-linear crystal to split the photon into
two equal-but-opposite photons, as in tests of Bell's theorem.
\item We send one of these to a reference clock to provide a start point
for a time-of-flight measurement. 
\item We use the other as a narrow gate in time, per above.
\item We arrange for this photon to scatter an electron via Coulomb scattering.
\item We measure the electron's time-of-arrival at a detector. 
\item We compute the time-of-flight as the difference between the time-of-arrival
and the reference time.
\item We do this enough times to build up a time-of-flight distribution
which will either conform to the predictions of TQM or falsify those.
\end{enumerate}
Dr. Klag was kind enough to point out that this will not, in fact,
work. The photons that are split by non-linear crystals in Bell's
theorem tests are much longer in time than those Lindner used. As
we probably need ultra-short photons, this is likely to be a problem.

Still this does show the essential elements of a realistic experiment:
\begin{enumerate}
\item We need a time-of-flight, so need in general both a start time and
a time of arrival.
\item We need something that can act as a narrow gate in time.
\item We are likely to need many many data points.
\end{enumerate}
We next propose a ``figure of merit'' to pick among candidate solutions.

\subsection{Figure of merit}
\begin{verse}
``by recording single electron detection events diffracting through
a double-slit, a diffraction pattern was built up from individual
events.'' -- Bach et al \cite{Bach:2013aa}
\end{verse}
\label{subsec:Comparison-of-TQM-3}

The primary effect of dispersion in time will be to increase the uncertainty
in time from what would otherwise be predicted by standard QED. The
fundamental metric is equation \ref{eq:time-signal}, repeated here
for convenience:

\begin{equation}
\left({\Delta t}\right)_{D}^{\left(T\right)2}\equiv\left({\Delta t}\right)_{D}^{2}-\left({\Delta t}\right)_{D}^{\left(S\right)2}\label{eq:time-signal-1}
\end{equation}
The term on the left is the signal $\left({\Delta t}\right)_{D}^{\left(T\right)2}$:
the uncertainty squared predicted by TQM $\left({\Delta t}\right)_{D}^{2}$
minus the uncertainty squared predicted by SQM $\left({\Delta t}\right)_{D}^{\left(S\right)2}$. 

If we are looking at time-of-flight as the prediction, then the distribution
in time-of-flight will build up one event at a time. If we assume,
say, Gaussian predictions for both TQM and SQM we will expect in general
a wider, flatter Gaussian for TQM.

Assume we pick a degree of confidence, say the traditional five sigmas
(giving a one in 3.5 million chance that the distribution was assigned
incorrectly). There are many different statistical tests for making
this sort of discrimination between two Gaussian distributions; we
will assume one appropriate to the specific situation has been chosen.

The chosen degree of confidence will in turn imply a minimal sample
size $N$ to achieve it. Now suppose our apparatus can run $T$ tests
per second. The number of seconds to achieve the targeted level of
confidence is then:

\begin{equation}
S=\frac{N}{T}
\end{equation}
Our proposed figure of merit is the log to the base ten of the number
of seconds required to achieve a five sigma level of confidence that
TQM is falsified:

\begin{equation}
M\equiv{\log_{10}}\left(S\right)
\end{equation}
If we need 100,000 tests and can run one test per second, then our
figure of merit is:

\begin{equation}
{\log_{10}}\left({100000}\right)=5
\end{equation}
We choose a log scale because we expect there will be considerable
variation in the efficiency of various experimental arrangements.
In this way, arrangements that generate a stronger signal can be fairly
compared to those that generate more tests per second and so on. The
smaller $S$ is the better of course. If $S$ is longer than the mean-time-between-failure
(MTBF) of the apparatus, then the specific experiment is not practical.
And of course if it is longer than the duration of the associated
grant, that too will be a problem.

One advantage of having a reasonably well-defined figure of merit
is that this makes it easier for an AI system e.g. \cite{Nichols:2018aa}
to compare experimental possibilities. AI systems have the specific
advantage in this case that they know nothing about time and are therefore
less likely to be distracted by preconceptions.

\subsection{TQM as an experiment factory}
\begin{quote}
“Henceforth space by itself, and time by itself, are doomed to fade
away into mere shadows, and only a kind of union of the two will preserve
an independent reality'' -- Hermann Minkowski \cite{Minkowski:1908yi}
\end{quote}
\label{subsec:Further-experiments}

With respect to the falsifiability of TQM, the small size of the basic
effect may be compensated for by the large number of experimental
possibilities. If quantum mechanics should in fact be extended in
the time direction then essentially any time dependent apparatus with
time dependent detectors may provide a possible line of attack. By
hypothesis, \emph{all} quantum mechanical phenomena seen in space
-- interference, diffraction, uncertainty, entanglement, tunneling,
… -- apply in time as they do in space. 

\subparagraph{Effects of TQM on the legs}

We have focused here on the applications within QED. This is needed
to treat interactions correctly. However there are many interesting
effects at the single particle level. 

These may described using the FS/T equation for a spinless particle
(\ref{eq:spinzero-FS/T}):

\begin{equation}
\left({{{\left({E-q\Phi}\right)}^{2}}-{{\left({\vec{p}-q\vec{A}}\right)}^{2}}-{m^{2}}}\right){\psi_{\tau}}=-2E\imath\frac{{\partial{\psi_{\tau}}}}{{\partial\tau}}
\end{equation}
or the TQM Dirac equation (\ref{eq:fermions-FS/T}):

\begin{equation}
\left({\cancel{p}-q\cancel{A}-m}\right){\psi_{\tau}}=-\imath{\gamma_{0}}\frac{\partial}{{\partial\tau}}{\psi_{\tau}}
\end{equation}
We include the vector potential via the minimal substitution. We can
also make use of the TQM equation for the free vector potential (\ref{eq:photon-FS/T}):

\begin{equation}
\left({{w^{2}}-{{\vec{k}}^{2}}}\right){A^{\nu}}=-2w\imath\frac{\partial}{{\partial\tau}}{A^{\nu}}
\end{equation}
All three equations are defined with reference to the rest frame of
the vacuum $\mathcal{V}$, so are invariant. As noted, these can often
be usefully simplified by rewriting the energy in terms of the quantum
energy: $E={E^{S}}+\delta E$ where $E^{S}$ is the value of the energy
in SQM, and then assuming that $\delta E$ is small (as in the ``narrow
beam'' approximation).

\subparagraph{Reuse of existing SQM results}

We will often start with an existing solution in SQM, perhaps a hard-earned
part analytic/part numeric approximation specific to a complex mashup
of lab-built and off-the-shelf tech. In many cases we will be able
to reuse these pre-existing solutions.

Assume we can use such a pre-existing solution to compute a time-of-arrival
distribution. The uncertainty in time-of-arrival is defined as:

\begin{equation}
\Delta_{\tau}^{\left(S\right)2}\equiv\int\limits _{-\infty}^{\infty}{d\tau{\left({\tau-{{\left\langle \tau\right\rangle }^{S}}}\right)^{2}}\rho_{\tau}^{S}},{\left\langle \tau\right\rangle ^{S}}\equiv\int\limits _{-\infty}^{\infty}{d\tau\tau\rho_{\tau}^{S}}
\end{equation}
We can also use the existing SQM solution to compute a first order
estimate of the corresponding GTF in coordinate time/energy by using
the entropic estimate. This will usually be much less sophisticated
than our existing SQM solution. But for falsification we are interested
primarily in order-of-magnitude numbers. We therefore use the estimated
GTF in time to infer the corresponding probability distribution in
time ${\rho_{\tau}}\left(t\right)$. 

The total uncertainty in coordinate time will be given by:

\begin{equation}
\Delta_{t}^{2}\equiv\int\limits _{-\infty}^{\infty}{dt\rho\left(t\right){{\left({t-\left\langle t\right\rangle }\right)}^{2}}},\left\langle t\right\rangle \equiv\int\limits _{-\infty}^{\infty}{dt\rho\left(t\right)}t,\rho\left(t\right)\equiv\int\limits _{-\infty}^{\infty}{d\tau{\rho_{\tau}}\left(t\right)}
\end{equation}
The difference between these two metrics is our signal. By this method
existing SQM results for time-of-flight and the like may be converted
into tests of TQM.

\subparagraph{Use of combinations of approaches}

Further as suggested in the Samurai and Pirate experiment \ref{par:Samurai-versus-pirate},
a combination of approaches may be useful. Post-filtering of the results
of an interaction to make the effects of dispersion in time more obvious;
pre-filtering to get a cleaner result are possible approaches.

\subsubsection{Three classes of effects}

The primary effects of TQM may be categorized as the common, the dramatic,
and the subtle.

\paragraph{Common}

By the fundamental hypothesis the effects of dispersion in time are
omnipresent. In a detector they will show as additional uncertainty
in time-of-flight measurements. Within a quantum system, they will
act as forces of ``anticipation and regret'': causing interactions
to start sooner and last longer than otherwise would be the case.

But by the initial estimate they are expected small, of order attoseconds
or less. There are several ways to address this problem:
\begin{enumerate}
\item Statistical approaches are implicit in the figure of merit. 
\item Scattering through a crystal is another approach; all of the atoms
of the crystal can act in combination to achieve an effect. Therefore
we can try sending a beam through a time crystal \cite{Shapere:2012wz,Wilczek:2012uq,Else:2016aa,Giergiel:2018uq,Sacha:2018aa,Sacha:2018tm,Khemani:2019aa,Nakatsugawa:2017aa,Niemi:2021vm};
look for diffraction effects specifically from TQM.
\end{enumerate}

\paragraph{Dramatic}

Diffraction effects, especially those associated with the Heisenberg
uncertainty principle in time/energy, would appear to present the
most dramatic possibilities. In SQM a single slit in time \emph{clips}
the wave function in time: the narrower the gate, the less the dispersion
in time-of-flight. But in TQM a single slit in time \emph{diffracts}
the wave function in time: the narrower the gate, the greater the
dispersion in time-of-flight. The effects go in opposite directions,
so the contrast may, in principle, be set \emph{arbitrarily} great.

\paragraph{Subtle}

While it is natural to propose tests using SQM wave functions as carrier,
the TQM part as signal; by the fundamental hypothesis a wave function
is always to be understood as fully entangled in time and space. Several
lines of attack arise out of this including:
\begin{enumerate}
\item Effects of duality: 
\begin{enumerate}
\item The single particle Lagrangian we started with (equation \ref{eq:4D-Lagrangian-1})
is symmetric under the interchange of $t,\Phi\leftrightarrow x,{A_{x}}$.
So the effects of sending a charged particle though a time-varying
electric field are dual to sending a charged particle through a space-varying
magnetic field. For instance we can start with the Aharonov-Bohm effect
with respect to magnetic fields \cite{Aharonov:1959jd} and ask if
the Aharonov-Bohm effect with respect to the electric field \cite{Weder:2011-Aharonov-Bohm}
has possibilities.
\item In general, Maxwell's equations are symmetric under an interchange
of electric and magnetic fields, of the time-space and space-space
components of the Maxwell stress tensor $F_{\mu\nu}$. We can look
for effects associated with this. 
\item And of course we can look at any experiment in space and ask if we
interchange $t\leftrightarrow x$, do any interesting possibilities
present themselves? 
\end{enumerate}
\item Effects of anti-symmetry in time:
\begin{enumerate}
\item For bosons: create a wave function which is anti-symmetric in both
time and space parts, then look for anti-symmetry under reflection
solely in the space part.
\item For fermions: create a wave function which is anti-symmetric in the
time part but symmetric in the space part, then look for symmetry
under reflection solely in the space part.
\item And in both cases we can look directly for anti-symmetry in time.
\end{enumerate}
\item We can look at EPR effects in time, Bell's theorem tests in time,
Greenberger--Horne--Zeilinger experiments, and so on.
\item We can look for the effects of tunneling in time, which may have interesting
practical applications in steganography (the art of concealing the
true message within an apparently innocent one).
\end{enumerate}

\subsubsection{Variations on existing fundamental tests of quantum mechanics}

Another, bottom up, line of attack is to take existing lists of foundational
tests of quantum mechanics, to see if there is an ``in time'' variation
for specific tests. Possible starting points are Lamoreaux \cite{Lamoreaux:1992nh},
Ghose \cite{Ghose:1999zy}, and the three hundred or so experiments
detailed in Auletta \cite{Auletta:2000vj}. Examples are the ``single
slit in time'' (as in the text), the ``double slit in time'' (as
Lindner op cit), and so on.

And of course any experiment that explicitly mentions time, i.e. the
Delayed Choice Quantum Eraser \cite{Ma:2014aa,Marlow:1978xy} is a
possible starting point.

\subsubsection{What if TQM is confirmed but with qualifications?}

This in our own opinion is the most likely result. We are pushing
classical mechanics in a quantum direction, historically a profitable
line of attack. And we are making aggressive use of established principles,
also often effective. 

But we are extrapolating from a 3D theory to a 4D one, from shadow
to substance. Any such extrapolation has ambiguities, whether done
by a CAT scan or a philosopher.

The figure of merit provides a convenient way to categorize deviations
from the simple extrapolation proposed by TQM. The points of maximal
deviation provide a natural guide for followup experiments.

\subsubsection{What if TQM is falsified?}

On the other hand, if TQM is falsified, then the falsification itself
should suggest further experimental possibilities. The most obvious
falsification would be that we do not in fact see the dispersion in
time/energy that is the central prediction. There are two main possibilities
here: 
\begin{enumerate}
\item There is a frame in which the violation is maximal. This would be
a preferred frame, anathema to relativity, and of great interest to
the contrarian experimenter. 
\item The absence is uniform across frames. The restriction of the paths
to on-shell paths is confirmed. This would perhaps not be entirely
in the spirit of relativity. But it would at a minimum help clarify
the relationship between the HUP in energy/time and the HUP in space/momentum.
For instance it would show how to transform between these two in one
frame and in another. As the precise relationship between these two
has been the subject of a considerable literature (e.g.\cite{Heisenberg:1930kb,Aharonov:1961ft,Hilgevoord:1996bh,Hilgevoord:1998qu,Busch-2001,Aharonov:2005fg})
this would be interesting as well.
\end{enumerate}

\subsubsection{There are no null experimental results}

Therefore there are no null experiments. Either we will have a variety
of novel phenomena to explore or our understanding of the role of
time/energy in QED will be deepened.

\section{Discussion}
\begin{quotation}
``Anything that is not compulsory is forbidden.'' -- Murray Gell-Mann
\cite{Gell-Mann:1956uw}
\end{quotation}
\label{sec:Discussion}

The main problem here has been to extend QED to include time while
keeping it consistent with all that has gone before. The approach
has been to use the path integral formulation but keep everything
but the paths themselves the same. We then extended the paths in a
way that is manifestly covariant. 

While there may be alternative ways to the same end, the requirement
of manifest covariance should cause the results to be consistent to
the first order of magnitude. Since the relevant scales, of attoseconds,
are now accessible by experiment the hypothesis that the wave function
should be extended in time is therefore falsifiable.

\paragraph{Evidence for confirmation}

Obviously no one writes a paper as long as this to define a hypothesis
that has no chance of being confirmed. We argue that the odds of it
being confirmed are perhaps even better than that given various advantages:
\begin{enumerate}
\item The treatment of the time/energy and space/momentum coordinates is
manifestly symmetric. Simply being able to do this is interesting.
\item We have a clear explanation of why time normally appears asymmetric
at the level of the observer (due to statistical effects at the scale
of Avogadro's number) while still at the particle level being completely
symmetric.
\item We have a treatment which goes smoothly from the single to the multiple
particle cases.
\item We do not see the ultra-violet divergences. We still have to normalize
the loops, but we no longer have to regularize them: that drops out
of the formalism.
\item And we have Gell-Mann's principle: what is not forbidden is compulsory.
If there is not a conservation principle or symmetry rule forbidding
dispersion in time, it would be surprising \emph{not} to see dispersion
in time.
\end{enumerate}

\paragraph{Implications of falsification}

As noted, if we falsify TQM we would at a minimum get a clearer understanding
of the relationship between the HUP in energy/time and the HUP in
space/momentum, especially with regard to the way they transform from
one frame to another.

\paragraph{Implications of confirmation}

High speed chemical and biological interactions, i.e. attosecond scale,
should show effects of time dispersion. For instance, if molecules
can sense into the future, it may affect their ability to find optimal
configurations. 

There are potential applications for quantum communication and quantum
computing. With TQM we have an additional channel to use for calculation/communication
but also an additional channel to act as a source of decoherence.

The implications for quantum gravity are particularly interesting:
with manifest covariance, elimination of the ultra-violet divergences,
some recent work by Horwitz, and earlier work by Verlinde, we appear
to have all the pieces needed to construct a complete, covariant,
and convergent theory of quantum gravity. Leaving the question of
the odds of this being correct to one side, we note that recent advances
in technique mean such a theory has a reasonable chance of being falsifiable
as well. We explore this in slightly more detail in the  \ref{subsec:Implications-for-quantum-gravity}.

\paragraph{Conclusion}

Any time dependent quantum phenomena viewed at a sufficiently short
time scale (attoseconds or less) and with sufficiently time sensitive
detectors should either display novel phenomena along the time/energy
axis or at a minimum deepen and make more precise our understanding
of the role of time in quantum mechanics.

\section*{Acknowledgments}

\label{sec:Acknowledgments}

I thank my long time friend Jonathan Smith for invaluable encouragement,
guidance, and practical assistance.

I thank Ferne Cohen Welch for extraordinary moral and practical support.

I thank Martin Land, L. P. Horwitz, James O'Brien, Tepper Gill, Petr
Jizba, Matthew Trump, and the other organizers of the International
Association for Relativistic Dynamics (IARD) 2018, 2020, and 2022
Conferences for encouragement, useful discussions, and hosting talks
on the papers in this series in the IARD conference series. I also
thank my fellow participants in the 2022 conference -- especially
Pascal Klag, Bruce Mainland, Luca Smaldone, Howard Perko, Philip Mannheim,
Alexey Kryukov, Ariel Edery, and others -- for many excellent questions
and discussions.

I thank Steven Libby for several useful conversations and in particular
for insisting on the extension of the original ideas to the high energy
limit and therefore to QED.

I thank Larry Sorensen for many helpful references. I thank Ashley
Fidler for helpful references to the attosecond physics literature.

I thank Avi Marchewka for an interesting and instructive conversation
about various approaches to time-of-arrival measurements.

I thank Hou Yau for an interesting discussion of variations on these
themes and for directing my attention to his paper\cite{Yau:2021ab}.

I thank Asher Yahalom for insisting that a better explanation of the
clock time than simply what clocks measure was required

I thank Thomas Cember for useful clarifications of several points
in the argument.

I thank the reviewer who drew my attention to Horwitz's work on gravity
\cite{Horwitz:2018aa}.

I think Danko Georgiev of the journal Quanta for very practical suggestions
and advice.

I thank Y. S. Kim for organizing the invaluable Feynman Festivals,
for several conversations, and for general encouragement and good
advice.

I thank Catherine Asaro, Julian Barbour, Gary Bowson, Howard Brandt,
Daniel Brown, Ron Bushyager, John G. Cramer, J. Ferret, Robert Forward,
Fred Herz, J. Peřina, Linda Kalb, A. Khrennikov, David Kratz, Andy
Love, Walt Mankowski, O. Maroney, John Myers, Paul Nahin, Marilyn
Noz, R. Penrose, Stewart Personick, V. Petkov, H. Price, Matt Riesen,
Terry Roberts, J. H. Samson, Lee Smolin, L. Skála, Arthur Tansky,
R. Tumulka, Joan Vaccaro, L. Vaidman, A. Vourdas, H. Yadsan-Appleby,
S. Weinstein, and Anton Zeilinger for helpful conversations over the
years.

I thank the organizers of several QUIST, DARPA, Perimeter Institute
conferences I've attended and the very much on topic conferences \textit{Quantum
Time} in Pittsburgh in 2014 and \textsl{Time and Quantum Gravity}
in San Diego in 2015.

And none of the above are in any way responsible for any errors of
commission or omission in this work.

\appendix

\section{Notation}

\label{sec:Notations}

We are using natural units throughout. 

Since the text usually alternates between SQM and TQM sections, the
meaning of an object should often be clear from context. In general
SQM objects will have three vectors (i.e. $\vec{p}$) as arguments
and TQM objects will have four vectors (i.e. $p$) as arguments. Where
necessary we use a superscript $S$ to mark an SQM object, a superscript
$T$ to mark a purely time object, and absence of an explicit mark
to indicate a fully  relativistic (i.e. TQM object). For example:

\begin{equation}
\begin{gathered}{\psi_{\tau}}\left({t,\vec{x}}\right)=\psi_{\tau}^{T}\left(t\right)\psi_{\tau}^{S}\left({\vec{x}}\right)\hfill\\
{\psi_{\tau}}\left({E,\vec{p}}\right)=\psi_{\tau}^{T}\left(E\right)\psi_{\tau}^{S}\left({\vec{p}}\right)\hfill
\end{gathered}
\end{equation}
In paper A we used an over-bar and over-tilde for the same markings,
but we had to abandon that usage because it conflicted with the use
of an over-bar to mark adjoint spinors in the Dirac equation. We still
use the occasional over-bar to indicate average as $\bar{E}\equiv\left\langle {\sqrt{{m^{2}}+{\vec{p}^{2}}}}\right\rangle $.
We are putting the clock time at the bottom right in the position
of a traditional index, in honor of its frequent use as an index in
the time-slicing used to compute path integrals. To further streamline
the frequent references to clock time we replace $\tau_{1}$ with
just the $1$. And if we are looking at differences between two times
we just put the two indexes in:

\begin{equation}
{K_{{\tau_{1}}}}\to{K_{1}};{K_{{\tau_{2}};{\tau_{1}}}}\to{K_{2;1}}\to{K_{21}}
\end{equation}
The complementary variable for clock time is always $\omega$; the
complementary variable for coordinate time may be $E,w,k_{0},p_{0}$
depending on context. It is usually obvious when a particular object
refers to momentum or to coordinate space. When it might not be obvious,
we use an over-hat to mark the momentum space form, i.e. $\hat{\phi}\left(p\right)$
for a four dimensional plane wave or ${{\hat{\sigma}}_{x}}\equiv{\sigma_{{p_{x}}}}$
for the dispersion in $p_{x}$. We use the Greek letter $\varpi$
for the clock energy/clock frequency (energy and frequency are the
same in natural units of course):

\begin{equation}
{\varpi_{k}}\equiv-\frac{{{w^{2}}-{{\vec{k}}^{2}}-{\mu^{2}}}}{{2w}}
\end{equation}
We define ${E_{\vec{p}}}$  as the relativistic mass ${E_{\vec{p}}}\equiv\sqrt{{m^{2}}+{{\vec{p}}^{2}}}$. 

We use a superscript $A$ to tag specific propagators as Attosecond,
primarily meant for use at attosecond scales (\ref{eq:spinzero-atto-prop}).
We use $\phi$ for plane waves, $\varphi$ for Gaussians, and $\psi$
for general wave functions.

\section{Gaussian Test Functions}

\label{sec:Gaussian-Test-Functions}
\begin{quote}
“That’s a great deal to make one word mean,” Alice said in a thoughtful
tone. 

“When I make a word do a lot of work like that,” said Humpty Dumpty,
“I always pay it extra.” 

-- Lewis Carroll \emph{Through the Looking Glass} 1871 
\end{quote}

\subsection{Uses of GTFs}

\label{subsec:Uses-of-GTFs}

By Gaussian test functions (GTFs) we mean functions of the general
form:

\begin{equation}
{\varphi_{0}}\left(x\right)=\sqrt[4]{{\frac{1}{{\pi{\sigma^{2}}}}}}{e^{\imath{p_{0}}x-\frac{{{\left({x-{x_{0}}}\right)}^{2}}}{{2{\sigma^{2}}}}}}
\end{equation}
We generally take them as normalized to one. We refer to the $\sigma$
as the dispersion. The uncertainty in the associated dimension is
given by the dispersion divided by $\sqrt{2}$:

\begin{equation}
\Delta x\equiv\sqrt{\left\langle {{\left({x-\left\langle x\right\rangle }\right)}^{2}}\right\rangle }=\frac{{\sigma}}{{\sqrt{2}}}
\end{equation}
We can get a rough approximation of any normalizable wave function
by using the GTF with the same uncertainty:

\begin{equation}
\varphi_{0}^{\left(\text{TYPICAL}\right)}\left(x\right)\equiv\sqrt[4]{{\frac{1}{{2\pi{{\left({\Delta x}\right)}^{2}}}}}}{e^{\imath\left\langle {p_{0}}\right\rangle x-\frac{{{\left({x-\left\langle {x_{0}}\right\rangle }\right)}^{2}}}{{4{{\left({\Delta x}\right)}^{2}}}}}}
\end{equation}

\begin{enumerate}
\item We are making this use of GTFs when we estimate the initial wave functions
using the entropic estimate of the uncertainty in subsection \ref{subsec:Birth}.
This corresponds loosely to a statistician's use of the mode to summarize
a population.
\item By using Morlet wavelet analysis, we can represent any normalizable
wave function as a sum over Gaussians (see \cite{Ashmead:2012kx}).
\item Since the GTFs in momentum space are exact solutions of the various
free equations in TQM (and in SQM for that matter) we can use them
as starting functions in perturbation expansions.
\item A final and perhaps surprising use here is that the use of normalizable
functions -- whether typical GTFs or sums over GTFs -- is critical
for ensuring convergence of path integrals in general and loop diagrams
in particular. 
\end{enumerate}

\subsection{Starting GTFs}

\label{subsec:Sample-GTFs}

\subsubsection{In momentum and space}

In $p_{x}$:

\begin{equation}
{{\varphi}_{0}}\left({p_{x}}\right)=\sqrt[4]{{\frac{1}{{\pi\sigma_{{p_{x}}}^{2}}}}}{e^{-\imath\left({{p_{x}}-p_{x}^{0}}\right){x_{0}}-\frac{{{\left({{p_{x}}-p_{x}^{0}}\right)}^{2}}}{{2\sigma_{{p_{x}}}^{2}}}}}
\end{equation}
In $x$:

\begin{equation}
{{\varphi}_{0}}\left(x\right)=\sqrt[4]{{\frac{1}{{\pi\sigma_{x}^{2}}}}}{e^{\imath{p_{0}}x-\frac{{{\left({x-{x_{0}}}\right)}^{2}}}{{2\sigma_{x}^{2}}}}}
\end{equation}
where ${\sigma_{{p_{x}}}}=\frac{1}{{\sigma_{x}}}$. We can simplify
slightly by taking $\hat{\sigma}_{x}\equiv\sigma_{{p_{x}}}$. The
$y,z$ GTFs are the same, replacing $x\rightarrow y,z$ and $p_{x}\rightarrow p_{y},p_{z}$.
In momentum space:

\begin{equation}
\varphi_{0}^{S}\left({\vec{p}}\right)=\sqrt[4]{{\frac{1}{{{\pi^{3}}\det\left({{\hat{\Sigma}}^{S}}\right)}}}}{e^{-\imath\left({\vec{p}-{{\vec{p}}_{0}}}\right)\cdot{{\vec{x}}_{0}}-\left({\vec{p}-{{\vec{p}}_{0}}}\right)\cdot\frac{1}{{2{{\hat{\Sigma}}^{S}}}}\cdot\left({\vec{p}-{{\vec{p}}_{0}}}\right)}},{{\hat{\Sigma}}^{S}}\equiv\left({\begin{array}{ccc}
{\hat{\sigma}_{x}^{2}} & 0 & 0\\
0 & {\hat{\sigma}_{y}^{2}} & 0\\
0 & 0 & {\hat{\sigma}_{z}^{2}}
\end{array}}\right)
\end{equation}
and in coordinate space:

\begin{equation}
\varphi_{0}^{S}\left({\vec{x}}\right)=\sqrt[4]{{\frac{1}{{{\pi^{3}}\det\left({\Sigma^{S}}\right)}}}}{e^{\imath{{\vec{p}}_{0}}\cdot\vec{x}-\left({\vec{x}-{{\vec{x}}_{0}}}\right)\cdot\frac{1}{{2{\Sigma^{S}}}}\cdot\left({\vec{x}-{{\vec{x}}_{0}}}\right)}},{\Sigma^{S}}=\left({\begin{array}{ccc}
{\sigma_{x}^{2}} & 0 & 0\\
0 & {\sigma_{y}^{2}} & 0\\
0 & 0 & {\sigma_{z}^{2}}
\end{array}}\right)=\frac{1}{{{\hat{\Sigma}}^{S}}}
\end{equation}

\subsubsection{In energy and time}

\label{subsec:gtf-In-time-energy}

We get the time and energy forms taking $p_{x}\to E,x\to t$ and complex
conjugating. If we take the wave function in energy as:
\begin{equation}
{{\varphi}_{0}}\left(E\right)\equiv\sqrt[4]{{\frac{1}{{\pi\sigma_{E}^{2}}}}}{e^{\imath\left({E-{E_{0}}}\right){t_{0}}-\frac{{{\left({E-{E_{0}}}\right)}^{2}}}{{2\sigma_{E}^{2}}}}}
\end{equation}
we have the wave function in time as:

\begin{equation}
{{\varphi}_{0}}\left(t\right)\equiv\sqrt[4]{{\frac{1}{{\pi\sigma_{t}^{2}}}}}{e^{-\imath{E_{0}}t-\frac{{{\left({t-{t_{0}}}\right)}^{2}}}{{2\sigma_{t}^{2}}}}}
\end{equation}
where ${\sigma_{E}}=\frac{1}{{\sigma_{t}}}$. 

\subsubsection{In time/energy and space/momentum}

\label{subsec:gtf-time-space}

We can get four dimensional wave functions by taking the direct product
of the wave functions in time/energy and space/momentum. In four momentum
space:

\begin{equation}
\varphi_{0}\left({E,\vec{p}}\right)=\varphi_{0}^{T}\left(E\right)\varphi_{0}^{S}\left({\vec{p}}\right)\label{eq:gtf-energy-momentum}
\end{equation}
spelled out:

\begin{equation}
{\varphi_{0}}\left(p\right)=\sqrt[4]{{\frac{1}{{{\pi^{4}}\det\left({\hat{\Sigma}}\right)}}}}{e^{\imath{\left({p-{p_{0}}}\right)_{\mu}}x_{0}^{\mu}-{{\left({p-{p_{0}}}\right)}^{\mu}}\frac{1}{{2{{\hat{\Sigma}}^{\mu\nu}}}}{{\left({p-{p_{0}}}\right)}^{\nu}}}},{{\hat{\Sigma}}^{S}}\equiv\left({\begin{array}{cccc}
{{\hat{\sigma}}_{t}} & 0 & 0 & 0\\
0 & {\hat{\sigma}_{x}^{2}} & 0 & 0\\
0 & 0 & {\hat{\sigma}_{x}^{2}} & 0\\
0 & 0 & 0 & {\hat{\sigma}_{z}^{2}}
\end{array}}\right)\label{eq:sample-tqm-gtf}
\end{equation}
In coordinate space:

\begin{equation}
\varphi_{0}\left({t,\vec{x}}\right)=\varphi_{0}^{T}\left(t\right)\varphi_{0}^{S}\left({\vec{x}}\right)
\end{equation}
spelled out:

\begin{equation}
{\varphi_{0}}\left(x\right)=\sqrt[4]{{\frac{1}{{{\pi^{4}}\det\left(\Sigma\right)}}}}{e^{-\imath p_{0}^{\mu}{x_{\mu}}-{{\left({x-{x_{0}}}\right)}^{\mu}}\frac{1}{{2{\Sigma^{\mu\nu}}}}{{\left({x-{x_{0}}}\right)}^{\nu}}}},\Sigma=\left({\begin{array}{cccc}
{\sigma_{t}} & 0 & 0 & 0\\
0 & {\sigma_{x}^{2}} & 0 & 0\\
0 & 0 & {\sigma_{x}^{2}} & 0\\
0 & 0 & 0 & {\sigma_{z}^{2}}
\end{array}}\right)=\frac{1}{{\hat{\Sigma}}}
\end{equation}
We are treating time and space as disentangled. We can entangle by
generalizing the dispersion matrix $\Sigma$ to be an arbitrary positive
definition matrix. 

\subsection{GTFs as a function of clock time}

\label{subsec:GTFs-clock-time}

We look at the evolution of the GTFs as a function of clock time.

\subsubsection{Momentum space}

\label{subsec:gtf-Momentum-space}

\paragraph{Non-relativistic and SQM GTFs}

The behavior of the momentum space GTFs is simple. In the non-relativistic
case:

\begin{equation}
\varphi_{\tau}^{NR}\left({\vec{p}}\right)=\exp\left({-\imath\frac{{{\vec{p}}^{2}}}{{2m}}\tau}\right)\varphi_{0}^{S}\left({\vec{p}}\right)
\end{equation}
The GTFs in SQM are similar:

\begin{equation}
\varphi_{\tau}^{S}\left({\vec{p}}\right)={e^{-\imath\sqrt{{m^{2}}+{{\vec{p}}^{2}}}\tau}}\varphi_{0}^{S}\left({\vec{p}}\right)
\end{equation}
If the dispersion in three space momenta is not too great we may write:

\begin{equation}
\sqrt{{m^{2}}+{{\vec{p}}^{2}}}\approx{{\bar{E}}_{0}}+\frac{{{\left({\vec{p}-{{\vec{p}}_{0}}}\right)}^{2}}}{{2{E_{0}}}},{{\bar{E}}_{0}}\equiv\sqrt{{m^{2}}+\vec{p}_{0}^{2}},{{\vec{p}}_{0}}\equiv\left\langle {\vec{p}}\right\rangle 
\end{equation}
Note we are taking the average relativistic mass ${\bar{E}}_{0}$
as the reference point, not the bare mass $m$. This means that the
utility of the approximation can survive to much greater velocities;
all that is required is that the dispersion of the momentum be small
relative to the average momentum. As ${\bar{E}}_{0}\to m$ we get
the non-relativistic form, modulo a constant and therefore uninteresting
overall factor of $\exp\left({-\imath{{\bar{E}}_{0}}\tau}\right)$.
Note also that the diagonal form of the clock time dependence is a
result of using direct product GTFs in the three space dimensions. 

For TQM the behavior is, of course, a bit more complex. We have:

\begin{equation}
{\varphi_{\tau}}\left(p\right)=\exp\left({\imath\frac{{{E^{2}}-{{\vec{p}}^{2}}-{m^{2}}}}{{2E}}\tau}\right){\varphi_{0}}\left(p\right)=\exp\left({\imath\frac{{{E^{2}}-{{\vec{p}}^{2}}-{m^{2}}}}{{2E}}\tau}\right)\varphi_{0}^{T}\left(E\right)\varphi_{0}^{S}\left({\vec{p}}\right)
\end{equation}
So even though the energy and the three momentum parts start disentangled,
they become entangled as a result of the $\frac{{{\vec{p}}^{2}}}{{2E}}$
term. For now, we will deal with this by again assuming that the dispersions
in energy/momentum are not that great, so that it makes sense to write
$\delta E=E-{{\bar{E}}_{0}}$ and therefore:

\begin{equation}
-\frac{{{E^{2}}-{{\vec{p}}^{2}}-{m^{2}}}}{{2E}}={\varpi_{p}}=-\frac{{{{\left({{{\bar{E}}_{0}}+\delta E}\right)}^{2}}-{{\left({{{\vec{p}}_{0}}+\delta\vec{p}}\right)}^{2}}-{m^{2}}}}{{2\left({{{\bar{E}}_{0}}+\delta E}\right)}}\approx-\delta E+\frac{{{{\left({\delta E}\right)}^{2}}-2\vec{p}\cdot\delta\vec{p}-{{\left({\delta\vec{p}}\right)}^{2}}}}{{2{{\bar{E}}_{0}}}}
\end{equation}
We can then divide the TQM GTF into the energy part:

\begin{equation}
{\varphi_{\tau}}\left(E\right)=\sqrt[4]{{\frac{1}{{\pi\sigma_{E}^{2}}}}}\exp\left({\imath\delta E\left({\tau+{t_{0}}}\right)-\imath\frac{{{\left({\delta E}\right)}^{2}}}{{2{{\bar{E}}_{0}}}}\tau-\frac{{{\left({\delta E}\right)}^{2}}}{{2\sigma_{E}^{2}}}}\right)
\end{equation}
and momentum space part:

\begin{equation}
\varphi_{\tau}^{TQM}\left({\vec{p}}\right)=\exp\left({\imath\frac{{2\vec{p}\cdot\delta\vec{p}+{{\left({\delta\vec{p}}\right)}^{2}}}}{{2{{\bar{E}}_{0}}}}\tau}\right)\varphi_{0}^{S}\left({\vec{p}}\right)
\end{equation}
The space part in TQM does not evolve in quite the same way as the
space part in SQM; it seems better in practice to compare SQM as a
whole with TQM as a whole.

\subsubsection{Coordinate space forms}

\label{subsec:gtf-Spacetime}

\paragraph{Non-relativistic GTFs}

We start with the non-relativistic form:

\begin{equation}
\varphi_{\tau}^{NR}\left({\vec{x}}\right)=\varphi_{\tau}^{NR}\left(x\right)\varphi_{\tau}^{NR}\left(y\right)\varphi_{\tau}^{NR}\left(z\right)
\end{equation}
We look at the $x$ direction \cite{Merzbacher:1998tc,Schulman:1981um}:

\begin{equation}
\varphi_{\tau}^{NR}\left(x\right)=\sqrt[4]{{\frac{1}{{\pi\sigma_{x}^{2}}}}}\sqrt{\frac{1}{{f_{\tau}^{\left(x\right)}}}}{e^{\imath p_{x}^{\left(0\right)}x-\frac{1}{{2\sigma_{x}^{2}f_{\tau}^{\left(x\right)}}}{{\left({x-{x_{\tau}}}\right)}^{2}}-\imath\frac{{p_{x}^{\left(0\right)2}}}{{2m}}\tau}},f_{\tau}^{\left(x\right)}=1-\imath\frac{\tau}{{m\sigma_{x}^{2}}}
\end{equation}
with average position in $x$:

\begin{equation}
{x_{\tau}}={x_{0}}+v_{x}^{\left(0\right)}\tau,v_{x}^{\left(0\right)}=\frac{{p_{x}^{\left(0\right)}}}{m}
\end{equation}
We have $y,x$ the same. The corresponding probability density is:

\begin{equation}
\rho_{\tau}^{NR}\left({\vec{x}}\right)=\rho_{\tau}^{NR}\left(x\right)\rho_{\tau}^{NR}\left(y\right)\rho_{\tau}^{NR}\left(z\right)
\end{equation}
And again focusing on the $x$ direction:

\begin{equation}
\rho_{\tau}^{NR}\left(x\right)=\sqrt{\frac{1}{{\pi\sigma_{x}^{2}\left({1+\frac{{\tau^{2}}}{{{m^{2}}\sigma_{x}^{4}}}}\right)}}}\exp\left({-\frac{{{\left({x-{x_{\tau}}}\right)}^{2}}}{{\sigma_{x}^{2}\left({1+\frac{{\tau^{2}}}{{{m^{2}}\sigma_{x}^{4}}}}\right)}}}\right)
\end{equation}
Notice the kink in behavior at the point where $\frac{{\tau}}{{{m}\sigma_{x}^{2}}}\approx1$.
At this turning point, the uncertainty goes from being proportional
to $\sigma$ to being proportional to $\tau/\sigma$.

\begin{equation}
{\left({\Delta x}\right)^{2}}\equiv\left\langle {x_{\tau}^{2}}\right\rangle -{\left\langle {x_{\tau}}\right\rangle ^{2}}=\frac{{\sigma_{x}^{2}}}{2}\left|{1+\frac{{\tau^{2}}}{{{m^{2}}\sigma_{x}^{4}}}}\right|
\end{equation}
So if the dispersion starts small it will end large. This may be understood
in terms of the HUP in space/momentum. If the initial uncertainty
in space is small, the uncertainty in momentum goes as $1/\sigma_{x}$
so is correspondingly large. Given a bit of time (clock time here)
the large dispersion in momentum causes the wave function to spread
out in space, creating correspondingly large dispersion in space.
We may think of this as diffraction at work. The behavior for the
SQM GTFs is essentially the same.

\paragraph{TQM GTFs}

\label{par:TQM-GTFs}

The time part is in close parallel to the $x$ part:

\begin{equation}
\varphi_{\tau}^{T}\left(t\right)=\sqrt[4]{{\frac{1}{{\pi\sigma_{t}^{2}f_{\tau}^{t}}}}}{e^{\imath{E_{0}}{t_{0}}-\frac{{{(t-{t_{\tau}})}^{2}}}{{2\sigma_{t}^{2}f_{\tau}^{t}}}}},f_{\tau}^{t}\equiv1+\frac{{\imath\tau}}{{{E_{0}}\sigma_{t}^{2}}}
\end{equation}
with:

\begin{equation}
{t_{\tau}}={t_{0}}+v_{t}\tau,v_{t}=\gamma=\frac{E_{0}}{m}
\end{equation}
Probability density:

\begin{equation}
{\rho_{\tau}^{T}}\left(t\right)=\sqrt{\frac{1}{{\pi\sigma_{t}^{2}\left({1+\frac{{\tau^{2}}}{{{E_{0}^{2}}\sigma_{t}^{4}}}}\right)}}}\exp\left({-\frac{{{\left({t-t_{\tau}}\right)}^{2}}}{{\sigma_{t}^{2}\left({1+\frac{{\tau^{2}}}{{{E_{0}^{2}}\sigma_{t}^{4}}}}\right)}}}\right)
\end{equation}
Again note the shift in uncertainty in time at the point where $\frac{\tau}{E_{0}\sigma_{t}^{2}}\approx1$.

\begin{equation}
{\left({\Delta t}\right)^{2}}\equiv\left\langle {t^{2}}\right\rangle -{t_{\tau}^{2}}=\frac{{\sigma_{t}^{2}}}{2}\left|{1+\frac{{\tau^{2}}}{{{E_{0}^{2}}\sigma_{t}^{4}}}}\right|
\end{equation}
For TQM, the HUP in time/energy is fully equivalent to the HUP for
space/momentum.

\paragraph{TQM Kernels}

\label{par:TQM-Kernels}

In the analysis of the mass loop correction we will work directly
with the kernels in the narrow beam approximation. We ignore the normalization
factor of $\frac{1}{{2{\bar{E}_{0}}}}$ here. We developed this in
some detail in paper A. In momentum space:

\begin{equation}
{{\hat{K}}_{\tau}}\left({p;p'}\right)={\delta^{\left(4\right)}}\left({p-p'}\right)\exp\left({-\imath\frac{{{E^{2}}-{{\vec{p}}^{2}}-{m^{2}}}}{{2{\bar{E}_{0}}}}\tau}\right)
\end{equation}
and in coordinate space:

\begin{equation}
{K_{\tau}}\left({x;x'}\right)=-\imath\frac{{\bar{E}_{0}^{2}}}{{4{\pi^{2}}{\tau^{2}}}}{e^{-\imath{\bar{E}_{0}}\frac{{{\left({t-{t^{\prime}}}\right)}^{2}}}{{2\tau}}+\imath{\bar{E}_{0}}\frac{{{\left({\vec{x}-\vec{x}'}\right)}^{2}}}{{2\tau}}-\imath\frac{m}{2}\tau}}
\end{equation}

\section{Feynman rules}

\label{sec:Feynman-rules}

In the text we develop the Feynman rules for a few simple cases; here
we show we can extend the Feynman rules to all orders. This is not
a given. The problem is that in most treatments of the path integral,
the Hamiltonian part acts like a kind of locomotive pulling the sum
over paths forward. With TQM there is no natural equivalent to the
Hamiltonian, since there is no dependence of the Lagrangian on clock
time and therefore no canonical momenta and therefore no non-trivial
Hamiltonian. 

To see this in the simplest case, consider the Lagrangian for a free
spin zero massive particle above (equation \ref{eq:spin-zero-sqm-lag}).
The obvious choice for the corresponding TQM Lagrangian is:

\begin{equation}
\mathcal{L}\left[{\phi}\right]=\frac{1}{2}{\partial_{t}}\phi{\partial_{t}}\phi-\frac{1}{2}\nabla\phi\nabla\phi-{\frac{m^{2}}{2}}{\phi^{2}}\label{eq:feynman-tqm-lag}
\end{equation}
In fact our requirement of complete covariance does not give us any
real alternatives. The derivatives with respect to time must be with
respect to coordinate time if they are to form part of a four vector
with the $\nabla$ operator. We can write the action as either the
integral of $d^{4}x$ or the integral of $d\tau d^{4}x$ over this.
But we still have no dependence on clock time in the Lagrangian. Therefore
this Lagrangian goes not give us an equation of motion. It has no
kinetic energy term; it is all potential. Therefore while we can usually
develop path integrals with the Hamiltonian or the Lagrangian approaches,
with TQM we \emph{must} use the Lagrangian formulation.

Other major differences are that:
\begin{enumerate}
\item Most 3D objects are promoted to 4D.
\item The treatment of clock time has a different character; it still orders
the Dyson series and the like, but most of the dependence on time
is via the dependence on coordinate time.
\item We are primarily interested in interactions over short times: the
effects we are interested in are of order attoseconds. Over picoseconds
-- or still more glacial intervals -- the effects of time dispersion
are likely to be averaged out.
\end{enumerate}

\subsection{Dyson series}

\label{subsec:Dyson-series}

\subsubsection{Dyson series in SQM}

\label{subsec:vacuum-dyson-sqm}

We start with the familiar case of the Dyson expansion for the $S$
matrix in SQM, in the interaction picture:

\begin{equation}
{S_{fi}}=\left\langle f\right|\exp\left({-\imath\int\limits _{{\tau_{i}}}^{{\tau_{f}}}{d\tau\int{d\vec{x}}{\mathcal{H}_{I}}}}\right)\left|i\right\rangle \label{eq:action expansion}
\end{equation}
with interaction Hamiltonian:

\begin{equation}
{H_{I}}\equiv\int{d\vec{x}}{\mathcal{H}_{I}}\label{eq:interaction Hamiltonian}
\end{equation}
We expand the exponential in a power series:

\begin{equation}
{S_{fi}}=\sum\limits _{n=0}^{\infty}{\frac{{{\left({-\imath}\right)}^{n}}}{{n!}}\int\limits _{{\tau_{i}}}^{{\tau_{f}}}{d{\tau_{1}}\int\limits _{{\tau_{i}}}^{{\tau_{f}}}{d{\tau_{2}}\ldots\int\limits _{{\tau_{i}}}^{{\tau_{f}}}{d{\tau_{n}}}}}}{H_{I}}\left({\tau_{1}}\right){H_{I}}\left({\tau_{2}}\right)\ldots{H_{I}}\left({\tau_{n}}\right)
\end{equation}
and then time order the individual terms, which disposes of the $1/n!$:

\begin{equation}
{S_{fi}}=\sum\limits _{n=0}^{\infty}{{{\left({-\imath}\right)}^{n}}\int\limits _{{\tau_{i}}}^{{\tau_{f}}}{d{\tau_{1}}\int\limits _{{\tau_{i}}}^{{\tau_{f}}}{d{\tau_{2}}\ldots\int\limits _{{\tau_{i}}}^{{\tau_{f}}}{d{\tau_{n}}}}}}T\left\{ {{H_{I}}\left({\tau_{1}}\right){H_{I}}\left({\tau_{2}}\right)\ldots{H_{I}}\left({\tau_{n}}\right)}\right\} 
\end{equation}
We then use Wick's theorem to replace the difficult-to-work-with time
ordered terms with normal ordered terms plus ``contractions'' --
the Feynman propagators. The sum over all topologically distinct ways
of doing this is the sum over all Feynman diagrams.

The handling of clock time requires particular attention in this context.
For instance at each vertex, the integrals over three space induce
in a natural way a $\delta$ function in three momentum. The usual
$\delta$ function over clock energy is produced by the taking the
$S$ matrix to run from $\tau=-\infty\to+\infty$. This guarantees
conservation of clock energy, treated as the fourth component of a
four vector, at each vertex and also for the $S$ matrix overall.

But this conservation law is purchased at the expense of limiting
the domain of applicability in a serious way. If we want to apply
the theory to short times (as in this investigation) we have to accept
that conservation of laboratory energy may be approximate. So we have
an unnatural limitation of the domain of applicability of the theory
or else have to take our chances with conservation of laboratory energy.
In the text we are able to finesse the problem by focusing on the
interaction of individual wave packets. They are set to interact for
only short periods of time, so we can let the limits of clock time
integrals go to $\pm\infty$ without changing the numeric value of
the integral. But the problem remains in the general case.

\subsubsection{Dyson series in TQM}

\label{subsec:vacuum-dyson-tqm}

Wick's theorem is general, applying equally to SQM and TQM. It can
be used in TQM in the same way as in SQM, letting the time ordered
elements in the Dyson series be expressed as sums over products of
normal ordered operators and of propagators. The topology and symmetry
of the two series are the same. 

However the individual objects go in general from 3D to 4D. The initial
and final wave functions go from 3D plane waves to 4D and the propagators
change as described in the text. At each vertex we go from an integral
over clock time and three space dimensions to an integral over clock
time, coordinate time, and the three space dimensions. The coordinate
time and the three space dimensions are part of a four vector, the
clock time is treated as a separate object (but really the time coordinate
of the rest frame of the vacuum, per discussion).

For the most part the effect is as if we had simply added a 4th space
dimension to each object in the series, one that comes in with opposite
sign with respect to the three space dimensions. With this it is possible
to write out each term in the Dyson series and solve for any specified
problem to any required order, with one further change. We will assume,
as with the SQM case, that we are working in the interaction picture.
Therefore we need only consider the interaction potential; the rest
of the dependence on clock time is contained in the basis states. 

The TQM Dyson series for the $S$ matrix is then:

\begin{equation}
{S_{fi}}=\left\langle f\right|\exp\left({\imath\int\limits _{{\tau_{i}}}^{{\tau_{f}}}{d\tau\int{dtd\vec{x}}{\mathcal{L}_{I}}}}\right)\left|i\right\rangle 
\end{equation}
The interaction Lagrangian is:

\begin{equation}
{\mathcal{L}_{I}}=-V_{I}\label{eq:interaction Lagrangian}
\end{equation}
so:

\begin{equation}
\mathcal{H}=-\mathcal{L}=V_{I}={\mathcal{H}_{I}}\label{eq:interaction Hamiltonian-1}
\end{equation}
The two sign flips of the potential cancel out and we are left with
the same arguments for the exponential, albeit with four dimensions
to integrate over rather than three:

\begin{equation}
\int{d\vec{x}}{\mathcal{H}_{I}}\to\int{dtd\vec{x}}{\mathcal{H}_{I}}\label{eq:3D to 4D}
\end{equation}

\subsection{Feynman rules in TQM}

\label{subsec:Feynman-rules-TQM}

Therefore we can write the Feynman rules for TQM down by inspection,
with the substitutions noted. The basic topology is unchanged, the
symmetry factors are unchanged. The external factors are adjusted
per discussion; the external $\delta$ functions go from three in
three momentum plus clock energy to four in four momentum plus clock
energy.

\subsubsection{S matrix}

\label{subsec:feynman-rules-S-matrix}

The $S$ matrix in SQM is:

\begin{equation}
{S_{fi}}={\delta_{fi}}+{{\left({2\pi}\right)}^{4}}\delta\left(\Omega_{f}-\Omega_{i}\right){\delta^{3}}\left({{\vec{P}_{f}}-{\vec{P}_{i}}}\right)\left({\prod\limits ^{\begin{subarray}{l}
{\text{all external}}\\
{\text{bosons}}
\end{subarray}}{\sqrt{\frac{1}{{2V\omega_{\vec{k}}}}}}}\right)\left({\prod\limits ^{\begin{subarray}{l}
{\text{all external}}\\
{\text{fermions}}
\end{subarray}}{\sqrt{\frac{m}{{VE_{\vec{p}}}}}}}\right)\mathcal{M^{S}}
\end{equation}
while the $S$ matrix in TQM is the formally similar:

\begin{equation}
{S_{fi}}={\delta_{fi}}+{{\left({2\pi}\right)}^{5}}\delta\left(\varpi_{f}-\varpi_{i}\right){\delta^{4}}\left({{P_{f}}-{P_{i}}}\right)\left({\prod\limits ^{\begin{subarray}{l}
{\text{all external}}\\
{\text{bosons}}
\end{subarray}}{\sqrt{\frac{1}{{2TVw}}}}}\right)\left({\prod\limits ^{\begin{subarray}{l}
{\text{all external}}\\
{\text{fermions}}
\end{subarray}}{\sqrt{\frac{m}{{TVE}}}}}\right)\mathcal{M}
\end{equation}
In both cases the amplitude $\mathcal{M}$ is a sum over all topologically
distinct terms:

\begin{equation}
\mathcal{M}=\sum\limits _{n=0}^{\infty}{\mathcal{M}^{\left(n\right)}}
\end{equation}
In TQM the raw interaction vertex is unchanged:

\begin{equation}
-e{\gamma^{\mu}}
\end{equation}
The sign rules are the same as in SQM. For our purposes, writing out
the usual Feynman diagrams while adding an additional -1 for each
flip of identical fermions suffices.

\subsubsection{Incoming and outgoing wave functions}

\label{subsec:Incoming-and-outgoing}

The handling of the incoming and outgoing wave functions is different.
The incoming wave functions in SQM are generally taken as plane waves.
In TQM we have to use GTFs. In SQM both incoming and outgoing legs
are taken as on-shell. In TQM the 4D wave functions oscillate around
the on-shell values, now taken as averages rather than absolutes,
more guidelines than rules.

To go from SQM to TQM we replace the relativistic mass $E_{\vec{p}}$
in each spinor with the coordinate energy $E$, but leave them otherwise
unchanged. So each spinor becomes a function of all four components
of the momentum:
\begin{enumerate}
\item Initial electron: ${{u}_{s}^{S}}\left(\vec{p}\right){\to{u}_{s}}\left(p\right)$
\item Each final electron: ${{\bar{u}}_{s}^{S}}\left(\vec{p}\right)\to{{\bar{u}}_{s}}\left(p\right)$
\item Each initial positron: ${{\bar{v}}_{s}^{S}}\left(\vec{p}\right)\to{{\bar{v}}_{s}}\left(p\right)$
\item Each final positron: ${v_{s}^{S}}\left(\vec{p}\right)\to{v_{s}}\left(p\right)$
\end{enumerate}
The polarization vectors for the photons are unchanged. We change
the argument from a three vector to a four vector to indicate we are
changing the context.
\begin{enumerate}
\item Each initial photon: ${\varepsilon_{r,\mu}}\left(\vec{k}\right)\to{\varepsilon_{r,\mu}}\left(k\right)$
\item Each final photon: ${\varepsilon_{r,\mu}}\left(\vec{k}\right)\to{\varepsilon_{r,\mu}}\left(k\right)$
\end{enumerate}

\subsubsection{Propagators}

\label{subsec:feynman-rules-Propagators}

The propagators are different, as noted in some detail in the text.
The apparently covariant character of the SQM propagators falls apart
under close examination: there is dispersion in the $x,y,z$ directions
but not in $t$. As a result in SQM the intermediate particles are
virtual; in TQM they are real. 

Photon propagator in SQM (\ref{eq:photons-sqm-prop-k}):

\begin{equation}
\imath D_{\omega}^{\left(S\right)\mu\nu}\left({\vec{k}}\right)=\frac{{-{\imath g^{\mu\nu}}}}{{{\omega^{2}}-{{\vec{k}}^{2}}+\imath\varepsilon}}
\end{equation}
Fermion propagator in SQM (\ref{eq:fermions-sqm-prop-p}):

\begin{equation}
\imath S_{\omega}^{S}\left({\vec{x}}\right)=\imath\frac{{\omega{\gamma_{0}}-\vec{p}\cdot\vec{\gamma}+m}}{{{\omega^{2}}-{{\vec{p}}^{2}}-{m^{2}}+\imath\varepsilon}}
\end{equation}
The $\imath\epsilon$ is used to pick out the contours so that positive
frequencies are associated with forwards in time; negative with backwards
in time. In TQM, at least in this initial analysis, we have found
it helpful to break out the forward and backward components and deal
with them separately. 

Photon propagator in TQM (\ref{eq:photons-tqm-prop-k}):

\begin{equation}
\imath D_{\omega}^{\mu\nu}\left(k\right)=\frac{{-\imath{g^{\mu\nu}}}}{{2w}}\left({\frac{\imath}{{{w^{2}}-{{\vec{k}}^{2}}+2w\omega+2w\imath\varepsilon}}+\frac{\imath}{{{w^{2}}-{{\vec{k}}^{2}}+2w\omega-2w\imath\varepsilon}}}\right)
\end{equation}
with:

\begin{equation}
{\varpi_{k}}\equiv-\frac{{{w^{2}}-{{\vec{k}}^{2}}}}{{2w}}\label{eq:4D photon clock frequency-1}
\end{equation}
and short time limit:

\begin{equation}
\imath D_{\omega}^{\left({A}\right)\mu\nu}\left(k\right)=-\imath{g^{\mu\nu}}\frac{\imath}{{{w^{2}}-{{\vec{k}}^{2}}}}
\end{equation}
Feynman propagator for fermions (\ref{eq:fermions-tqm-prop-p}):

\begin{equation}
\imath{S_{\omega}}\left(p\right)=\imath\frac{{\left({\cancel{p}+m}\right)}}{{{E^{2}}-{{\vec{p}}^{2}}-{m^{2}}+2\omega E+2E\imath\varepsilon}}+\imath\frac{{\left({\cancel{p}+m}\right)}}{{{E^{2}}-{{\vec{p}}^{2}}-{m^{2}}+2\omega E-2E\imath\varepsilon}}
\end{equation}
with short time limit:

\begin{equation}
\imath S_{\omega}^{\left(A\right)}\left(p\right)=\imath\frac{{\cancel{p}+m}}{{{E^{2}}-{{\vec{p}}^{2}}-{m^{2}}}}
\end{equation}

\subsubsection{Vertexes}

\label{subsec:feynman-rules-Vertexes}

\paragraph{SQM}

\label{par:feynman-rules-Vertexes-SQM}

SQM vertex (\ref{eq:sqm-vertex}):

\begin{equation}
-e\bar{\psi}^{S}\left({\vec{p}'}\right)A^{\left(S\right)\nu}\left({\vec{k}}\right){\gamma_{\nu}}\psi^{S}\left({\vec{p}}\right)
\end{equation}
The vertex is accompanied by $\delta$ functions in the three space
momenta:

\begin{equation}
{\delta^{3}}\left({{{\vec{p}}_{out}}-{{\vec{p}}_{in}}}\right)
\end{equation}
and in clock energy if we are taking the limit as $\tau\pm\infty$:

\begin{equation}
\delta\left({\sum{\Omega_{out}}-\sum{\Omega_{in}}}\right)
\end{equation}

\paragraph{TQM}

\label{par:feynman-rules-Vertexes-TQM}

TQM uses the same vertex except for the obvious replacement of three
vector by four vector (\ref{eq:tqm-vertex}):

\begin{equation}
-e\bar{\psi}\left({p'}\right)A^{\nu}\left(k\right){\gamma_{\nu}}\psi\left({p}\right)
\end{equation}
The vertex is accompanied by $\delta$ functions in the four momenta:

\begin{equation}
{\delta^{4}}\left({{p_{out}}-{p_{in}}}\right)
\end{equation}
and in clock energy if we are taking the limit as $\tau\pm\infty$:

\begin{equation}
\delta\left({{\varpi_{out}}-{\varpi_{in}}}\right)
\end{equation}

\section{The rest frame of the vacuum}

\label{sec:rest-frame-vacuum}
\begin{quotation}
“If you look long enough into the void, the void begins to look back
through you.” -- Frederick Nietzsche
\end{quotation}

\subsection{Energy-momentum of spacetime}

\label{subsec:vacuum-at-rest}

In the development in the text we have to pick a specific laboratory
frame to define the clock time $\tau$; we have therefore dependence
on that choice. For laboratories going at non-relativistic velocities,
the corrections to clock time will result in a correction to a correction,
therefore not of first order, therefor not essential to falsifiability.

Still we would like to define TQM in a completely frame independent
way. We noted in paper A that we can make an invariant choice of frame
by taking advantage of an observation from Weinberg \cite{Weinberg:1972un}.
Consider the Einstein field equations: 
\begin{equation}
{G_{\mu\nu}}\equiv{R_{\mu\nu}}-\frac{1}{2}{g_{\mu\nu}}R=-8\pi G{T_{\mu\nu}}\label{eq:einstein field equation}
\end{equation}
Rewrite as: 
\begin{equation}
{\left({{G^{\mu\nu}}+8\pi G{T^{\mu\nu}}}\right)_{;\nu}}=0\label{eq:Einstein field equation compactly}
\end{equation}
We may use this to associate an energy momentum tensor with local
spacetime. Define: 
\begin{equation}
{g_{\mu\nu}}={\eta_{\mu\nu}}+{h_{\mu\nu}}\label{eq:g equal nu plus h}
\end{equation}
where $h_{\mu\nu}$ vanishes at infinity but is not assumed small.
The part of the Ricci tensor linear in $h$ is: 
\begin{equation}
R_{\mu\nu}^{\left(1\right)}\equiv\frac{1}{2}\left({\frac{{{\partial^{2}}h_{\lambda}^{\lambda}}}{{\partial{x^{\mu}}\partial{x^{\nu}}}}-\frac{{{\partial^{2}}h_{\mu}^{\lambda}}}{{\partial{x^{\lambda}}\partial{x^{\nu}}}}-\frac{{{\partial^{2}}h_{\nu}^{\lambda}}}{{\partial{x^{\lambda}}\partial{x^{\mu}}}}+\frac{{{\partial^{2}}{h_{\mu\nu}}}}{{\partial{x^{\lambda}}\partial{x_{\lambda}}}}}\right)\label{eq:Ricci linear in h}
\end{equation}
The exact Einstein equations may be written as: 
\begin{equation}
R_{\mu\nu}^{\left(1\right)}-\frac{1}{2}{\eta_{\mu\nu}}R_{\lambda}^{\left(1\right)\lambda}=-8\pi G\left({{T_{\mu\nu}}+{t_{\mu\nu}}}\right)\label{eq:exact Einstein equation expanded}
\end{equation}
where $t_{\mu\nu}$ is defined as quadratic in $h$ and higher:
\begin{equation}
{t_{\mu\nu}}\equiv\frac{1}{{8\pi G}}\left({{R_{\mu\nu}}-\frac{1}{2}{g_{\mu\nu}}R_{\lambda}^{\lambda}-R_{\mu\nu}^{\left(1\right)}+\frac{1}{2}{\eta_{\mu\nu}}R_{\lambda}^{\left(1\right)\lambda}}\right)\label{eq:energy momentum of spacetime}
\end{equation}
Weinberg then argues we may interpret $t_{\mu\nu}$ as the energy-momentum
of the gravitational field itself.

If we can associate an energy momentum with spacetime, we may define
a local rest frame with respect to that energy momentum tensor. We
will refer to this as $\mathcal{V}$, the rest frame of the vacuum.
If we define clock time with respect to this frame, we have an invariant
definition of the clock time. Presumably this invariant frame is in
free fall. A laboratory in Near Earth Orbit would do. More practically,
we can make the associated calculations to correct for our earth bound
laboratory being in an accelerated frame.

\subsection{The four dimensional Schrödinger equation in the rest frame of the
vacuum}

\label{subsec:vacuum-4D-Schr=0000F6dinger}

Weinberg was working in terms of spacetime, an essentially classical
concept. Now let us replace spacetime with a quantum vacuum. We will
assume that it is full of real (rather than virtual) particles in
a statistical ensemble.

With this context we return to the 4D Schrödinger equation:

\begin{equation}
\imath\frac{\partial}{{\partial\tau}}\psi=-\frac{{{p^{2}}-{m^{2}}}}{{2m}}\psi\label{eq:4D seqn}
\end{equation}
We define the energy momentum operators in the rest frame of the vacuum
as:

\begin{equation}
\mathcal{E}\equiv\imath\frac{\partial}{{\partial{\tau_{\mathcal{V}}}}},\overrightarrow{\mathcal{P}}\equiv-\imath{\vec{\nabla}_{\mathcal{V}}}\label{eq:rest energy of vacuum}
\end{equation}
with four momentum:

\begin{equation}
{\rm \mathcal{P}}=\left({{\rm \mathcal{E}},\vec{{\rm \mathcal{P}}}}\right)\label{eq:four momentum of vacuum}
\end{equation}
We are using capital script letters for values and operators associated
with the vacuum. We form the invariant ${\mathcal{Q}}^{\ensuremath{2}}$
for the invariant difference between the foreground and the vacuum
momentum:

\begin{equation}
{\mathcal{Q}^{2}}\equiv{\left({\mathcal{P}-p}\right)^{2}}\label{eq:vacuum invariant}
\end{equation}
Suppose that the Klein-Gordon equation should really not be written
as an absolute but as relative to the local spacetime. This is clearly
reasonable. Then the Klein-Gordon equation becomes, in an invariant
form:

\begin{equation}
{\mathcal{Q}^{2}}\psi={m^{2}}\psi\label{eq:vacuum KG equation}
\end{equation}
Any realistic quantum vacuum will be some sort of quantum soup of
photons and other bosons, electrons and other fermions. But at core
the free equations for all of these are the Klein-Gordon equation
with a light flavoring of spin/polarization added. We therefore take
the vacuum as obeying an averaged Klein-Gordon equation of its own:

\begin{equation}
\left({{\mathcal{P}^{2}}-{\mathcal{M}^{2}}}\right)\left|{\mathcal{V}}\right\rangle =0\label{eq:KG for vacuum}
\end{equation}
using $\mathcal{P},\mathcal{M}$ as conveniently vague local averages
over the mass-energy of the vacuum. Write the system of vacuum plus
particle as a direct product:

\begin{equation}
\left|\psi\right\rangle \left|\mathcal{V}\right\rangle \label{eq:direct product of vacuum and particle}
\end{equation}
We expand the Klein-Gordon equations:

\begin{equation}
\left({{\mathcal{P}^{2}}-2p\mathcal{P}+{p^{2}}-{m^{2}}-{\mathcal{M}^{2}}}\right)\left|\psi\right\rangle \left|{\mathcal{V}}\right\rangle =0\label{eq:combined-vacuum-particle-KG-eqn}
\end{equation}
The purely vacuum part cancels by assumption $\left\langle {{\mathcal{P}^{2}}-{\mathcal{M}^{2}}}\right\rangle \approx0$.
Further we choose to work in the rest frame of the vacuum so that
$\mathcal{P}\to\mathcal{E},\vec{0}$:

\begin{equation}
\left\langle {2E{\rm \mathcal{E}}}\right\rangle =\left\langle {{p^{2}}-{m^{2}}}\right\rangle {\text{ }}\label{eq:NR for vacuum-1}
\end{equation}
The energy of the vacuum $\mathcal{E}$ is, in coordinate space, given
by the time operator of the vacuum $\mathcal{E}\equiv\imath\frac{\partial}{{\partial{\tau_{\mathcal{V}}}}}$:

\begin{equation}
\imath\frac{\partial}{{\partial{\tau_{\mathcal{V}}}}}=\frac{{{p^{2}}-{m^{2}}}}{{2E}}\label{eq:NR for seqn for vacuum}
\end{equation}
Since we are looking at the difference between the energy operator
of the particle and the vacuum we therefore also need to look at the
difference between the energy operator of the laboratory and the vacuum
(the laboratory itself is after all nothing but particles). Therefore
the clock time or laboratory time is to be understood as the \emph{negative}
of the time operator of the vacuum:

\begin{equation}
\frac{\partial}{{\partial{\tau_{\mathcal{\mathcal{V}}}}}}=-\frac{\partial}{{\partial\tau}}\label{eq:vacuum kicks like frog}
\end{equation}
Therefore the correct 4D Schrödinger equation is (if we are in the
$\mathcal{V}$ frame):

\begin{equation}
{p^{2}}-{m^{2}}=-2E\imath\frac{\partial}{{\partial\tau}}\label{eq:seqn for particle in a vacuum}
\end{equation}
So the slow drift of the observed system's wave function with respect
to the observer is tracked by the cross-term of observed and observer
wave functions. The 4D Schrödinger equation reduces to the non-relativistic
form (equation \ref{eq:single-FS/T-x}) using $E\to m$. 

Note use of the Machian hypothesis has recovered the FS/T, which we
built starting from the single particle path integral approach. It
is striking that these two distinct approaches agree.

\subsection{Implications for quantum gravity}

\label{subsec:Implications-for-quantum-gravity}

In the text proper we treat the Machian hypothesis as a formal hypothesis,
useful for extending QED in time in a self-consistent way. However
if we are prepared to accept this hypothesis as physical, at least
for purpose of argument, there are some interesting implications for
quantum gravity. 

In the text proper we show in TQM we have:
\begin{enumerate}
\item a fully covariant treatment of time in QED,
\item and the elimination of the ultra-violet divergences.
\end{enumerate}
These are two of the principal barriers in the way of getting to quantum
gravity.

Further, as noted in the introduction, TQM is a part of the Relativistic
Dynamics program so TQM can draw on the extensive Relativistic Dynamics
literature. In particular we can take advantage of Horwitz's extension
of Relativistic Dynamics to General Relativity \cite{Horwitz:2018aa,Horwitz:2020vh,Horwitz:2021wd}. 

The Horwitz approach does not itself supply a mechanism. However,
consider the conventional practice of dropping disconnected diagrams
in Feynman diagrams. What if these terms should be seen not as disconnected
but as connected to the vacuum? 

Consider the mass terms in particular. They typically have a form
like:

\begin{equation}
\frac{{m^{2}}}{2}\left({{a_{k}}{a_{-k}}+{a_{k}}a_{k}^{\dag}+a_{k}^{\dag}{a_{k}}+a_{k}^{\dag}a_{-k}^{\dag}}\right)
\end{equation}
All four of the terms conserve momentum. The leftmost describes pair
annihilation; the rightmost pair creation. These are typically thrown
away as part of the process of throwing out ``disconnected diagrams''.
What unconscionable waste!

Perhaps a pair annihilation term is really describing two particles
descending to the vacuum, to flit round there for a short time, then
return as particles spontaneously appearing from the vacuum via the
pair creation term. The discarded terms might represent a kind of
quantum friction with the vacuum.

This is consistent with Weinberg's approach, providing a mechanism
for the exchange of energy/momentum between foreground and vacuum.
And consistent with Verlinde's entropic gravity approach \cite{Verlinde:2011aa,Verlinde:2017tj},
which encourages us to treat spacetime as itself a statistical system. 

Of course, these provide only ``elements'' of a theory of quantum
gravity. There may be zero, one, or multiple acceptable ways of combining
these elements.

To motivate such an effort consider this:
\begin{enumerate}
\item explains the hierarchy problem: ``quantum friction'' is a qualitatively
different mechanism; it is not surprising it would be much weaker
than the electromagnetic, weak, or strong interactions.
\item treats the mass in foreground and background the same way: the pair
creation term in the foreground is also a pair annihilation term in
the vacuum, and vice-versa. This suggests an interesting perspective
on the equivalence principle. 
\item resolves the information paradox \cite{Susskind:2008zo}: as spacetime
is ``nothing but'' the quantum vacuum, information can transfer
to it, hide out for a bit, then escape as during the process of black
hole evaporation or the like. Total information in foreground plus
background is still expected constant, as per the various no-cloning,
no-deleting theorems.
\item provides a mechanism by which gravity could act as a source of decoherence,
as in the Penrose Interpretation \cite{Ekert:1998aa,Penrose:1996aa,Penrose:2014aa}.
\end{enumerate}
But the primary advantage that such a theory naturally couples interactions
at the quantum scale with effects at the scale of the universe. This
opens up interesting experimental and observational possibilities. 

For instance the foreground (i.e. the particles we observe) should
be at a higher energy and less disordered than the vacuum. So we expect
a continuous transfer of energy and information from foreground to
vacuum over time. We expect this transfer would be a monotonically
increasing function of the energy of the foreground, more rapid within
a supernova than in the gaps between the stars.

If we take the zero-zero component of the mass-energy tensor of the
vacuum as providing a time scale, this transfer could look like an
expansion of the vacuum (as it acquires a greater proportion of the
total energy). To lowest order this might therefore look like a general
expansion of spacetime (continuously increasing dark energy).

Further if the increased energy in the vacuum is identified as dark
matter we can predict that the amount of dark matter in a galaxy will
be proportional to the time the galaxy has existed (i.e. small for
new borns) and to the mass of the galaxy.

Therefore, for essentially no work, we have two qualitative predictions
for the evolution of the universe. And we have an interesting line
of attack on the problem of quantum gravity. The associated technical
problems are obviously non-trivial (e.g. what does the total Lagrangian
look like?), but any well-defined and testable hypothesis has considerable
value. At a minimum, this should suggest interesting experiments,
especially now that we can see quantum effects of gravity (as in Bothwell
et al \cite{Bothwell:2022wz}). For candidate quantum gravity experiments
see for instance \cite{Christodoulou:2020aa,Huggett:2022uh,Polino:2022va,Christodoulou:2022ux,Marshall:2003ma}.

\bibliographystyle{hplain}
\bibliography{/Volumes/Isis/Projects/bibliography/taqm}

\end{document}